\documentclass[journal,hidelinks]{new-aiaa}
\usepackage{subfiles}
\usepackage[utf8]{inputenc}
\usepackage{graphicx}
\usepackage{amsmath}
\usepackage{siunitx}
\usepackage{longtable,tabularx}
\usepackage[caption=false]{subfig}
\usepackage[normalem]{ulem}
\usepackage{xcolor}
\usepackage{pdflscape}
\usepackage{chemmacros}
\usepackage{mathtools}
\usepackage{ulem}
\usepackage{xr}

\begin{document}

\title{Modified Marrone-Treanor model: parameterization and benchmarking for five-species air\footnote{Portions of this work were presented as paper 2023-3489 at the AIAA Aviation forum held in San Diego, CA} \\}

\author{Erik Torres\footnote{Research Associate; Now at Aero-Thermo-Mechanical Laboratory, Universit\'e Libre de Bruxelles, Avenue F.D. Roosevelt 50, B-1050 Brussels, Belgium; erik.matthias.torres@ulb.be (Corresponding author)}, Thomas Gross\footnote{Research Associate, AIAA member}, Graham V. Candler\footnote{McKnight Presidential Endowed Chair, AIAA Fellow} and Thomas E. Schwartzentruber.\footnote{Professor, AIAA Associate Fellow}}
\affil{Department of Aerospace Engineering and Mechanics, University of Minnesota\\107 Akerman Hall, 110 Union St SE, Minneapolis, MN, 55455}

\maketitle

\begin{abstract}
 We present updated parameters for five-species air ($\mathrm{N_2}$, $\mathrm{O_2}$, $\mathrm{NO}$, $\mathrm{N}$ and $\mathrm{O}$) reactions to be used with the Modified Marrone-Treanor two-temperature model. The vibrational relaxation and chemical reaction rates are derived from quasiclassical trajectory calculations and direct molecular simulations using ab initio potential energy surfaces. The resulting model enables efficient computational fluid dynamics simulations of nonequilibrium air chemistry in hypersonic flows. We show that the model reproduces direct molecular simulation benchmark solutions with high accuracy in zero-dimensional heat baths representative of strong nonequilibrium post-shock conditions. The model’s analytical expressions for dissociation rate coefficient and vibrational energy change per reaction ensure that the correct amount of energy is transferred between the vibrational and trans-rotational modes. Detailed balance is imposed for three-body recombination reactions and our simulations exhibit quasi-steady-state dissociation rates and proper approach to thermochemical equilibrium. In direct comparison with the Park $TT_\mathrm{v}$ model, the Modified Marrone-Treanor model predicts significantly slower conversion of $\mathrm{N_2}$ into $\mathrm{N}$ below $10\,000 \, \mathrm{K}$ and significantly more $\mathrm{NO}$ production at all temperatures. This is likely due to its significantly higher Zeldovich reaction rates compared to Park.
\end{abstract}

\section*{Nomenclature} \label{sec:nomenclature2}

{\renewcommand\arraystretch{1.0}
\noindent\begin{longtable*}{@{}l @{\quad=\quad} l@{} l@{}}
$a_U$ & fitting constant in MMT model $\, [-]$ \\
$\mathcal{A}_s$ & species $s$ chemical symbol in reaction equation $\, [-]$ \\
$C$ & modified Arrhenius pre-exponential factor coefficient $\, [L^3 \, T^{-1} \, N^{-1} \, \Theta^{-n}]$ \\
$c_0$ & Proportionality factor $= 1 \, \mathrm{atm \cdot s}$ \\
$c_\mathrm{p} / c_\mathrm{v}$ & constant-pressure / constant-volume heat capacity per unit mass  $\, [L^2 \, T^{-2} \, \Theta^{-1}]$ \\
$D$ & mixture set of diatomic species $\, [-]$ \\
$e$ & $\sum_{s \in S} \{ \rho_s / \rho \, e_s \}$: mixture thermal energy per unit mass $\, [L^2 \, T^{-2}]$ \\
$e_s$ & species $s$ thermal energy per unit mass $\, [L^2 \, T^{-2}]$ \\
$e_{\mathrm{v},s}$ / $e_{\mathrm{v-el},s}$ & species $s$ vibrational / vib.-electronic energy per unit mass $\, [L^2 \, T^{-2}]$ \\
$E$ & mixture total energy per unit volume $\, [M \, L^{-1} \, T^{-2}]$ \\
$E_a$ & per-mole activation energy in modified Arrhenius rate coefficient expression $\, [M \, L^2 \, T^{-2} \, N^{-1}]$ \\
$E_\mathrm{v}$ & mixture vibrational / vib.-electronic energy per unit volume $\, [M \, L^{-1} \, T^{-2}]$ \\
$f_k^\mathrm{NB}$ & rate coefficient non-Boltzmann factor $[-]$ \\
$f_\varepsilon^\mathrm{NB}$ & vibrational energy change non-Boltzmann factor $[-]$ \\
$h_s$ & species $s$ static enthalpy per unit mass $\, [L^2 \, T^{-2}]$ \\
$h_s^{\mathrm{f, \standardstate}}$ & species $s$ standard formation enthalpy per unit mass $\, [L^2 \, T^{-2}]$ \\
$\underline{I}$ & identity tensor $\, [-]$ \\
$k^\mathrm{Arr}$ & modified Arrhenius rate coefficient $[L^3 \, T^{-1} \, N^{-1}]$ \\
$k_\mathrm{B}$  & Boltzmann constant $= 1.38065 \times 10^{-23} \, \mathrm{J \cdot K^{-1}}$ \\
$k_{\mathrm{diss/rec}}$ & effective dissociation/recombination rate coefficient $\, [L^3 \, T^{-1} \, N^{-1}] / [L^6 \, T^{-1} \, N^{-2}]$ \\
$k_r^{f/b}$ & forward / backward rate coefficient for generic reaction $r$ $\, [(L^{3} \, N^{-1})^{(\nu_r^{f/b} - 1)} \, T^{-1} \, ]$ \\
$K_\mathrm{c}^\mathrm{eq}$ & molar concentration-based equilibrium constant $\, [(L^{-3} \, N)^{\nu_r^\mathrm{T}}]$ \\
$L$ & physical dimension of length \\
$M$ & physical dimension of mass \\
$M_s$ & species $s$ molar mass $\, [M \, N^{-1}]$ \\
$m_{s,q}^\mathrm{high/low}$ & high/low-temperature limit vibrational relaxation time fit slope parameter for species pair $s,q$ $\, [\Theta^{1/3}]$ \\
$n_{s,q}^\mathrm{high/low}$ & high/low-temperature limit vibrational relaxation time fit offset parameter for species pair $s,q$ $\, [-]$ \\
$N$ & physical dimension of amount of substance \\
$n$ & modified Arrhenius pre-exponential factor temperature exponent $\, [-]$ \\
$n$ & number density $\, [L^{-3}]$ \\
$p$ & mixture static pressure $\, [M \, L^{-1} \, T^{-2}]$ \\
$R$ & set of chemical reactions $\, [-]$ \\
$\bar{R}$ & universal gas constant $= 8314.5 \, \mathrm{Pa \cdot m^3 \cdot kmol^{-1} \cdot K^{-1}}$ \\ 
$\mathcal{R}_r$ & net rate per unit volume of reaction $r$ $\, [L^{-3} \, T^{-1} \, N]$ \\
$\vec{q}_\mathrm{t-r}$ & trans-rotational energy heat flux $\, [M \, T^{-3}]$ \\
$\vec{q}_\mathrm{v}$ & vibrational / vib.-electronic energy heat flux $\, [M \, T^{-3}]$ \\
$\tilde{Q}_{\mathrm{v},s}$ & approximate vibrational partition function of species $s$ $\, [-]$ \\
$S$ & mixture set of chemical species $\, [-]$ \\
$t$ & time $\, [T]$ \\
$T$ & physical dimension of time \\
$T$ & mixture trans.-rotational temperature in CFD calculation $\, [\Theta]$ \\
$T_{\mathrm{D},s}$ & species $s$ characteristic dissociation temperature $\, [\Theta]$ \\
$T_F$ & pseudotemperature in MMT model $\, [\Theta]$ \\
$T_\mathrm{r}$ & mixture rotational temperature in DMS calculation $\, [\Theta]$ \\
$T_\mathrm{t}$ & mixture translational temperature in DMS calculation $\, [\Theta]$ \\
$T_\mathrm{t-r}$ & mixture trans.-rotational temperature in DMS calculation $\, [\Theta]$ \\
$T_\mathrm{v}$ & mixture vibrational / vib.-electronic temperature $\, [\Theta]$ \\
$\vec{u}, \vec{v}_s$ & bulk and species $s$ diffusion velocities $\, [M \, L^{-1}]$ \\
$U, U^*$ & pseudotemperatures in MMT model $\, [\Theta]$ \\
$w_s$ & rate of formation of species $s$ mass per unit volume $\, [M \, L^{-3} \, T^{-1}]$ \\
$w_{\mathrm{v}}^\mathrm{relax}$ & vibration-translational energy relaxation source term $\, [M \, L^{-1} \, T^{-3}]$ \\
$w_{\mathrm{v}}^\mathrm{chem}$ & vibrational energy-chemistry coupling source term $\, [M \, L^{-1} \, T^{-3}]$ \\
$x_s$ & mole fraction of species $s$ $\, [-]$ \\
$Z$ & vibrational nonequilibrium rate factor $\, [-]$ \\
$\langle \varepsilon_{\mathrm{v},s} \rangle_r$ & per-molecule vibrational energy removed for species $s$ due to reaction $r$ $\, [M \, L^2 \, T^{-2}]$ \\
$\langle \Delta \varepsilon_{\mathrm{v},s} \rangle_r$ & per-molecule vibrational energy \emph{change} for species $s$ due to reaction $r$ $\, [M \, L^2 \, T^{-2}]$ \\
$\zeta_r$ & nonequilibrium concentration ratio of reaction $r$ $\, [-]$ \\
$\eta_\mathrm{O_2}$ & $\mathrm{O_2}$ multi-electronic-surface enhancement factor (= 16/3) $\, [-]$ \\
$\Theta_{\mathrm{v},s}$ & species $s$ characteristic temperature of vibration $\, [\Theta]$\\
$\Theta$ & physical dimension of temperature \\
$\nu_{s,r}^{f/b}$ & species $s$ stoichiometric coefficient in forward / backward sense of reaction $r$ $\, [-]$ \\
$\nu_r^{f/b}$ & stoichiometric coefficient sum in forward / backward sense of reaction $r$: $\nu_r^{f/b} = \sum_{s \in S} \{ \nu_{s,r}^{f/b}\}$ \, $\mathrm{[-]}$ \\
$\nu_r^\mathrm{T}$ & total difference of stoichiometric coefficients in reaction $r$: $\nu_r^\mathrm{T} = \nu_r^{b} - \nu_r^{f}$ \, $\mathrm{[-]}$ \\
$\rho$ & mixture density $\, [M \, L^{-3}]$ \\
$\rho_s$ & species $s$ density $\, [M \, L^{-3}]$ \\
$\langle \tau_s^\mathrm{v} \rangle$ & species $s$ mixture-averaged vibrational relaxation time $\, [T]$ \\
$\underline{\tau}$ & viscous stress tensor $\, [M \, L^{-1} \, T^{-2}]$ \\
$\tau_{s,q}^\mathrm{v}$ & species $s$ vibrational relaxation time with collision partner species $q$ $\, [T]$ \\
CFD & Computational Fluid Dynamics \\
DMS & Direct Molecular Simulation \\
DSMC & direct simulation Monte Carlo \\
MMT & Modified Marrone-Treanor \\
M\&W & Millikan and White \\
PES & potential energy surface \\
QCT & quasi-classical trajectory \\
QSS & quasi-steady state \\
\end{longtable*}}


\section{Introduction} \label{sec:introduction2}

\lettrine{T}{his} article presents the most complete parameterization of the Modified Marrone-Treanor (MMT) model~\cite{chaudhry25a} for coupled vibrational relaxation and dissociation in five-species air ($\mathrm{N_2}$, $\mathrm{O_2}$, $\mathrm{NO}$, $\mathrm{N}$ and $\mathrm{O}$) currently available. Models like this are needed for performing computational fluid dynamics (CFD) simulations of high-enthalpy flow, where air exists in a state of thermal and chemical nonequilibrium. 

For almost the entire time that hypersonic flow has been a subject of active study, the estimation of rate data for nonequilibrium air chemistry models has relied on ground tests, primarily in shock tubes. Such experiments are costly and, in complex mixtures such as partially dissociated air, the quantities of interest needed to calibrate model parameters may be affected by significant uncertainty. In practice, Park's two-temperature model~\cite{park90a} with its $\sqrt{T T_\mathrm{v}}$ dependence for the dissociation rate coefficient is still the most widely used. It may not have been derived from first-principles microscopic calculations, but its simple formulation and ready availability of rate parameters for relevant mixtures~\cite{park93a, park94a} makes its use in CFD codes rather straightforward.

In recent decades an alternative approach has emerged, which relies on computational chemistry methods to generate reaction rate data essentially devoid of empiricism. It begins with the construction of ab initio potential energy surfaces (PESs) that dictate the instantaneous forces between colliding molecules arbitrarily arranged in physical space. At present, upward of 20 distinct PESs have been developed to predict the dynamics between the most relevant collision partners in five-species air in their respective ground electronic states. These PESs were purpose-built for simulating inelastic and reactive collisions, because they explicitly allow for stretching, breaking and re-forming of chemical bonds between the colliding atoms. All PESs employed to arrive at the rate parameters presented in this work were originally generated by the University of Minnesota Computational Chemistry group~\cite{paukku13a, bender15a, lin16b, varga16a, paukku17a, varga17a, paukku18a, li20a, varga21a, varga21b} and can be found online~\cite{potlib21}.

Time integration of the classical equations of motion on these surfaces from an initial to a final molecular arrangement (here either 4-atom, or 3-atom systems) is referred to as quasiclassical trajectory (QCT) calculations~\cite{truhlar79a}. After a sweep is conducted over a suitable range of initial internal and relative collision energies (for instance, sampled from prescribed single-temperature, or $(T,T_\mathrm{v})$ Boltzmann distributions) and the post-collision states are analyzed, reaction rate coefficients and average internal energy change per reaction, can be estimated~\cite{bender15a, chaudhry18b}. Statistical noise inherent to the Monte-Carlo sampling of initial conditions is driven down by performing trajectory batches of larger size. In Ref.~\cite{chaudhry25a} an extensive QCT dataset for oxygen and nitrogen dissociation with $\mathrm{N_2}$, $\mathrm{O_2}$, $\mathrm{N}$ and $\mathrm{O}$ as collision partners was analyzed, resulting in the formulation and initial parameterization of the Modified Marrone-Treanor (MMT) dissociation model.

A different use-case for such interaction potentials is to embed them in the framework of the direct simulation Monte-Carlo~\cite{bird94a} (DSMC) algorithm during the particle collision phase. This effectively produces a gas-kinetic flow simulation method, wherein the standard DMSC models for molecular scattering, internal energy relaxation and chemistry have been replaced by on-the-fly trajectory calculations on potential energy surfaces. The idea was originally proposed by Koura~\cite{koura98b} to study rotational nonequilibrium across normal shocks in nitrogen and later adopted by Bruno et al.~\cite{bruno15a} to estimate transport properties in oxygen at moderate temperatures. But it took until the emergence of suitable ab initio PESs, together with sufficient computational power for this method to become a viable option to simulate the high-temperature nonequilibrium chemistry of nitrogen~\cite{valentini15a, valentini16a, grover23a}, oxygen~\cite{grover19a, grover19b, torres22b} and finally five-species air mixtures~\cite{torres24a, torres24b, valentini24b}. In its current form it is referred to as Direct Molecular Simulation~\cite{schwartzentruber18a} (DMS) and represents a method able to simulate high-temperature nonequilibrium reacting flows essentially from first principles, including non-Boltzmann chemistry effects. Reference~\cite{schwartzentruber18a} gives an overview of the method and presents several reference test cases.

Due to its high computational cost, the DMS method has been mostly employed to study space-homogeneous adiabatic and isothermal heat baths. However, such a geometrically simple setup is ideal for studying nonequilibrium chemistry of air isolated from any interference due to flow effects (e.g. species diffusion, viscous dissipation, etc.). In this work we use DMS solutions in two complementary ways. First, to propose and calibrate the non-Boltzmann correction factors in the MMT model (see Sec.~V in Ref.~\cite{chaudhry25a} and Sec.~\ref{sec:new_mmt_parameters} in this work) as well as derive characteristic vibrational relaxation time parameters for air mixture components (see Ref.~\cite{torres24b} and Sec.~\ref{sec:tau_vt_fits_again} in this work). Second, to produce benchmark test cases against which CFD solutions with the MMT model will be compared. Since the DMS benchmarks and MMT parameters are both derived from the same set of ab initio PESs, close agreement between solutions obtained with the two methods constitutes the main metric of success. This approach replaces traditional model validation, because nowhere in the process do we rely on experimental data.


This work follows Ref.~\cite{chaudhry25a} in which the MMT model formulation was first derived and parameterized for partially dissociated $\mathrm{N_2/N}$ and $\mathrm{O_2/O}$ mixtures. Furthermore, it is preceded by two articles~\cite{torres24a, torres24b} which presented new reaction rates and vibrational relaxation times derived from QCT and DMS calculations on all the Minnesota ab initio PESs for air species currently available. Therefore, this paper is best read in conjunction with the cited references. The two main objectives of this paper are to present the most extensive set of MMT parameters so far in a format suitable for CFD implementation and to benchmark said implementation against DMS reference solutions at conditions representative of shock-heated air in hypersonic flows. As described in Sec.~\ref{sec:fluid_equations}, only the species conservation and vibrational energy balance equations are directly affected by the nonequilibrium chemistry model. Therefore, simple isothermal and adiabatic heat bath calculations are sufficient for benchmarking the model’s behavior.

This paper is organized as follows. Section~\ref{sec:mmt_model2} recalls the MMT model formulation originally presented in Ref.~\cite{chaudhry25a} and provides the most up-to-date listing of parameters to calculate chemical reaction rates, vibrational energy change per reaction and characteristic vibrational relaxation times. This set includes the most important dissociation reactions in air, as well as updated rate parameters for the Zeldovich exchange reactions. As discussed in Refs.~\cite{torres24a, torres24b}, the currently available PES set enables QCT and DMS calculations for the most relevant collision pairs in 5-species air, but does not yet encompass all the permutations possible. In particular, the absence of ab initio surfaces for simulating $\mathrm{N_2 - NO}$ and $\mathrm{O_2 - NO}$ trajectories makes it impossible to generate dissociation rate data and characteristic vibrational relaxation times for these collision pairs. Furthermore, partial gaps also exist in the PES set necessary to fully describe $\mathrm{NO - NO}$, $\mathrm{NO-N}$, $\mathrm{NO-O}$ and $\mathrm{N_2 - O}$ vibrational relaxation. We discuss these limitations with regard to the MMT model parameter set for 5-species air in Secs.~\ref{sec:new_mmt_parameters} and \ref{sec:tau_vt_fits_again} together with our proposed workarounds. Then, in Sec.~\ref{sec:testing_vs_dms_again} we benchmark our CFD implementation of the MMT model against several DMS reference solutions. Special emphasis is placed on ensuring that the model parameters for thermodynamics, vibrational relaxation and air chemistry are directly derived from, or fully consistent with the ab initio PESs that DMS also relies on. Next, in Sec.~\ref{sec:mmt_nb_vs_vnb} we examine the effect of employing constant versus variable non-Boltzmann correction factors with the MMT model as the mixture approaches thermochemical equilibrium. The reasoning behind these adaptations was originally laid out in Sec.~VI of Ref.~\cite{chaudhry25a}. In Sec.~\ref{sec:dms_vs_real} we compare MMT predictions to assess the influence of a theorized multi-surface electronic energy enhancement factor for oxygen dissociation  (see pp. 331-334 of Ref.~\cite{nikitin74a}). Finally, in Sec.~\ref{sec:mmt_vs_park} we compare MMT predictions to those obtained with Park’s standard $TT_\mathrm{v}$ model~\cite{park90a} and in Sec.~\ref{sec:conclusions} we formulate conclusions for this work. Additional figures and details relevant for ensuring consistency between the model’s implementation in CFD with the DMS reference calculations are discussed in the supplemental information.

\section{Modified Marrone-Treanor model for air} \label{sec:mmt_model2}

In this section we summarize the MMT model formulation and present the most complete list of model parameters currently available for calculating its reaction rate coefficients, vibrational energy-chemistry coupling terms and vibrational energy relaxation times for five-species air.


\subsection{Governing fluid equations for a two-temperature chemistry model} \label{sec:fluid_equations}

The compressible, viscous fluid equations for five-species reacting air described by a two-temperature nonequilibrium chemistry model may be written as:
\begin{align}
 \frac{\partial\rho_s}{\partial t} & + \nabla \cdot \Bigl( \rho_s \vec{u} + \rho_s \vec{v}_s \Bigr) = w_s, \qquad s \in S, \label{eq:species_mass_balance} \\
 \nonumber \\
 \frac{\partial\rho \vec{u}}{\partial t} & + \nabla \cdot \Bigl( \rho \vec{u} \otimes \vec{u} + p \underline{I} - \underline{\tau} \Bigr) = \vec{0} \label{eq:momentum_balance} \\
 \nonumber \\
 \frac{\partial E}{\partial t} & + \nabla \cdot \Bigl( \left( E + p \right) \vec{u} - \underline{\tau} \vec{u} + \vec{q}_\mathrm{t-r} + \vec{q}_\mathrm{v} + \sum_{s \in S} \rho_s \, h_{s} \vec{v}_s \Bigr) = 0 \label{eq:total_energy_balance} \\
 \nonumber \\
 \frac{\partial E_\mathrm{v}}{\partial t} & + \nabla \cdot \Bigl( E_\mathrm{v} \, \vec{u} + \vec{q}_\mathrm{v} + \sum_{s \in S} \rho_s \, e_{\mathrm{v-el}, s} \vec{v}_s \Bigr) = w_\mathrm{v}^\mathrm{relax} + w_\mathrm{v}^\mathrm{chem}, \label{eq:vib_energy_balance}
\end{align}
with balances of species mass, overall momentum, total energy and vibrational-electronic energy represented by Eqs.~(\ref{eq:species_mass_balance}), (\ref{eq:momentum_balance}), (\ref{eq:total_energy_balance}) and (\ref{eq:vib_energy_balance}) respectively. The set $S = \{ \mathrm{N_2}, \mathrm{O_2}, \mathrm{NO}, \mathrm{N}, \mathrm{O} \}$ encompasses the five mixture components and $D = \{ \mathrm{N_2}, \mathrm{O_2}, \mathrm{NO} \}$ represents the set of diatomic species. Total energy density $E = \frac{1}{2} \rho |\vec{u}|^2 + \sum_{s \in S} \{ \rho_s \, e_s (T,T_\mathrm{v}) \}$ in Eq.~(\ref{eq:total_energy_balance}) is the sum of bulk flow kinetic energy and composition-averaged thermal\footnote{Here we refer to it as \emph{thermal} instead of \emph{internal} energy so as to not cause confusion with internal energy (rotational, vibrational, electronic) mode contributions.} energy per unit volume. We calculate the two-temperature species thermal energies as:
\begin{equation}
 e_s (T,T_\mathrm{v}) = \left\lbrace \begin{array}{l l}
                     h_s^\mathrm{fit} (T_\mathrm{v}) + \frac{5}{2} R_s \, T - \frac{7}{2} R_s \, T_\mathrm{v} & \text{for} \quad s = \mathrm{N_2}, \mathrm{O_2}, \mathrm{NO} \\
                     h_s^\mathrm{fit} (T_\mathrm{v}) + \frac{3}{2} R_s \, T - \frac{5}{2} R_s \, T_\mathrm{v} & \text{for} \quad s = \mathrm{N}, \mathrm{O}.
                    \end{array} \right.  \label{eq:thermal_energy}
\end{equation}

Note that only the diatomic species $s \in D$ contribute to vibrational energy, but at temperatures above a few thousand kelvin the electronic modes of all 5 species increasingly become excited and must be accounted for in the energy balances. In the present model both modes are assumed to remain equilibrated at all times and to relax together at the characteristic time scale of the vibrational mode. Therefore, in Eq.~(\ref{eq:thermal_energy}) $e_s (T,T_\mathrm{v})$ includes the combined vibrational and electronic energy of the diatomic species at mixture vibrational temperature $T_\mathrm{v}$, whereas for atomic ones it includes only the electronic mode contribution. Furthermore, for the diatomic species $e_s$ contains contributions from the translational and rotational energy modes at combined trans-rotational temperature $T$ and the standard heat of formation. For the monatomic species the translational mode at $T$ and the heat of formation are the remaining contributors to thermal energy. As indicated by Eq.~(\ref{eq:thermal_energy}), we leverage polynomial fits in the NASA Lewis~\cite{mcbride93a} format to compute the enthalpies $h_s^\mathrm{fit}$ at temperature $T_\mathrm{v}$ and swap out the trans-rotational/translational contributions at $T_\mathrm{v}$ for those at $T$. The contribution of the  heats of formation are included in $h_s^\mathrm{fit}$ and are thus automatically accounted for in Eq.~(\ref{eq:thermal_energy}). For further context on how we calculate the thermodynamic properties for the MMT model, refer to Secs.~S1-S3 of the supplemental information.

Finally, Eq.~(\ref{eq:vib_energy_balance}) represents the balance equation for the mixture's combined vibrational-electronic energy per unit volume $E_\mathrm{v} = \sum_{s \in S} \{ \rho_s \, e_{\mathrm{v-el},s} (T_\mathrm{v}) \}$.

Nonequilibrium chemistry directly influences the source terms of Eqs.~(\ref{eq:species_mass_balance}) and (\ref{eq:vib_energy_balance}). First, in the calculation of the species mass production rates on the right-hand side of Eq.~(\ref{eq:species_mass_balance}):
\begin{equation}
 w_s = M_s \sum_{r \in R} ( \nu_{s,r}^b - \nu_{s,r}^f ) \, \mathcal{R}_r, \qquad s \in S, \label{eq:species_source_term_again}
\end{equation}
where $M_s$ is the molar mass of species $s$, $R$ represents the set of reversible reactions, $\nu_{s,r}^f$ and $\nu_{s,r}^b$ represent the stoichiometric coefficients of species $s$ in reaction $r$ to the left and to the right of the reaction equation written in its generic form:
\begin{equation}
 \sum_{s \in S} \nu_{s,r}^f \, \mathcal{A}_{s} \xrightleftharpoons[k_r^b]{k_r^f} \sum_{s \in S} \nu_{s,r}^b \, \mathcal{A}_{s}, \qquad r \in R. \label{eq:reaction_equation_again}
\end{equation}

The factor $\mathcal{R}_r$ in Eq.~(\ref{eq:species_source_term_again}) represents net rate of every reaction $r$:
\begin{equation}
 \mathcal{R}_r = \left( k_r^f \prod_{q \in S} \left( \frac{\rho_q}{M_q} \right)^{\nu_{q,r}^f} - k_r^b \prod_{q \in S} \left( \frac{\rho_q}{M_q} \right)^{\nu_{q,r}^b} \right), \label{eq:net_rate_again}
\end{equation}
where stoichiometric coefficients of species not participating in the given reaction are set to zero and $k_r^f$ and $k_r^b$ represent the forward and backward rate coefficients when Eq.~(\ref{eq:reaction_equation_again}) is read in the left-to-right sense. 

Backward and forward rate coefficients are linked through the detailed balance relation $k_r^b = k_r^f / K_{\mathrm{c},r}^\mathrm{eq}$, where $K_{\mathrm{c},r}^\mathrm{eq}$ represents the equilibrium constant for reaction $r$. Section~S1 of the supplemental information provides further detail on how these equilibrium constants are evaluated for the reactions listed in Table~\ref{tab:mmt_2024_reactions}. For the dissociation-recombination reaction pairs whose dissociation rate coefficients are computed from the MMT functional form (reactions 1-12 and 14-15 in Table~\ref{tab:mmt_2024_reactions}), the forward (dissociation) rate coefficient is evaluated according to Eq.~(\ref{eq:mmt_rate_coefficient_again}) at the local $(T,T_\mathrm{v})$ temperature combination, whereas the backward (recombination) rate coefficient is evaluated as $k_\mathrm{rec} = k_\mathrm{diss}^\mathrm{MMT-NB} (T,T) / K_{\mathrm{c}, [\mathrm{diss-rec}]}^\mathrm{eq} (T)$ (see Sec.~VI of Ref.~\cite{chaudhry25a} for further details). For the remaining ``non-preferential'' reaction pairs (reactions 13 and 16-21 in Table~\ref{tab:mmt_2024_reactions}), the MMT functional form is not employed. In these cases, the forward rate coefficient is evaluated directly using Eq.~(\ref{eq:karr_again}) at the local trans-rotational temperature $T$, i.e. $k_\mathrm{non-pref}^f = k^\mathrm{Arr} (T)$. The exact same value is then used when applying the detailed balance relations to obtain the corresponding backward rate coefficient, i.e. $k_\mathrm{non-pref}^b = k^\mathrm{Arr} (T) / K_{\mathrm{c}, [\mathrm{non-pref}]}^\mathrm{eq} (T)$. In Sec.~\ref{sec:mmt_rates_formulation} we provide additional details on the functional form for the reaction rate coefficients and list the most up-to-date parameters in the MMT model for air.

The second aspect of the fluid equations directly affected by the nonequilibrium model are the two source terms on the right hand side of Eq.~(\ref{eq:vib_energy_balance}). The functional form of the term $w_\mathrm{v}^\mathrm{relax}$ is given by Eq.~(\ref{eq:vib_relax_equation_again}) and the parameters for calculating the associated vibrational relaxation rates are presented later in Sec.~\ref{sec:tau_vt_fits_again}. The other source term, $w_\mathrm{v}^\mathrm{chem}$, governs the conversion rate between vibrational-electronic and chemical potential energy. Its functional form is given by Eq.~(\ref{eq:vib_chem_source}).

None of the inviscid, or viscous fluxes in the Navier-Stokes equations are influenced by the choice of gas-phase chemistry model and it becomes most convenient to test the implementation of such models separate from any interfering flow effects. Thus, if one assumes zero spatial gradients for all state variables, Eqs.~(\ref{eq:species_mass_balance})-(\ref{eq:vib_energy_balance}) reduce to a set of ordinary differential equations. The expressions for momentum and total energy balances become trivial, i.e. $\mathrm{d} ( \rho \vec{u} ) / \mathrm{d} t = \vec{0}$ and $\mathrm{d} E / \mathrm{d} t = 0$ respectively, implying that $\rho \vec{u}$ and $E$ remain constant over time and need not be solved for explicitly as part of the remaining ODE system:
\begin{align}
 \frac{\mathrm{d} \rho_s}{\mathrm{d} t} & = w_s, \qquad s \in S, \label{eq:species_mass_balance_0d} \\
 \nonumber \\
 \frac{\mathrm{d} E_\mathrm{v}}{\mathrm{d} t} & = w_\mathrm{v}^\mathrm{relax} + w_\mathrm{v}^\mathrm{chem}. \label{eq:vib_energy_balance_0d}
\end{align}

These become the equations governing the time evolution of a constant-volume (isochoric), constant-total-energy (adiabatic) heat bath. This system of equations can be integrated in time for given initial conditions, typically specified in terms of an initial composition $\rho_{s, (t=0)}, \forall s \in S$ and initial temperatures $T_{(t=0)} \ne T_{v, (t=0)}$, which end up fixing the initial value of the total energy density $E = \frac{1}{2} \rho | \vec{u} |^2 + \sum_{s \in S} \{ \rho_s e_s (T_{(t=0)},T_{v, (t=0)}) \}$\footnote{Without loss of generality, we choose to set $\vec{u} = \vec{0}$ in all our heat bath calculations.} and the vibrational-electronic energy density $E_\mathrm{v} = \sum_{s \in S} \{ \rho_s e_{\mathrm{v-el},s} (T_{v, (t=0)}) \}$. On the other hand, when the constraint $\mathrm{d} E / \mathrm{d} t = 0$ is dropped in favor of imposing a fixed value for the translation-rotational temperature $T = T_{(t=0)}$ at all times, Eqs.~(\ref{eq:species_mass_balance_0d})-(\ref{eq:vib_energy_balance_0d}) describe the time evolution of a gas mixture in what we call a constant-volume (isochoric), constant-temperature (isothermal) heat bath. Both scenarios represent rather simplified and somewhat abstract nonequilibrium ``flows fields''. They are useful nonetheless, because they allow us to study the effect of the chemistry model in isolation. Testing the CFD implementation of the MMT model in this way is especially convenient, because the DMS benchmark solutions are obtained in exactly the same type of space-homogeneous (0D) isothermal or adiabatic heat baths.

\clearpage
\subsection{Formulation for dissociation rate coefficient in the MMT model} \label{sec:mmt_rates_formulation}

As shown in Ref.~\cite{chaudhry25a}, in the Modified Marrone-Treanor model two-temperature dissociation rate coefficients are evaluated as:
\begin{equation}
 k_\mathrm{diss}^\mathrm{MMT-NB} (T,T_\mathrm{v}) = k^\mathrm{Arr} (T) \, Z (T,T_\mathrm{v}) \, f_k^\mathrm{NB}, \label{eq:mmt_rate_coefficient_again}
\end{equation}
where $k^\mathrm{Arr} (T)$ represents the \emph{thermal equilibrium} reaction rate coefficient in modified Arrhenius form:
\begin{equation}
 k^\mathrm{Arr} (T) = C \, T^n\exp(-E_a/ \bar{R} T), \label{eq:karr_again}
\end{equation}
while $Z (T,T_\mathrm{v})$ is the MMT model's vibrational nonequilibrium factor, as derived by Marrone and Treanor~\cite{marrone63a}:
\begin{equation}
 Z(T,T_\mathrm{v}) = \frac{\tilde{Q}_{\mathrm{v},s}(T)\,\tilde{Q}_{\mathrm{v},s}(T_F)}{\tilde{Q}_{\mathrm{v},s}(T_\mathrm{v})\,\tilde{Q}_{\mathrm{v},s}(-U)}, \label{eq:Zfact_again}
\end{equation}
and $f_k^\mathrm{NB}$ is a non-Boltzmann correction factor discussed in more detail in Secs.~\ref{sec:non_boltzmann_factors} and \ref{sec:mmt_nb_vs_vnb} of this article, as well as in Secs.~V and VI of Ref.~\cite{chaudhry25a}. The different factors $\tilde{Q}_{\mathrm{v},s}$ appearing in Eq.~(\ref{eq:Zfact_again}) are approximate vibrational partition functions of the form:
\begin{equation}
 \tilde{Q}_{\mathrm{v},s}(T) = \frac{1 - \exp (-T_{\mathrm{D},s}/T) }{1 - \exp(-\Theta_{\mathrm{v},s}/T)},
\end{equation}
and $T_F$ and $U$ are pseudo-temperatures defined as:
\begin{equation}
 T_F (T,T_\mathrm{v}) = \left( \frac{1}{T_\mathrm{v}}-\frac{1}{T}-\frac{1}{U(T)} \right)^{-1}, \label{eq:TF_again}
\end{equation}
and:
\begin{equation}
 U(T) = \left( \frac{a_U}{T}+\frac{1}{U^*} \right)^{-1}, \label{eq:NewFit_U_again}
\end{equation}
with $a_U$ and $U^*$ as reaction-specific MMT parameters.


\subsection{Formulation of vibrational energy-chemistry coupling term in the MMT model} \label{sec:mmt_devib_formulation}

The net rate of change of vibrational-electronic energy due to chemical reactions is accounted for in the source term $w_\mathrm{v}^\mathrm{chem}$ on the right hand side of Eq.~(\ref{eq:vib_energy_balance}):
\begin{equation}
 w_\mathrm{v}^\mathrm{chem} = \sum\limits_{s \in S} \sum\limits_{r \in R} \mathcal{R}_r \left( \nu_{s,r}^b - \nu_{s,r}^f \right) \langle \varepsilon_{v,s} \rangle_r \label{eq:vib_chem_source}
\end{equation}

This source term collects energy loss/gain contributions from all species possessing vibrational-electronic energy (in our 5-species air model only the three diatomic species $\mathrm{N_2}, \mathrm{O_2}$ and $\mathrm{NO}$ contribute vibrational energy, but all 5 of them contribute electronic energy) for every reaction $r$ that they participate in. Note that by the convention followed in this work $\langle \varepsilon_{\mathrm{v},s} \rangle_r$ is always a non-negative quantity. However, the associated \emph{change} in vibrational energy of species $s$ due to reaction $r$, which we define as $\langle \Delta \varepsilon_{\mathrm{v},s} \rangle_r = ( \nu_{s,r}^b - \nu_{s,r}^f ) \langle \varepsilon_{\mathrm{v},s} \rangle_r$, \emph{does} account for the sign of vibrational energy production. It will be negative if vibrational energy is removed from the gas due to net destruction of species $s$ (by virtue of $\nu_{s,r}^b - \nu_{s,r}^f$ being negative), but positive if a net amount of species $s$ is being produced in the given reaction. The net reaction rate $\mathcal{R}_r$ is governed by the forward and backward rate coefficients through Eq.~(\ref{eq:net_rate_again}). Thus, in addition to the reaction rate coefficients, the nonequilibrium chemistry model must supply values for $\langle \varepsilon_{\mathrm{v},s} \rangle_r$ for every species and every reaction as a function of the local conditions, characterized by the local translational and/or vibrational temperature.

Modeling the $\langle \varepsilon_{\mathrm{v},s} \rangle_r$ term in a general fashion is not trivial, because the two main reaction types (dissociation, exchange) exhibit different degrees of vibrational energy-chemistry coupling. In previous work~\cite{torres24a} we have found that for the exchange processes vibrational energy removed/replenished per reaction roughly matches that of the diatomic species' local average in the gas, i.e:
\begin{equation}
 \langle \varepsilon_{v,s} \rangle_r^\mathrm{non-pref.} = e_{\mathrm{v},s} (T_\mathrm{v}) / M_s \label{eq:devib_nonpref}
\end{equation}

We refer to this as ``non-preferential'' vibrational energy-chemistry coupling. However, it is also known that this assumption is not well suited for dissociation reactions, which exhibit greater-than-thermal-average, or ``preferential'' removal of vibrational energy per reaction. Several forms for $\langle \varepsilon_{\mathrm{v},s} \rangle_\mathrm{diss}$ have been proposed over the years, in particular for use together with Park's reaction rates. In an early publication~\cite{park88a} the functional form $\langle \varepsilon_{\mathrm{v},s} \rangle_\mathrm{diss} = R \, ( T_{\mathrm{D}, s} - T)$ was proposed, where $T_{\mathrm{D}, s}$ is the dissociation temperature of the species in question. This formulation partially takes into account the local thermodynamic state of the gas through the translation-rotational temperature, but ignores the vibrational energy of the dissociating species. Sharma et al~\cite{sharma92a} proposed an even simpler expression $\langle \varepsilon_{\mathrm{v},s} \rangle_\mathrm{diss} = 0.3 \, R T_{\mathrm{D}, s}$, which does not take into account the local thermodynamic state of the gas at all. Neither of these functional forms reproduce the preferential vibrational energy-chemistry coupling of dissociation observed in our QCT/DMS studies~\cite{torres24a} particularly well. 

One major advantage of the MMT model is that it includes an analytical expression for $\langle \varepsilon_{\mathrm{v},s} \rangle_\mathrm{diss}$ as a function of the local nonequilibrium thermodynamic state together with the characteristic temperatures $\Theta_{\mathrm{v}, s}$ and $T_{\mathrm{D},s}$ of the dissociating species:
\begin{equation}
 \langle \varepsilon_{v,s} \rangle_r^\mathrm{MMT-NB} = f_\varepsilon^\mathrm{NB} \left( \frac{k_\mathrm{B} \, \Theta_{\mathrm{v},s}}{\exp( \Theta_{\mathrm{v},s}/T_F) - 1} - \frac{k_\mathrm{B} \, T_{\mathrm{D},s}}{\exp(T_{\mathrm{D},s}/T_F)-1} \right) \label{eq:devib_mmt}
\end{equation}

As discussed in Sec.~II.B of Ref.~\cite{chaudhry25a}, reaction-specific fit parameters $a_U$ and $U^{*}$ necessary to compute $\langle \varepsilon_{\mathrm{v},s} \rangle_\mathrm{diss} (T, T_\mathrm{v})$ can be extracted from QCT-derived estimates of the same quantity at thermal equilibrium (i.e. when $T\!=\!T_\mathrm{v}$). Furthermore, Eq.~(\ref{eq:devib_mmt}) contains another non-Boltzmann correction factor $f_{\varepsilon}^\mathrm{NB}$, which was introduced to account for the reduction in vibrational energy removed during the quasi-steady-state (QSS) dissociation phase. As will be shown next, this simple correction factor can be calibrated to reproduce $\langle \varepsilon_{\mathrm{v},s} \rangle_\mathrm{diss}$ observed in DMS calculations for $\mathrm{N_2}$, $\mathrm{O_2}$ and $\mathrm{NO}$ dissociation~\cite{torres24a} with a range of collision partners.

\subsection{Updated MMT parameters for air} \label{sec:new_mmt_parameters}

In this section we present the most up-to-date kinetic rate parameters to be used with the Modified Marrone-Treanor model. Note that what is referred to as ``the MMT model'' actually consists of multiple parts. First, the collection of Arrhenius rate parameters for 5-species air reactions fit to our most recent QCT calculations based on ab initio PESs. Second, all relevant expressions and fit parameters, i.e. $a_U, U^*$ in Eq.~(\ref{eq:NewFit_U_again}), necessary to compute the vibrational nonequilibrium factor $Z ( T, T_\mathrm{v})$ according to Eq.~(\ref{eq:Zfact_again}), average vibrational energy removed per dissociation $\langle \varepsilon_{\mathrm{v}, s} \rangle_\mathrm{diss}$ (see Eq.~(\ref{eq:devib_mmt}) in Sec.~\ref{sec:mmt_devib_formulation}) as well as the non-Boltzmann correction factors $f_k^\mathrm{NB}$ and $f_{\varepsilon}^\mathrm{NB}$ during QSS dissociation. Only when all these components are integrated within the MMT model can it accurately predict nonequilibrium dissociation in a post-shock environment. It should further be mentioned that the MMT model, as presented in Ref.~\cite{chaudhry25a}, was calibrated taking into account only a subset of all air reactions. This original dataset relied on QCT calculations by Bender et al.~\cite{bender15a} and Chaudhry et al.~\cite{chaudhry18b, chaudhry18d} for collision-induced $\mathrm{N_2}$-dissociation by impact with $\mathrm{N_2}$, $\mathrm{N}$ and $\mathrm{O_2}$, as well as $\mathrm{O_2}$-dissociation by impact with $\mathrm{N_2}$, $\mathrm{O}$ and $\mathrm{O_2}$. The Arrhenius and MMT parameters derived from the original calibration by Chaudhry et al. are listed for reference in Table~1 of Ref.~\cite{chaudhry25a}.

New, more complete QCT and DMS calculations~\cite{torres24a} incorporating the most up-to-date ab initio potentials complement these earlier results and, in some instances, now supersede them. Table~\ref{tab:mmt_2024_reactions} in this paper lists the updated reaction set, split into dissociation-type reactions for the three diatomic species (rows 1 - 15), exchange-type, among them the two Zeldovich reactions $\mathrm{N_2} + \mathrm{O} \rightleftharpoons \mathrm{NO} + \mathrm{N}$ and $\mathrm{NO} + \mathrm{O} \rightleftharpoons \mathrm{O_2} + \mathrm{N}$ (rows 16 - 18), plus three remaining reactions of ``mixed'' dissociation-exchange type. The second-to-last column indicates whether a given reaction's rate coefficient and vibrational energy removal are computed using the MMT expressions, i.e. Eqs.~(\ref{eq:mmt_rate_coefficient_again}) and (\ref{eq:devib_mmt}), or whether a thermal rate coefficient is assumed together with vibrationally non-preferential energy removal, i.e. the combination of Eqs.~(\ref{eq:karr_again}) and (\ref{eq:devib_nonpref}). The last column of Table~\ref{tab:mmt_2024_reactions} lists the provenance of the associated QCT data. The actual model parameters are listed in two separate tables. Our recommended modified Arrhenius parameters for all reactions are listed in Table~\ref{tab:arrhenius_2024_parameters_complete}, whereas MMT reaction-specific parameters are collected in Table~\ref{tab:mmt_2024_parameters_complete}. 

A careful comparison of both sub-tables in Table~1 of Ref.~\cite{chaudhry25a} with Tables~\ref{tab:arrhenius_2024_parameters_complete} and \ref{tab:mmt_2024_parameters_complete} in this paper reveals different numerical values for the fit parameters of all common reactions listed. For the ``pure'' nitrogen and oxygen dissociation reactions (i.e. reactions 1, 4, 7 and 10 in Table~\ref{tab:arrhenius_2024_parameters_complete}) this is a consequence of having re-fitted the original QCT data of Bender et al.~\cite{bender15a} and Chaudhry~\cite{chaudhry18d} to a two-parameter modified Arrhenius expression with dissociation temperatures ($T_D = E_a / \bar{R}$) held fixed to the standard values $T_{\mathrm{D}, \mathrm{N_2}} = 113\,200 \, \mathrm{K}$ and $T_{\mathrm{D}, \mathrm{O_2}} = 59\,330 \, \mathrm{K}$ respectively. The associated MMT parameters $a_U$ and $U^*$ in Table~\ref{tab:mmt_2024_parameters_complete} also differ slightly from those in Table~1 of Ref.~\cite{chaudhry25a}, because updated values for the characteristic vibrational temperatures ($\Theta_{v, \mathrm{N_2}} = 3414 \, \mathrm{K}$ and $\Theta_{v, \mathrm{O_2}} = 2251 \, \mathrm{K}$) were used in the latter. In addition to these minor changes it became necessary to update the Arrhenius and MMT parameters for both ``mixed'' $\mathrm{N_2}-\mathrm{O_2}$ dissociation reactions (i.e. compare reactions 2 and 6 in Tables~\ref{tab:arrhenius_2024_parameters_complete} and \ref{tab:mmt_2024_parameters_complete} vs. rows 3 and 6 of both sub-tables in Table~1 of Ref.~\cite{chaudhry25a}), because the $\mathrm{N_2O_2}$ potential was recently re-fitted (most recent PES source employed is \verb|N2O2_3A_MB-PIP-MEG2| vs. older source \verb|PES_N2O2_triplet_umn_v1| on potlib~\cite{potlib21}). This resulted in small, but still noticeable changes to the QCT rate coefficients relative the ones obtained by Chaudhry et al.~\cite{chaudhry18b} with the older version of the PES.


\begin{table}
 \centering
 
 \caption{Reaction types included in updated MMT chemistry model for air.}
 \label{tab:mmt_2024_reactions}
 
 \begin{tabular}{c c c c c}
     & Type  & Reaction                                                  & Model    & QCT rates source \\ \hline \hline 
   1 & diss  & $\mathrm{N_2  + N_2 \rightleftharpoons N   + N   + N_2}$  & MMT      & Bender et al~\cite{bender15a} \\
   2 & diss  & $\mathrm{N_2  + O_2 \rightleftharpoons N   + N   + O_2}$  & MMT      & Torres et al~\cite{torres24a} \\
   3 & diss  & $\mathrm{N_2  + NO  \rightleftharpoons N   + N   + NO}$   & re-use (1)~[a]   & no dedicated QCT data \\
   4 & diss  & $\mathrm{N_2  + N   \rightleftharpoons N   + N   + N}$    & MMT      & Bender et al\cite{bender15a} \\
   5 & diss  & $\mathrm{N_2  + O   \rightleftharpoons N   + N   + O}$    & MMT      & Torres et al~\cite{torres24a} (partial PESs) \\ \hline
   6 & diss  & $\mathrm{O_2  + N_2 \rightleftharpoons O   + O   + N_2}$  & MMT~[b]  & Torres et al~\cite{torres24a} \\
   7 & diss  & $\mathrm{O_2  + O_2 \rightleftharpoons O   + O   + O_2}$  & MMT~[b]  & Chaudhry et al~\cite{chaudhry18b} \\
   8 & diss  & $\mathrm{O_2  + NO  \rightleftharpoons O   + O   + NO}$   & re-use (7)~[a,b]  & no dedicated QCT data \\
   9 & diss  & $\mathrm{O_2  + N   \rightleftharpoons O   + O   + N}$    & MMT~[b]  & Torres et al~\cite{torres24a} \\
  10 & diss  & $\mathrm{O_2  + O   \rightleftharpoons O   + O   + O}$    & MMT~[b]  & Chaudhry et al~\cite{chaudhry18b} \\ \hline 
  11 & diss  & $\mathrm{NO   + N_2 \rightleftharpoons N   + O   + N_2}$  & re-use (13)~[a] & no dedicated QCT data \\
  12 & diss  & $\mathrm{NO   + O_2 \rightleftharpoons N   + O   + O_2}$  & re-use (13)~[a] & no dedicated QCT data \\
  13 & diss  & $\mathrm{NO   + NO  \rightleftharpoons N   + O   + NO}$   & thermal/non-pref.  & Torres et al~\cite{torres24a} \\
  14 & diss  & $\mathrm{NO   + N   \rightleftharpoons N   + O   + N}$    & MMT      & Torres et al~\cite{torres24a} (partial PESs) \\
  15 & diss  & $\mathrm{NO   + O   \rightleftharpoons N   + O   + O}$    & MMT      & Torres et al~\cite{torres24a} (partial PESs) \\ \hline
  16 & exch  & $\mathrm{N_2  + O   \rightleftharpoons NO   + N}$         & thermal/non-pref.  & Torres et al~\cite{torres24a} \\
  17 & exch  & $\mathrm{NO   + O   \rightleftharpoons O_2  + N}$         & thermal/non-pref.  & Torres et al~\cite{torres24a} \\
  18 & exch  & $\mathrm{N_2  + O_2  \rightleftharpoons 2NO}$             & thermal/non-pref.  & Torres et al~\cite{torres24a} \\ \hline
  19 & mixed & $\mathrm{NO   + NO  \rightleftharpoons O    + O  + N_2}$  & thermal/non-pref.  & Torres et al~\cite{torres24a} (partial PESs) \\
  20 & mixed & $\mathrm{NO   + NO  \rightleftharpoons N    + N  + O_2}$  & thermal/non-pref.  & Torres et al~\cite{torres24a} (partial PESs) \\ 
  21 & mixed & $\mathrm{N_2  + O_2 \rightleftharpoons NO + N + O}$       & thermal/non-pref.  & Torres et al~\cite{torres24a} \\ \hline
  \multicolumn{5}{l}{[a] Reaction ignored when comparing against DMS benchmarks} \\
  \multicolumn{5}{l}{[b] Multi-electronic-surface rate enhancement factor $\eta = 16/3$ for $\mathrm{O_2}$-dissociation is not applied} \\
  \multicolumn{5}{l}{when comparing against DMS (MMT-benchmark), but may be applied in the full model (MMT)} \\
  \multicolumn{5}{l}{intended for realistic conditions.}
 \end{tabular}

\end{table}


\begin{table}
 \centering
 \caption{Modified Arrhenius parameters for air reactions}
 \label{tab:arrhenius_2024_parameters_complete}
 
 \begin{tabular}{r c c c c c c}
    &       &                                                     & $C$         & $n$   & $E_a/ \bar{R}$ & \\
    & Type  &       Reaction                                      & $\mathrm{\left[\frac{cm^{3}}{mol \cdot s \cdot K^n}\right]}$ & $\mathrm{[-]}$ & $\mathrm{[K]}$ & note \\ \hline \hline 
  1 & diss  & $\mathrm{2N_2 \rightleftharpoons 2N   + N_2}$       & $1.5259e+17$  & $-0.40654$ & $113\,200$              & \\
  2 & diss  & $\mathrm{N_2  + O_2 \rightleftharpoons 2N   + O_2}$ & $1.6671e+20$  & $-1.1484$  & $113\,200$              & \\
  3 & diss  & $\mathrm{N_2  + NO  \rightleftharpoons 2N   + NO}$  & \multicolumn{3}{c}{- Re-use parameters from (1) -} & [a] \\
  4 & diss  & $\mathrm{N_2  + N   \rightleftharpoons 2N   + N}$   & $4.6825e+17$  & $-0.51270$ & $113\,200$              & \\
  5 & diss  & $\mathrm{N_2  + O   \rightleftharpoons 2N   + O}$   & $3.4678e+17$  & $-0.50732$ & $113\,200$              & \\ \hline
  6 & diss  & $\mathrm{O_2  + N_2 \rightleftharpoons 2O   + N_2}$ & $2.3279e+17$  & $-0.53233$ & $59\,330$               & [b] \\
    &       &                                                     & $1.2415e+18$  &          &                     & $\times 16/3$ \\
  7 & diss  & $\mathrm{2O_2 \rightleftharpoons 2O   + O_2}$       & $6.4741e+17$  & $-0.59391$ & $59\,330$               & [b] \\
    &       &                                                     & $3.4529e+18$  &          &                     & $\times 16/3$ \\
  8 & diss  & $\mathrm{O_2  + NO  \rightleftharpoons 2O   + NO}$  & \multicolumn{3}{c}{- Re-use parameters from (7) -} & [a] \\
  9 & diss  & $\mathrm{O_2  + N   \rightleftharpoons 2O   + N}$   & $1.7446e+17$  & $-0.44587$ & $59\,330$               & [b] \\
    &       &                                                     & $9.3045e+17$  &          &                     & $\times 16/3$ \\
  10 & diss  & $\mathrm{O_2  + O   \rightleftharpoons 3O}$        & $1.5537e+17$  & $-0.47445$ & $59\,330$               & [b] \\
     &       &                                                    & $8.2864e+17$  &          &                     & $\times 16/3$ \\ \hline
  11 & diss  & $\mathrm{NO   + N_2 \rightleftharpoons N   + O   + N_2}$ & \multicolumn{3}{c}{- Re-use parameters from (13) -} & [a] \\
  12 & diss  & $\mathrm{NO   + O_2 \rightleftharpoons N   + O   + O_2}$ & \multicolumn{3}{c}{- Re-use parameters from (13) -} & [a] \\
  13 & diss  & $\mathrm{2NO  \rightleftharpoons N   + O   + NO}$   & $3.5889e+14$ & $-0.07945$ & $52\,644$                & \\
  14 & diss  & $\mathrm{NO   + N   \rightleftharpoons 2N   + O}$   & $2.5759e+17$  & $-0.59741$ & $75\,360$               & \\
  15 & diss  & $\mathrm{NO   + O   \rightleftharpoons N   + 2O}$   & $7.0310e+16$  & $-0.52778$ & $75\,360$               & \\ \hline 
  16 & exch  & $\mathrm{N_2  + O   \rightleftharpoons NO   + N}$   & $1.6339e+11$  & $0.76972$  & $37\,850$               & \\
  17 & exch  & $\mathrm{NO   + O   \rightleftharpoons O_2  + N}$   & $1.99183e+13$ & $0.10007$  & $24\,427$               & \\
  18 & exch  & $\mathrm{N_2  + O_2  \rightleftharpoons 2NO}$       & $1.15292e+18$ & $-0.83459$ & $97\,485$               & \\  \hline 
  19 & mixed & $\mathrm{2NO  \rightleftharpoons 2O  + N_2}$        & $1.1786e+16$ & $-0.54728$ & $53\,783$                & \\
  20 & mixed & $\mathrm{2NO  \rightleftharpoons 2N  + O_2}$        & $1.3688e+21$ & $-1.6611$  & $106\,199$    & \\
  21 & mixed & $\mathrm{N_2  + O_2 \rightleftharpoons NO + N + O}$ & $2.9053e+14$  & $0.15022$ & $89\,282$    & \\
  \\
  \multicolumn{7}{l}{[a] Reaction ignored when benchmarking against DMS} \\
  \multicolumn{7}{l}{[b] Multi-electronic-surface rate enhancement factor $\eta = 16/3$ for $\mathrm{O_2}$-dissociation is not} \\
  \multicolumn{7}{l}{applied to $C$ when comparing against DMS (MMT-benchmark), but may be applied} \\
  \multicolumn{7}{l}{in the full model (MMT) intended for realistic conditions. (see row below)}
 \end{tabular}
 
\end{table}


\begin{table}
 \centering
 \caption{MMT parameters for air reactions}
 \label{tab:mmt_2024_parameters_complete}
 
 \begin{tabular}{r c c c c c c c c}
   &       &                                                          & $a_U$   & $U^*$    & $\Theta_v$ & $f_k^\mathrm{NB}$ [*] & $f_\varepsilon^\mathrm{NB}$ [*] & \\
   & Type  &       Reaction                                           & $\mathrm{[-]}$ & $\mathrm{[K]}$ & $\mathrm{[K]}$ & $\mathrm{[-]}$ & $\mathrm{[-]}$ & note \\ \hline \hline 
   1 & diss  & $\mathrm{2N_2 \rightleftharpoons 2N   + N_2}$       & $0.49725$ & $265\,482$   & $3\,414$       & $0.360$ & $0.900$ & \\
   2 & diss  & $\mathrm{N_2  + O_2 \rightleftharpoons 2N   + O_2}$ & $0.36900$ & $376\,750$   & $3\,414$       & $0.400$ & $0.830$ & \\
   3 & diss  & $\mathrm{N_2  + NO  \rightleftharpoons 2N   + NO}$  & \multicolumn{5}{c}{--Re-use parameters from (1)--} & [a] \\
   4 & diss  & $\mathrm{N_2  + N   \rightleftharpoons 2N   + N}$   & $0.39547$ & $805\,087$   & $3\,414$       & $0.480$ & $0.845$ & \\
   5 & diss  & $\mathrm{N_2  + O   \rightleftharpoons 2N   + O}$   & $0.35075$ & $-720\,429$  & $3\,414$       & $0.440$ & $0.800$ & \\ \hline
   6 & diss  & $\mathrm{O_2  + N_2 \rightleftharpoons 2O   + N_2}$ & $0.71386$ & $-107\,879$  & $2\,251$       & $0.285$ & $0.835$ & \\
   7 & diss  & $\mathrm{2O_2 \rightleftharpoons 2O   + O_2}$       & $0.46110$ & $54\,974$    & $2\,251$       & $0.415$ & $0.870$ & \\
   8 & diss  & $\mathrm{O_2  + NO  \rightleftharpoons 2O   + NO}$  & \multicolumn{5}{c}{--Re-use parameters from (7)--} & [a] \\
   9 & diss  & $\mathrm{O_2  + N   \rightleftharpoons 2O   + N}$   & $0.37965$ & $-313\,988$  & $2\,251$       & $0.465$ & $0.810$ & \\
  10 & diss  & $\mathrm{O_2  + O   \rightleftharpoons 3O}$         & $0.36157$ & $96\,143$    & $2\,251$       & $0.445$ & $0.850$ & \\ \hline
  11 & diss  & $\mathrm{NO   + N_2 \rightleftharpoons N   + O   + N_2}$ & \multicolumn{5}{c}{--Re-use parameters from (13)--} & [a] \\
  12 & diss  & $\mathrm{NO   + O_2 \rightleftharpoons N   + O   + O_2}$ & \multicolumn{5}{c}{--Re-use parameters from (13)--} & [a] \\
  13 & diss  & $\mathrm{2NO  \rightleftharpoons N   + O   + NO}$   & \multicolumn{5}{c}{-Treat as non-preferential-}        & \\
  14 & diss  & $\mathrm{NO   + N   \rightleftharpoons 2N   + O}$   & $0.51679$ & $3\,876\,179$  & $2\,744$      & $0.460$ & $0.840$ & \\
  15 & diss  & $\mathrm{NO   + O   \rightleftharpoons N   + 2O}$   & $0.30100$ & $-3\,007\,416$ & $2\,744$      & $0.555$ & $0.825$ & \\ \hline 
  16 & exch  & $\mathrm{N_2  + O   \rightleftharpoons NO   + N}$   & \multicolumn{5}{c}{-Treat as non-preferential-}        & \\
  17 & exch  & $\mathrm{NO   + O   \rightleftharpoons O_2  + N}$   & \multicolumn{5}{c}{-Treat as non-preferential-}        & \\
  18 & exch  & $\mathrm{N_2  + O_2  \rightleftharpoons 2NO}$       & \multicolumn{5}{c}{-Treat as non-preferential-}        & \\  \hline 
  19 & mixed & $\mathrm{2NO  \rightleftharpoons 2O  + N_2}$        & \multicolumn{5}{c}{-Treat as non-preferential-}        & \\
  20 & mixed & $\mathrm{2NO  \rightleftharpoons 2N  + O_2}$        & \multicolumn{5}{c}{-Treat as non-preferential-}        & \\
  21 & mixed & $\mathrm{N_2  + O_2 \rightleftharpoons NO + N + O}$ & \multicolumn{5}{c}{-Treat as non-preferential-}        & \\
  \\
  \multicolumn{9}{l}{[a] Reaction ignored when benchmarking against DMS} \\
  \multicolumn{9}{l}{[*] Uniform values of $f_k^\mathrm{NB} = 0.5$ and $f_\varepsilon^\mathrm{NB} = 0.85$ are used for all MMT-type reactions in this work.}
 \end{tabular}
 
\end{table}


\subsubsection{Model parameters for comparison against DMS results} \label{sec:mmt_params_for_testing}

As discussed in Ref.~\cite{torres24a}, a fraction of the ab inito PESs necessary to generate the full set of ground-electronic-state 5-species air reactions from QCT calculations currently does not exist. This problem partially affects reactions 5, 14, 15 and 19-20 in Tables~\ref{tab:mmt_2024_reactions}-\ref{tab:mmt_2024_parameters_complete}, but especially the entirely missing PESs for reactions involving $\mathrm{N_2 - NO}$ and $\mathrm{O_2 - NO}$ collision pairs (reactions 3, 8, 11 and 12). We deal with these limitations differently depending on whether we are comparing the model against DMS results, or proposing the model for realistic air simulations. In Sec.~\ref{sec:testing_vs_dms_again}, where we perform several calculations to test the MMT model's ability to reproduce a series of DMS reference results, we adjust the CFD reaction set from Table~\ref{tab:mmt_2024_reactions} to exclude reactions 3, 8, 11 and 12, as these do not take place in the DMS calculations. Furthermore, the DMS reference calculations currently are not set up to simulate any reactions involving three simultaneous collision partners as reactants. Thus, none of the reverse reactions of the dissociation-type (reactions 1-15) nor of the ``mixed'' exchange-dissociation (reactions 19-21) are taken into account in the DMS reference results. As a consequence, in Sec.~\ref{sec:testing_vs_dms_again} we explicitly model all aforementioned reactions as irreversible in the forward direction. The only exceptions are the three exchange reactions (reactions 16-18 in Table~\ref{tab:mmt_2024_reactions}). Since both their forward and backward partial reactions involve only two colliding particles as reactants, they both are included in our DMS solutions and consequently also in the corresponding CFD calculations. For these three exchange reactions the reverse rate coefficients are calculated using the detailed balance relations for non-preferential reactions (see Sec.~\ref{sec:fluid_equations}) together with the relevant equilibrium constants (see Eqs.~(S9)-(S11) in the supplemental information). We should also note that during the process of comparing our MMT results against DMS we evaluate the rate coefficients of all oxygen dissociation reactions (6-10 in Table~\ref{tab:arrhenius_2024_parameters_complete}) explicitly \emph{without} the multi-electronic-surface enhancement factor $\eta = 16/3$ (See Refs.~\cite{grover19b} and \cite{nikitin74a} (pp. 331-334) for the reasoning behind proposing this factor). During the model benchmarking stage this is the correct choice, since our DMS calculations only account for reactions involving oxygen molecules in their ground electronic state.

\subsubsection{Model parameters proposed for full 5-species air simulations} \label{sec:mmt_params_for_real}

The model parameters for realistic dissociating air simulations must include all possible collision pairs in a 5-species air mixture. Here we fill in the aforementioned gaps for dissociation reactions involving nitric oxide by re-using rate data from other, similar reactions. Specifically, reaction 3 in Tables~\ref{tab:mmt_2024_reactions}-\ref{tab:mmt_2024_parameters_complete} uses the same parameters as reaction 1, while reaction 8 relies on the same parameters as reaction 7. Furthermore, both for reactions 11 and 12 we assume the same rate parameters as for reaction 13. We now also include the possibility for all reactions in Tables~\ref{tab:mmt_2024_reactions}-\ref{tab:mmt_2024_parameters_complete} to be reversible by imposing detailed balance relations, as discussed in Sec.~VI of Ref.~\cite{chaudhry25a}). Finally, we must now take into account the effect of electronically excited states of molecular oxygen on the dissociation rate. Essentially, we consider two limiting cases. One assumption, discussed in Refs.~\cite{grover19b} and \cite{nikitin74a} (pp. 331-334), states that behind strong shocks all electronic excited states of $\mathrm{O_2}$ lying between its ground state energy and its dissociation threshold of $5.213 \, \mathrm{eV}$ become Boltzmann-populated and enhance the overall dissociation rates well beyond that of $\mathrm{O_2}$ populating the ground state alone. The scaling proposed is to assume the ground state $\mathrm{O_2}(X^3 \Sigma_g^-)$ and all excited states in question contribute according to their respective degeneracies $3+2+1+1+6+3$ (first six rows in Table~I of Ref.~\cite{grover19b}) to augment the ground-state-only dissociation rate (hence $\eta = 16/3$). Thus, in this limit we modify all reaction rate coefficients involving $\mathrm{O_2}$ as the dissociating species (reactions 6 - 10 in Table~\ref{tab:arrhenius_2024_parameters_complete}) to have their pre-exponential Arrhenius terms augmented by the multi-electronic surface factor $\eta = 16/3$ when simulating realistic flow conditions. The other limiting case is to assume none of the excited stated are populated and the overall $\mathrm{O_2}$ dissociation rate will be equal that of the ground state alone. Thus, no enhancement factor would be used. To the authors' best knowledge there are at present no ab inito results in the literature to fully support either assumption and we merely present both options as the two limiting cases. We therefore note that the MMT model can be used with, or without these enhancement factors. In our comparison calculations against the Park model in Sec.~\ref{sec:mmt_vs_park}, we present MMT results with and without the factor applied.


\subsubsection{MMT-specific non-Boltzmann correction factors} \label{sec:non_boltzmann_factors}

Table~\ref{tab:mmt_2024_parameters_complete} collects the additional parameters necessary to evaluate Eqs.~(\ref{eq:mmt_rate_coefficient_again}) and (\ref{eq:devib_mmt}) for all those reactions that employ the MMT functional form. This comprises the five nitrogen and oxygen dissociation reactions (reactions 1-10), as well as two of the nitric oxide dissociation reactions (reactions 14-15). 

The last two columns in particular list non-Boltzmann factors for rate coefficient and vibrational energy removal, $f_k^\mathrm{NB}$ and $f_\varepsilon^\mathrm{NB}$ respectively. These values were derived from an analysis of the combined QCT and DMS data in Refs.~\cite{chaudhry18d} and \cite{torres24a} and are specific to each individual reaction. When comparing these values it becomes apparent that, except for a few outliers, they both cluster within fairly narrow bands. For the rate coefficient factor they mostly range between $0.4$ and $0.5$ and for the energy removal factor all values lie somewhere between $0.8$ and $0.9$. In principle, one may use these tabulated values directly when evaluating Eqs.~(\ref{eq:mmt_rate_coefficient_again}) and (\ref{eq:devib_mmt}) for each MMT reaction. However, we have opted for simplifying the model and resort only to a single pair of values $f_k^\mathrm{NB} = 0.5$ and $f_\varepsilon^\mathrm{NB} = 0.85$ common to all reactions. The underlying reason is that statistical uncertainty in the DMS data used to extract the non-Boltzmann factors for some of the reactions is large enough to be compatible with a wider range of values. This becomes apparent when examining the rate coefficient and vibrational energy change plots for each individual reaction in Sec.~S4 of the supplemental information. Thus, we think recommending use of the exact non-Boltzmann factor values in Table~\ref{tab:mmt_2024_parameters_complete} could cause the false impression that the listed values are known to much higher accuracy than is actually the case.


\subsection{Characteristic behavior of rate coefficient and vibrational energy removal terms} \label{sec:comparison_rate_devib_examples}

In what follows, we pick out a few examples from the full reaction list to highlight differences in characteristic behavior observed for dissociation-type, exchange-type and mixed-type reactions. The complete collection of Arrhenius and vibrational energy change plots for all reactions is available in Figs.~S3-S19 of the supplemental information.

\begin{figure}
 \centering

 \subfloat[Rate coefficient]{\label{fig:NO-O_diss_arrhenius_fits2}
 \includegraphics[width=0.49\textwidth]{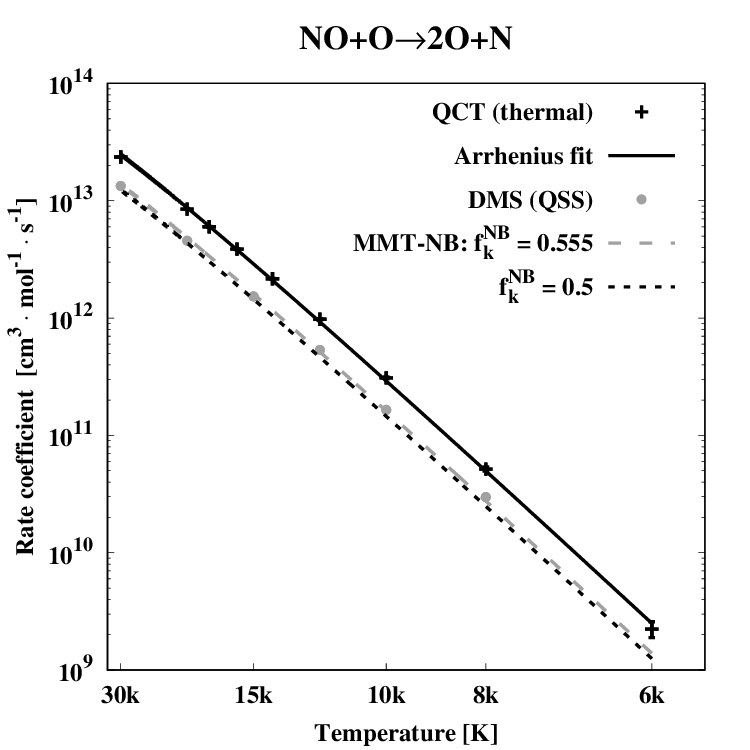}}~
 \subfloat[Vibrational energy change per reaction]{\label{fig:mmt_evib_fit_NO-O_diss2}
 \includegraphics[width=0.49\textwidth]{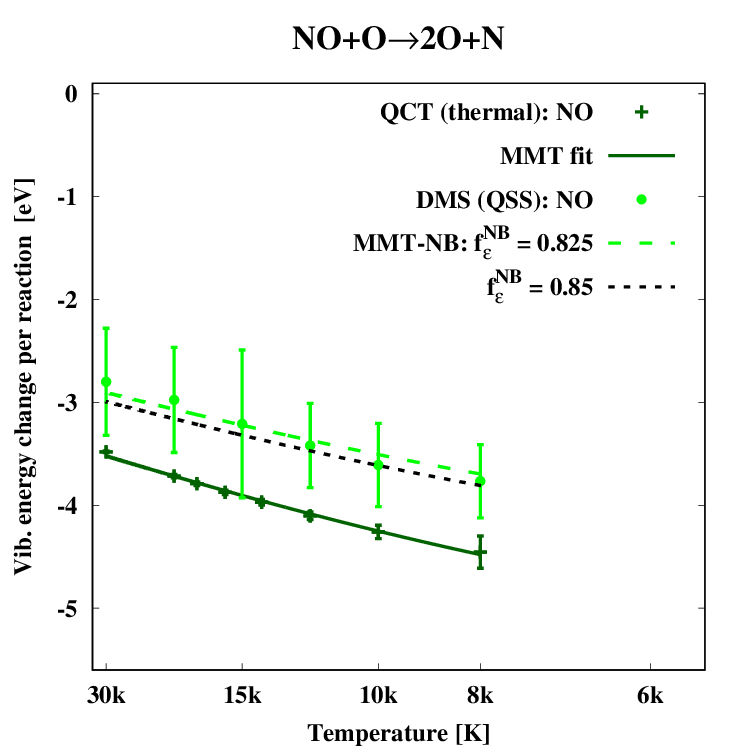}}
 
 \caption{Example of an MMT dissociation-type reaction: $\boldsymbol{\mathrm{NO + O \rightleftharpoons 2O + N}}$}
 \label{fig:NO-O_diss2}
\end{figure}

Dissociation-type reactions that are modeled using the MMT formulation are represented by reaction 15 of Tables~\ref{tab:mmt_2024_reactions}-\ref{tab:mmt_2024_parameters_complete}, i.e. nitric oxide dissociation induced by collision with atomic oxygen. Fig.~\ref{fig:NO-O_diss_arrhenius_fits2} shows the Arrhenius plot for this reaction. In it the QCT-derived \emph{thermal equilibrium} rate coefficients are shown as black crosses, with corresponding DMS-derived QSS-dissociation-regime values plotted as gray circles. The modified Arrhenius fit to the QCT data (evaluated using Eq.~(\ref{eq:karr_again}) and parameters from Table~\ref{tab:arrhenius_2024_parameters_complete}) is shown as a continuous black line. A dashed gray line represents the same curve, but offset downward by a factor of $f_k^\mathrm{NB} = 0.555$ to provide the closest fit to the DMS data points. As mentioned previously, for this reaction, as well as for nearly all other dissociation-type reactions we end up using the MMT model formulations with a common non-Boltzmann factor of $f_k^\mathrm{NB}=0.5$ to evaluate the rate coefficient. Thus, the black dotted line, labeled ``MMT-NB: $f_k^\mathrm{NB}=0.5$'', shows the result of evaluating Eq.~(\ref{eq:mmt_rate_coefficient_again}) at $T_\mathrm{v}\!=\!T$ with this value. 

Figure~\ref{fig:mmt_evib_fit_NO-O_diss2} provides the corresponding plot of vibrational energy change for the same reaction. Here we explicitly plot the net energy \emph{change} per reaction, i.e. $\langle \Delta \varepsilon_{\mathrm{v}, \mathrm{NO}} \rangle_\mathrm{diss}$, which will be negative for $\mathrm{NO}$ as the species being consumed in this particular reaction. The original thermal-equilibrium QCT data (see Table~\ref{tab:mmt_2024_reactions} for data sources) are plotted using dark green crosses for $\mathrm{NO}$, with error bars representing the one-standard-deviation error. The corresponding DMS data at the same translational temperature, but extracted during the QSS-dissociation regime, are plotted using light green dots, with the error bars representing the standard error. The solid dark green line represents the vibrational energy change per dissociation of $\mathrm{NO}$ at thermal equilibrium, which provides the best fit to the QCT data. It is evaluated using Eq.~(\ref{eq:devib_mmt}) together with the MMT parameters for reaction 15 from Table~\ref{tab:mmt_2024_parameters_complete} and $f_\varepsilon^\mathrm{NB} = 1$. The dashed line in light green represents the corresponding adjusted curve for Eq.~(\ref{eq:devib_mmt}) with the non-Boltzmann-factor $f_{\varepsilon}^\mathrm{NB} = 0.825$. Analogous to what was done for the rate coefficient this represents the curve that most closely matches the DMS data points for this reaction. An additional dotted curve in black shows Eq.~(\ref{eq:devib_mmt}) evaluated using our chosen common factor of $f_{\varepsilon}^\mathrm{NB} = 0.85$ for all MMT-type reactions.

Pairs of plots like in Fig.~\ref{fig:NO-O_diss2} are available for all other dissociation reactions in the supplemental information. Nearly all of them follow the same general pattern established with the example shown in Fig.~\ref{fig:NO-O_diss2}. The dissociation rate coefficient in the QSS regime is consistently smaller than that for thermal equilibrium by a nearly constant factor over a wide temperature range and the behavior is similar regarding the vibrational energy change. These trends align with prior studies, for example Singh and Schwartzentruber~\cite{singh20b, singh20c} showed that directly modeling non-Boltzmann internal energy distribution functions using a surprisal formulation~\cite{singh18a} resulted in near-constant factors in the QSS regime for nitrogen. Note that, even though the original calibration of $f_k^\mathrm{NB} \approx 0.5$ and  $f_{\varepsilon}^\mathrm{NB} \approx 0.85$ in Sec.~V of Ref.~\cite{chaudhry25a} only involved nitrogen and oxygen dissociation in $\mathrm{N_2/N}$ and $\mathrm{O_2/O}$ systems, the same non-Boltzmann correction factor also turns out to be a reasonable estimate for most of the other dissociation reactions. In hindsight, this provides a justification of sticking with these two values for all MMT-type reactions. 

The remaining reactions (reactions 11-13 and 16-21 in Table~\ref{tab:mmt_2024_reactions}) exhibit either fully thermal, or ``mixed'' behavior in terms of non-Boltzmann QSS vs. vibrationally non-preferential trends. As discussed in detail in Ref.~\cite{torres24a}, rather than decreasing with $T$, the average vibrational energy removed, or gained in most of these reactions tends to increase along with translational temperature, which is more in line with the non-preferential energy coupling assumption discussed in Sec.~II.A of Ref.~\cite{chaudhry25a}. The clearest representatives of this behavior are the exchange reactions shown in Figs.~S14-S16 of the supplemental information. 

\begin{figure}
 \centering
 
 \subfloat[Rate coefficient]{\label{fig:NO-O_exch_arrhenius_fits2}
 \includegraphics[width=0.49\textwidth]{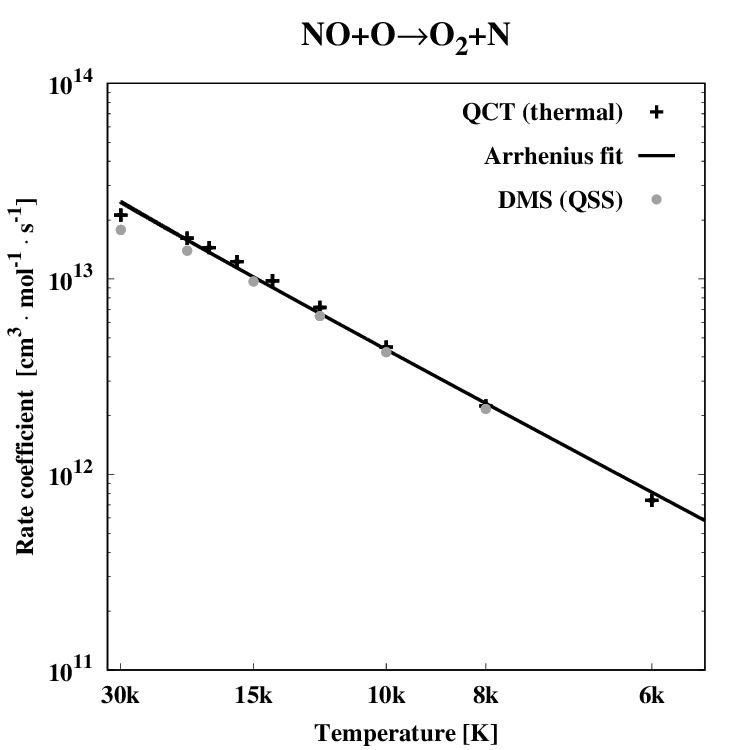}}~
 \subfloat[Vibrational energy change per reaction]{\label{fig:mmt_evib_fit_NO-O_exch2}
 \includegraphics[width=0.49\textwidth]{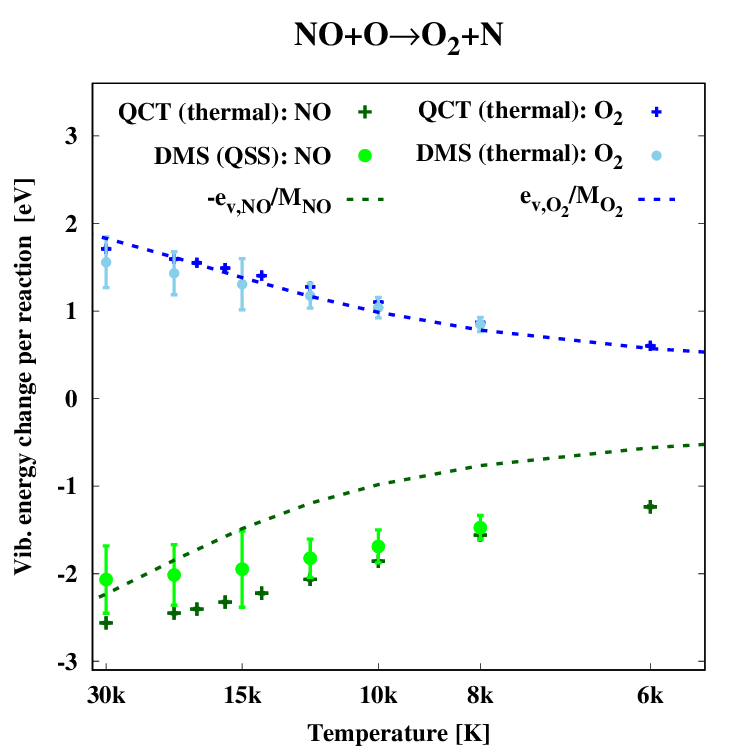}}
 
 \caption{Example of an exchange-type reaction: $\boldsymbol{\mathrm{NO + O \rightleftharpoons O_2 + N}}$}
 \label{fig:NO-O_exch2}
\end{figure}

As an example, we show here the Arrhenius plots for the second Zeldovich reaction, i.e. reaction 17 in Table~\ref{tab:mmt_2024_reactions}, for closer discussion. Figure~\ref{fig:NO-O_exch_arrhenius_fits2} shows that the QCT-derived thermal equilibrium and DMS-derived QSS rate coefficients are nearly identical over the entire temperature range considered. Our preceding DMS studies in Ref.~\cite{torres24a} confirmed that these exchange reactions exhibit only weak, if any, vibrational bias and proceed at near-thermal rates regardless of whether the gas mixture as a whole is in thermal equilibrium or not. Therefore, the single-temperature modified Arrhenius fit (solid black line) likely provides an accurate enough representation of the rate coefficient. It can also be seen that the QCT- and DMS-predicted vibrational energy change values for nitric oxide (dark and light green symbols in Fig.~\ref{fig:mmt_evib_fit_NO-O_exch2} respectively) and molecular oxygen (dark and light blue symbols) largely follow the same temperature trend as that of the respective average vibrational energy in the gas, i.e. $\langle \Delta \varepsilon_{\mathrm{v}, \mathrm{NO}} \rangle_\mathrm{exch} \approx -e_{\mathrm{v}, \mathrm{NO}} (T) / M_\mathrm{NO}$ (dark green dotted line) and $\langle \Delta \varepsilon_{\mathrm{v}, \mathrm{O_2}} \rangle_\mathrm{exch} \approx e_{\mathrm{v}, \mathrm{O_2}} (T) / M_\mathrm{O_2}$ (dark blue dotted line). Recall that, since nitric oxide is being consumed, $\langle \Delta \varepsilon_{\mathrm{v}, \mathrm{NO}} \rangle_\mathrm{exch}$ is negative, whereas $\langle \Delta \varepsilon_{\mathrm{v}, \mathrm{O_2}} \rangle_\mathrm{exch}$ is positive for the molecular oxygen being produced. Although there remains a small gap between the QCT and DMS-predicted values for $\langle \Delta \varepsilon_{\mathrm{v}, \mathrm{NO}} \rangle_\mathrm{exch}$ and that of $- e_{\mathrm{v}, \mathrm{NO}} (T) /  M_\mathrm{NO}$, this non-preferential coupling assumption provides a much better choice than if one had attempted to fit the data to Eq.~(\ref{eq:devib_mmt}). A similar observation can be made for the first Zeldovich reaction, which involves $\mathrm{N_2}$ and $\mathrm{NO}$ instead.

For the remaining ``mixed'' exchange-dissociation type reactions (reactions 19-21 in Table~\ref{tab:mmt_2024_reactions}) we do not observe a consistent pattern regarding their vibrational bias. The rate coefficient comparisons in Figs.~S17-S19 of the supplemental information show that for some of these reactions the DMS-derived QSS rates are nearly identical to the QCT-derived thermal equilibrium ones, whereas for others a clear gap between the two can be observed, suggesting a non-Boltzmann reaction rate during QSS. It is also difficult to classify these reactions as entirely MMT-type or non-preferential regarding the vibrational energy change of the diatomic species involved. Whereas non-preferential vibration-chemistry coupling seems to fit the QCT/DMS data reasonably well for one species, for another species in the same reaction vibration-preferential ``MMT-type'' coupling might be a better assumption. In order to avoid over-complicating the overall chemistry model, we make the choice of assuming non-preferential vibrational energy-chemistry coupling for all diatomic species participating in any of these reactions (see comment in Tables~\ref{tab:mmt_2024_reactions} and \ref{tab:mmt_2024_parameters_complete}).

\begin{figure}
 \centering
 
 \subfloat[Rate coefficient]{\label{fig:NONO_Diss_Exch_N2_Product_arrhenius_fits2}
 \includegraphics[width=0.49\textwidth]{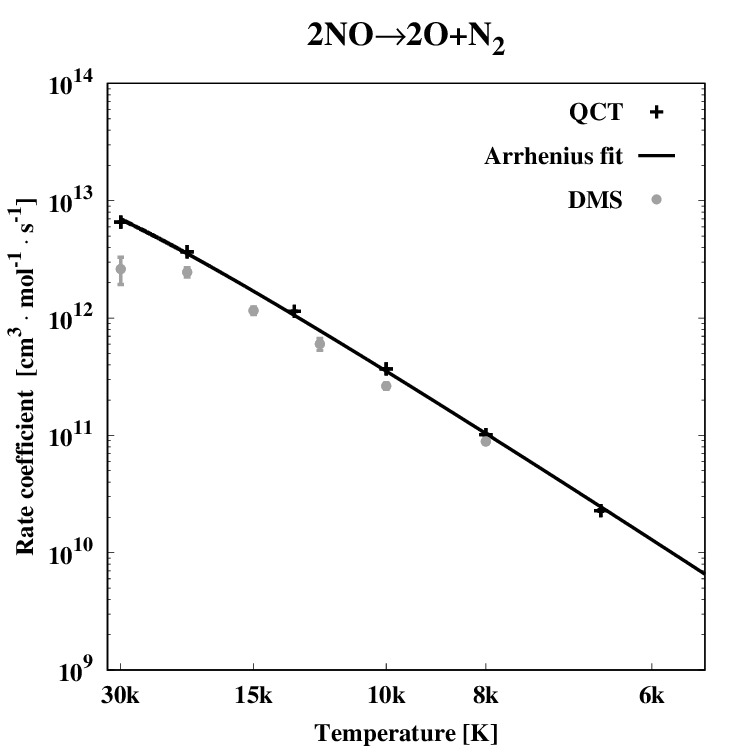}}~
 \subfloat[Vibrational energy change per reaction]{\label{fig:mmt_evib_fit_NONO_Diss_Exch_N2_Product2}
 \includegraphics[width=0.49\textwidth]{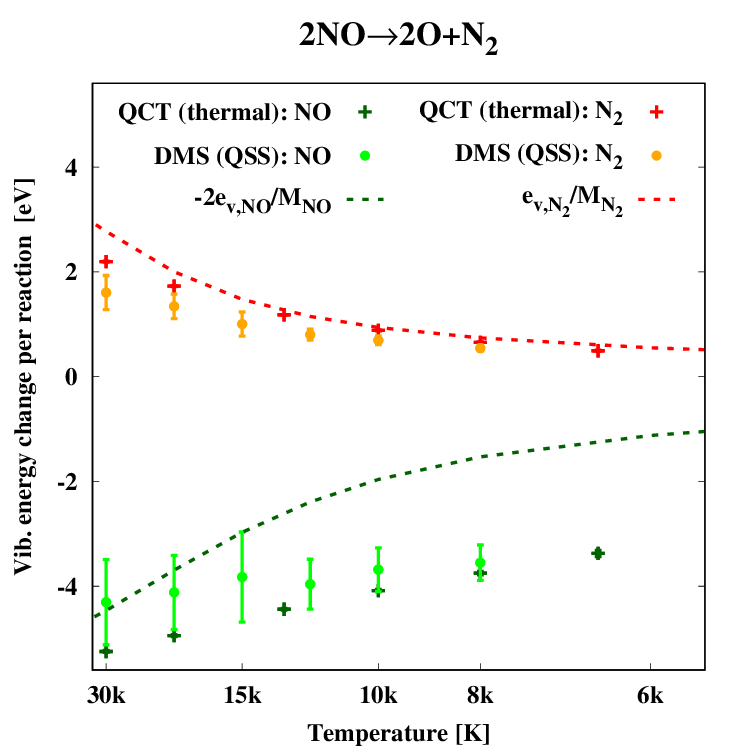}}
 
 \caption{Example of a mixed-type reaction: $\boldsymbol{\mathrm{2NO \rightleftharpoons 2O + N_2}}$}
 \label{fig:NONO_Diss_Exch_N2_Product2}
\end{figure}

As an example case, in Fig.~\ref{fig:NONO_Diss_Exch_N2_Product2} we show the rate coefficient and vibrational energy change for reaction 19 of Table~\ref{tab:mmt_2024_reactions}, in which two $\mathrm{NO}$ molecules are consumed to produce two oxygen atoms, as well as one nitrogen molecule. The DMS/QCT results in Fig.~\ref{fig:mmt_evib_fit_NONO_Diss_Exch_N2_Product2} suggest that the average vibrational energy of $\mathrm{N_2}$ molecules being produced is roughly equal to that species' average vibrational energy in the gas at temperature $T$ (orange and red symbols vs. red dotted line). Thus, non-preferential vibrational energy-chemistry coupling is a good fit for molecular nitrogen in this reaction. Simultaneously, the QCT and DMS data (dark and light green symbols in Fig.~\ref{fig:mmt_evib_fit_NONO_Diss_Exch_N2_Product2}) for the $\mathrm{NO}$ being consumed suggest that these molecules possess a slightly greater-than-average amount of vibrational energy compared to those present in the gas. Further note that in this reaction not just one, but two $\mathrm{NO}$ molecules are consumed at a time. Therefore, the vibrational energy change of $\mathrm{NO}$ per reaction with the non-preferential model ends up being $\langle \Delta \varepsilon_{\mathrm{v}, \mathrm{NO}} \rangle_\mathrm{exch-diss}^\mathrm{non-pref.} = - 2 \, e_{\mathrm{v}, \mathrm{NO}} / M_\mathrm{NO}$.

%
\subsection{Vibrational relaxation times} \label{sec:tau_vt_fits_again}

In two-temperature nonequilibrium models relaxation of translational and vibrational modes toward thermal equilibrium is described by the source term $w_\mathrm{v}^\mathrm{relax}$ in Eq.~(\ref{eq:vib_energy_balance}):
\begin{equation}
 w_\mathrm{v}^\mathrm{relax} = \sum_{s \in D} \rho_{s} \frac{e_{\mathrm{v}, s} (T) - e_{\mathrm{v}, s} (T_\mathrm{v})}{\langle \tau_{s}^\mathrm{v} \rangle}, \label{eq:vib_relax_equation_again}
\end{equation}
consisting of a mixture-weighted sum over all vibrationally relaxing diatomic species $D = \{ \mathrm{N_2}, \mathrm{O_2}, \mathrm{NO} \}$. Each species' contribution is modeled as a Landau-Teller relaxation term, where the numerator represents departure of that species' vibrational energy $e_{\mathrm{v}, s} (T_\mathrm{v})$ at local vibrational temperature $T_\mathrm{v}$ from the corresponding hypothetical equilibrium value at the local gas trans-rotational temperature $e_{\mathrm{v}, s} (T)$. The denominators in Eq.~(\ref{eq:vib_relax_equation_again}) represent each of the three relaxing diatomic species' average vibrational relaxation times, themselves defined as averages over reciprocals of pair-wise relaxation times $\tau_{s,q}^\mathrm{v}$ weighted by mole fraction $x_q$ of every collision partner $q \in S$. 
\begin{equation}
 \langle \tau_{s}^\mathrm{v} \rangle = \left( \sum_{q \in S} \frac{x_{q}}{\tau_{s,q}^\mathrm{v}} \right)^{-1}, \qquad s \in D, \label{eq:species_relaxation_time_mw_again}
\end{equation}

For air species one would traditionally rely on the Millikan \& White (M\&W) correlation~\cite{millikan63a} in conjunction with Park's high-temperature correction~\cite{park93a} to determine these pair-wise relaxation times. However, experimental evidence~\cite{ibraguimova13a, streicher20c, streicher21a, streicher22a, streicher22b} as well as recent first-principles calculations~\cite{grover19a, grover19b, torres24b} have revealed significant discrepancies for some species pairings, especially in the high-temperature limit. In this paper we present an alternative expression together with the necessary fit parameters to calculate characteristic vibrational relaxation times for $\mathrm{N_2}$, $\mathrm{O_2}$ and $\mathrm{NO}$, which are derived from recent DMS calculations~\cite{torres24b} using only ab initio PESs.

The analytical expression chosen to curve-fit the DMS-derived pair-wise vibrational relaxation times (repeated from Eq.~(19) in Ref.~\cite{chaudhry25a}) has the form:
\begin{equation}
 \tau_{s,q}^\mathrm{v} = \frac{c_0}{p} \left( \exp \left( m_{s,q}^\mathrm{low} \, T^{-1/3} + n_{s,q}^\mathrm{low} \right) + \exp \left( m_{s,q}^\mathrm{high} \, T^{-1/3} + n_{s,q}^\mathrm{high} \right) \right), \qquad s \in D, q \in S, \label{eq:two_slope_fit_again}
\end{equation}
where $s$ and $q$ are species indices, the factor $c_0 = 1\, \mathrm{atm \cdot s}$, $p$ is the pressure in atmospheres and $T$ is the trans-rotational temperature. This expression assumes that there are two distinct ``low-temperature''  and ``high-temperature'' limiting behaviors for each $\tau_{s,q}^\mathrm{v}$, which can be parameterized by separate straight-line segments on a $\log ( \tau_{s,q}^\mathrm{v} \cdot p )$ vs. $T^{-1/3}$ plot.

\begin{figure}
 \centering
 
 \includegraphics[width=0.7\textwidth]{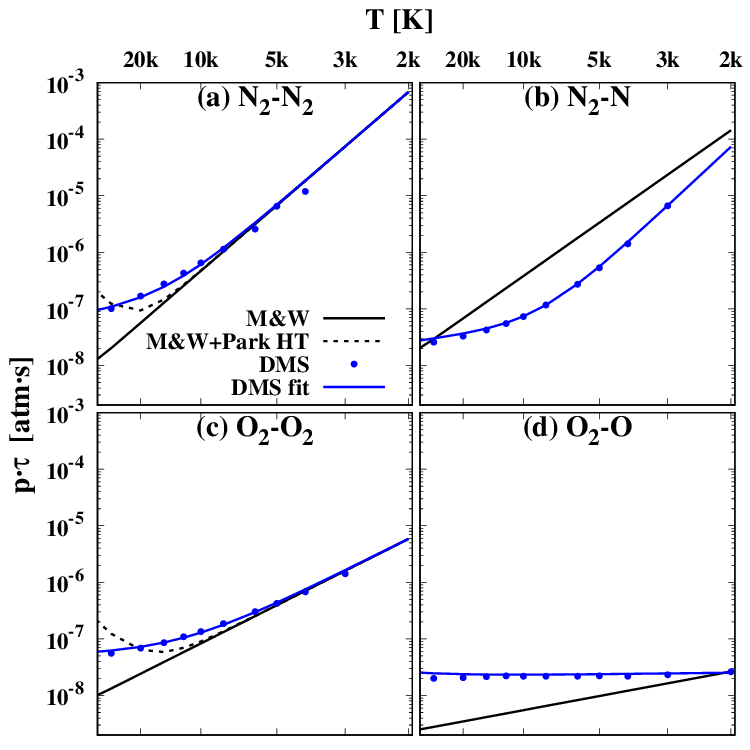}
 
 \caption{Vibrational relaxation times (1/3). DMS data~\cite{torres24b} and curve fits (blue), M\&W correlation~\cite{millikan63a} (black) plus Park's high-temperature correction~\cite{park93a} (dotted black).}
 \label{fig:tau_vt_fits_1}
\end{figure}

\begin{figure}
 \centering
 
 \includegraphics[width=0.7\textwidth]{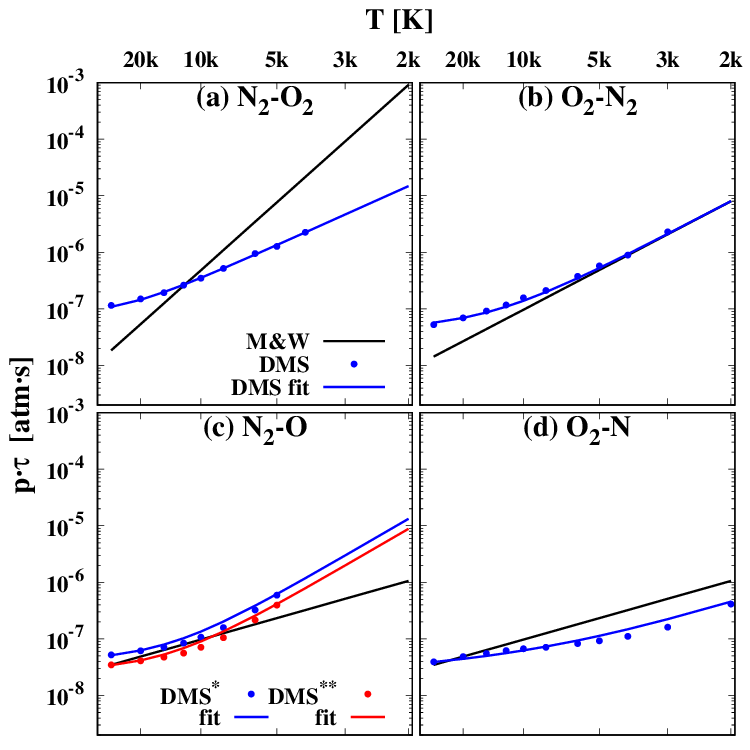}
 
 \caption{Vibrational relaxation times (2/3). Original (blue) / adjusted (red) DMS data~\cite{torres24b} and curve fits, M\&W correlation~\cite{millikan63a} (black).}
 \label{fig:tau_vt_fits_2}
\end{figure}

\begin{figure}
 \centering
 
 \includegraphics[width=0.7\textwidth]{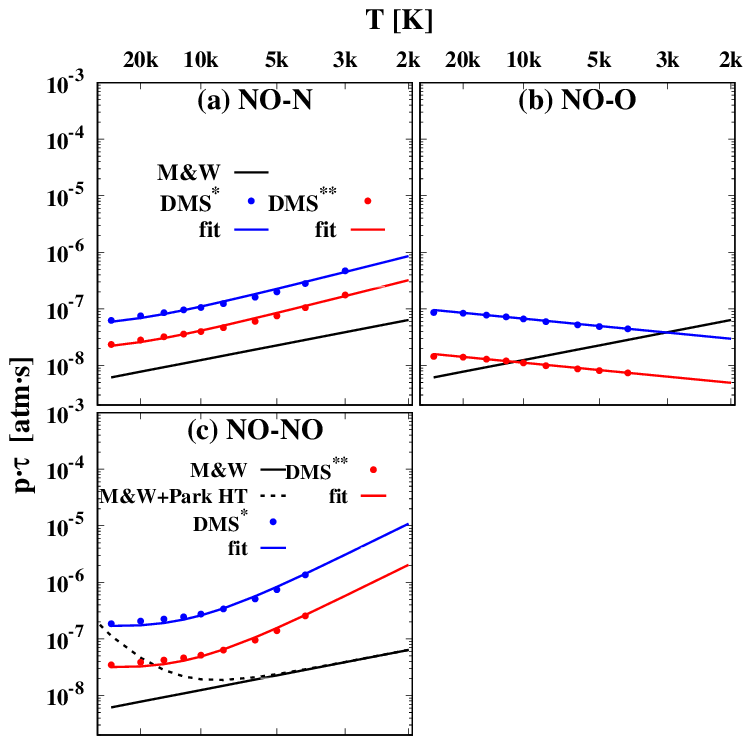}
 
 \caption{Vibrational relaxation times (3/3). Original (blue) / adjusted (red) DMS data~\cite{torres24b} and curve fits, M\&W correlation~\cite{millikan63a} (black) plus Park's high-temperature correction~\cite{park93a} (dotted black).}
 \label{fig:tau_vt_fits_3}
\end{figure}

Numerical parameters required to evaluate Eq.~(\ref{eq:two_slope_fit_again}) for each individual relaxation pair are listed in Table~\ref{tab:tau_v_fit_combined} and the results are summarized in Figs.~\ref{fig:tau_vt_fits_1}-\ref{fig:tau_vt_fits_3}. Blue symbols represent the original DMS data from Ref.~\cite{torres24b}, while the blue lines show our corresponding curve fits. For comparison, the continuous black line in each sub-plot of Figs.~\ref{fig:tau_vt_fits_1}-\ref{fig:tau_vt_fits_3} shows the corresponding prediction with the M\&W correlation~\cite{millikan63a}. For some species pairs (e.g. for $s-q = \mathrm{N_2 - N_2}$ and $\mathrm{O_2 - O_2}$), the DMS predictions tend to agree well with the M\&W correlations at lower temperatures, whereas for others the DMS behavior and experiments differ significantly from Millikan \& White.

For all species pairs the high-temperature relaxation behavior predicted by the M\&W correlation deviates significantly from that observed in our DMS calculations. Dotted black curves representing the effect of adding Park's high-temperature correction to the Millikan \& White prediction are shown in Figs.~\ref{fig:tau_vt_fits_1}(a), \ref{fig:tau_vt_fits_1}(c) and ~\ref{fig:tau_vt_fits_3}(c) for the three single-species diatom-diatom interactions $s-q = \mathrm{N_2 - N_2}$, $\mathrm{O_2 - O_2}$ and $\mathrm{NO - NO}$. This correction, as defined in the section entitled ``Vibrational Relaxation Parameters'' of Ref.~\cite{park93a}, is applied to each of the three diatomic species' overall relaxation times, i.e.: $1/\langle \tau_s^\mathrm{v} \rangle^\mathrm{overall} = 1/\langle \tau_s^\mathrm{v} \rangle^\mathrm{M\&W} + 1/\langle \tau_s^\mathrm{v} \rangle^\mathrm{Park-HT}$, after the Millikan \& White portion itself has been calculated by weighting the pair-wise contributions according to Eq.~(\ref{eq:species_relaxation_time_mw_again}). Each species-specific overall relaxation time is thus dependent on the given mixture composition. Conversely, the Park high-temperature correction cannot be applied to individual pair-wise relaxation times, but only to the combined one as a whole. It therefore only makes sense to plot the correction for a particular mixture composition, or as done in this case, for the single-species diatom-diatom interactions that simultaneously represent the overall relaxation time in a pure gas, i.e. $\langle \tau_s^\mathrm{v} \rangle (x_s=1) = \tau_{s,s}^\mathrm{v}$. For all three single-species interactions, the Park high-temperature correction and the corresponding DMS-derived curve fits both depart upward from the linear-slope $T$-dependence at very high temperatures. However, the effect of the Park correction is far more pronounced than in any of the DMS-derived curve fits.

For species pairings $s-q = \mathrm{N_2 - O}$, $\mathrm{NO - N}$, $\mathrm{NO - O}$ and $\mathrm{NO - NO}$ a second set of red symbols and corresponding lines are shown in Figs.~\ref{fig:tau_vt_fits_2}(c) and Figs.~\ref{fig:tau_vt_fits_3}(a) - \ref{fig:tau_vt_fits_3}(c). As discussed in Ref.~\cite{torres24b}, only a fraction of the PESs necessary for computing the relaxation times of these particular species pairings is currently available. At present we estimate them based on the subset of available potentials. The two curves shown represent our estimates based on the available data. The larger time constant estimates (blue symbols, blue line) were obtained under the assumption that only trajectories on the available PESs contribute to vibrational relaxation for the species pair in question. The lower estimates (red symbols, red line) are obtained under the assumption that trajectories on the unavailable PESs contribute in the same manner to vibrational relaxation as the available ones. Note that for each species pairing the two sets of curves exhibit the exact same temperature dependence, only offset by a constant value along the ordinate axis. Both estimates are relevant to the work discussed in this paper. The  blue lines represent the relaxation behavior that is consistent with the reference DMS calculations used for model benchmarking purposes in Sec.~\ref{sec:testing_vs_dms_again}. Since these DMS calculations are subject to the same PES availability gaps, we employ these upper-end estimates when comparing CFD results against pure DMS (see footnote [b] in Table~\ref{tab:tau_v_fit_combined}). By contrast, the red symbols and lines represent our best-guess estimate for the vibrational relaxation times in a realistic air mixture. These alternative values are used in Sec.~\ref{sec:mmt_vs_park}, where we make adjustments to the nonequilibrium model to improve its behavior for realistic hypersonic flows (see footnote [c] in Table~\ref{tab:tau_v_fit_combined}) and compare its predictions against simulations employing M\&W relaxation times together with the Park $TT_\mathrm{v}$ model~\cite{park93a}.

Further note that in Figs.~\ref{fig:tau_vt_fits_1}-\ref{fig:tau_vt_fits_3} plots for pairings $s-q = \mathrm{N_2 - NO}$, $\mathrm{O_2 - NO}$, $\mathrm{NO - N_2}$ and $\mathrm{NO - O_2}$ are missing entirely. No DMS data for these pairings could be generated in Ref.~\cite{torres24b}, as none of the prerequisite $\mathrm{NO_3}$ and $\mathrm{N_3O}$ ab initio PESs currently exist. Thus, in the model comparisons against DMS in Sec.~\ref{sec:testing_vs_dms_again} we will assume that the respective relaxation times tend toward infinity (see footnote [a] in Table~\ref{tab:tau_v_fit_combined}), mimicking the behavior of the DMS reference solutions. Note again that this approach makes sense only during the model benchmarking phase, but not so when the intention is to use CFD to simulate realistic hypersonic flows. For that scenario we re-purpose data from other available pairings as surrogates. The underlying assumption is that vibrational relaxation times of a particular species $s$ will depend to a lesser degree on the collision partner species $q$. Thus, for pairing $s-q = \mathrm{N_2 - NO}$ we choose to replace $\mathrm{NO}$ with $\mathrm{N_2}$ as the collision partner and use the same curve fit values as for $\mathrm{N_2 - N_2}$. In analogous fashion, for pairing $\mathrm{O_2 - NO}$ we swap out $\mathrm{NO}$ for $\mathrm{O_2}$ as collision partner and end up using the same parameters as for $\mathrm{O_2 - O_2}$. Finally, for the two remaining nitric oxide relaxation times $\mathrm{NO - N_2}$ and $\mathrm{NO - O_2}$, we apply the same principle and assume both relaxation times behave the same as the lower-bound estimate for $\mathrm{NO - NO}$.

The preceding discussion in this section made reference only to relaxation times of the vibrational mode. The sum in Eq.~(\ref{eq:vib_relax_equation_again}) only comprises the diatomic species. And this makes sense, as these are the only mixture components that may possess vibrational energy. However, as mentioned in Sec.~\ref{sec:fluid_equations}, when the electronic energy is also taken into account, we assume that the vibrational and electronic mode relax in lockstep at the characteristic time scale of the former. In that case, a slightly modified version of Eq.~(\ref{eq:vib_relax_equation_again}) is employed, which includes contributions from all species. Consult Sec.~S3.C of the supplemental information for further details.


\begin{table}
 \centering
 \caption{DMS-derived fit parameters for pressure-weighted characteristic vibrational relaxation times $\mathrm{[atm \cdot s]}$ evaluated according to Eq.~(\ref{eq:two_slope_fit_again})}
 \label{tab:tau_v_fit_combined}
 
 \begin{tabular}{l c c c c l}
  Species               & $m^\mathrm{low}$ & $n^\mathrm{low}$ & $m^\mathrm{high}$ & $n^\mathrm{high}$ & note \\
  pair $s-q$            & $\mathrm{[K^{1/3}]}$ & $\mathrm{[-]}$ & $\mathrm{[K^{1/3}]}$ & $\mathrm{[-]}$ & \\ \hline \hline
  $\mathrm{N_2-N_2}$    & 221.0 & -24.83 &  33.30 & -17.31 & \\
  $\mathrm{N_2-O_2}$    & 115.0 & -20.25 & -51.42 & -15.26 & \\
  $\mathrm{N_2-NO}$     & \multicolumn{4}{c}{--- Re-use parameters for $\mathrm{N_2-N_2}$ ---} & [a] \\   
  $\mathrm{N_2-N}$      & 239.4 & -28.52 &  32.34 & -18.39 & \\
  $\mathrm{N_2-O}^{*}$  & 149.5 & -23.10 & -1.201 & -17.00 & [b] \\
  $\mathrm{N_2-O}^{**}$ & 149.5 & -23.51 & -1.201 & -17.40 & [c] \\ \hline
  $\mathrm{O_2-N_2}$    & 134.0 & -22.37 & -0.7123 & -16.95 & \\
  $\mathrm{O_2-O_2}$    & 129.0 & -22.29 & -4.384  & -16.70 & \\
  $\mathrm{O_2-NO}$     & \multicolumn{4}{c}{--- Re-use parameters for $\mathrm{O_2-O_2}$ ---} & [a] \\
  $\mathrm{O_2-N}$      & 81.22 & -21.16 & 9.111   & -17.61 \\
  $\mathrm{O_2-O}$      & 3.018 & -17.74 & -121.6  & -15.82 \\ \hline
  $\mathrm{NO-N_2}$     & \multicolumn{4}{c}{--- Re-use parameters for $\mathrm{NO-NO}^{**}$ ---} & [a] \\
  $\mathrm{NO-O_2}$     & \multicolumn{4}{c}{--- Re-use parameters for $\mathrm{NO-NO}^{**}$ ---} & [a] \\
  $\mathrm{NO-NO}^{*}$  & 127.2  & -20.72 &  -27.61 & -14.06 & [b] \\
  $\mathrm{NO-NO}^{**}$ & 127.2  & -23.20 &  -27.61 & -16.55 & [c] \\
  $\mathrm{NO-N}^{*}$   &  64.31 & -19.08 &  -62.33 & -15.82 & [b] \\
  $\mathrm{NO-N}^{**}$  &  64.31 & -20.06 &  -62.33 & -16.80 & [c] \\
  $\mathrm{NO-O}^{*}$   & -24.76 & -15.37 & -113.0  & -18.91 & [b] \\
  $\mathrm{NO-O}^{**}$  & -24.76 & -17.16 & -113.0  & -20.70 & [c] \\
  \\
  \multicolumn{6}{l}{[a] Relaxation time $\tau_{s,q}^\mathrm{v} \rightarrow \infty$ when comparing against DMS} \\
  \multicolumn{6}{l}{benchmarks by setting $m_l = m_h = 0$ and $n_l = n_h$ to ``very large''} \\
  \multicolumn{6}{l}{constant.} \\
  \multicolumn{6}{l}{[b] Used only when comparing against DMS (MMT-benchmark)} \\
  \multicolumn{6}{l}{[c] Used with full model for realistic conditions (MMT)}
 \end{tabular} 
 
\end{table}


\section{MMT model comparison against Direct Molecular Simulations} \label{sec:testing_vs_dms_again}

In Sec.~VII of Ref.~\cite{chaudhry25a}, we presented a first set of comparison calculations for our CFD implementation of the MMT model against DMS benchmark solutions. Those calculations dealt exclusively with ``pure'' $\mathrm{N_2/N}$ and $\mathrm{O_2/O}$ mixtures and thus only involved dissociation reactions of the respective diatomic species. In this section we extend this effort to the updated MMT model for 5-species air (i.e. $\mathrm{N_2}$, $\mathrm{O_2}$, $\mathrm{NO}$, $\mathrm{N}$ and $\mathrm{O}$), which now includes formation and consumption of nitric oxide via the Zeldovich exchange reactions, among others. Thus, we now employ the more complete set of kinetic rate parameters from Sec.~\ref{sec:new_mmt_parameters} and curve-fit parameters required to evaluate Landau-Teller type vibrational relaxation times from Sec.~\ref{sec:tau_vt_fits_again} in this paper. As mentioned in Sec.~VII of Ref.~\cite{chaudhry25a}, such CFD comparisons must be performed consistently by including only collision-pair interactions corresponding to the available PESs used in the DMS calculations. This also means that the thermodynamic properties employed in the CFD-MMT calculations of this section explicitly \emph{do not} account for contributions of molecular or atomic species' electronic excited states. Refer to Tables~\ref{tab:arrhenius_2024_parameters_complete}, \ref{tab:mmt_2024_parameters_complete} and \ref{tab:tau_v_fit_combined} (including footnotes) for the interactions modeled and parameters used for MMT results presented in this section. Further recall that constant non-Boltzmann factors of $f_k^\mathrm{NB} = 0.5$ and $f_\varepsilon^\mathrm{NB} = 0.85$ are employed in Eq.~(\ref{eq:mmt_rate_coefficient_again}) and (\ref{eq:devib_mmt}) respectively throughout all MMT calculations presented in this section. Finally, note that the CFD solutions employing the MMT or Park models in this section and in Secs.~\ref{sec:mmt_nb_vs_vnb}-\ref{sec:mmt_vs_park} were all obtained using a standard explicit fourth-order Runge-Kutta ODE time integration scheme.

%
\subsection{Vibrational excitation and dissociation of 5-species air mixture in isothermal heat baths}\label{sec:isothermal_air5_mmt_vs_dms}

In analogous fashion to Ref.~\cite{chaudhry25a}, the first set of comparison calculations are carried out in isochoric, isothermal heat baths. In the CFD calculations we keep the mixture trans-rotational temperature fixed over time, while the mixture vibrational temperature is initially set to $300 \, \mathrm{K}$ and the initial number density to $10^{24} \, \mathrm{m^{-3}}$. The gas now consists of a molecular nitrogen/oxygen mixture with initial mole fractions $x_{\mathrm{N_2}} = 0.8$ and $x_{\mathrm{O_2}} = 0.2$ respectively. As was the case in Sec.~VII.A of Ref.~\cite{chaudhry25a}, the initial conditions are slightly different for the DMS reference calculations. Here, every species' respective translational temperature is kept constant by re-sampling their center-of-mass velocities from their respective Maxwellians at the heat bath's temperature after every time step. Meanwhile, the internal energy states of all molecules initially present are sampled from the respective Boltzmann rovibrational distribution at common $T_\mathrm{r, N_2}(t=0) = T_\mathrm{v, N_2}(t=0) = 300 \, \mathrm{K}$ and $T_\mathrm{r, O_2}(t=0) = T_\mathrm{v, O_2}(t=0) = 300 \, \mathrm{K}$ respectively. Thus unlike the CFD calculations, where translational and rotational modes are assumed to be equilibrated at all times, in the DMS calculation there exists an additional early phase of rotational nonequilibrium. Beyond $\mathrm{N_2}$- and $\mathrm{O_2}$- dissociation responsible for the formation of atomic nitrogen and oxygen, these test cases simulate the production of nitric oxide, primarily as a result of the Zeldovich exchange reactions. In addition to this, a range of other reactions involving $\mathrm{NO}$ molecules affect the overall evolution of the 5-species mixture composition. 

\begin{figure}
 
 \subfloat[Isothermal heat bath at $T = 8\,000 \, \mathrm{K}$]{\label{fig:iso_8k_dms_vs_cfd}
 \includegraphics[width=0.5\textwidth]{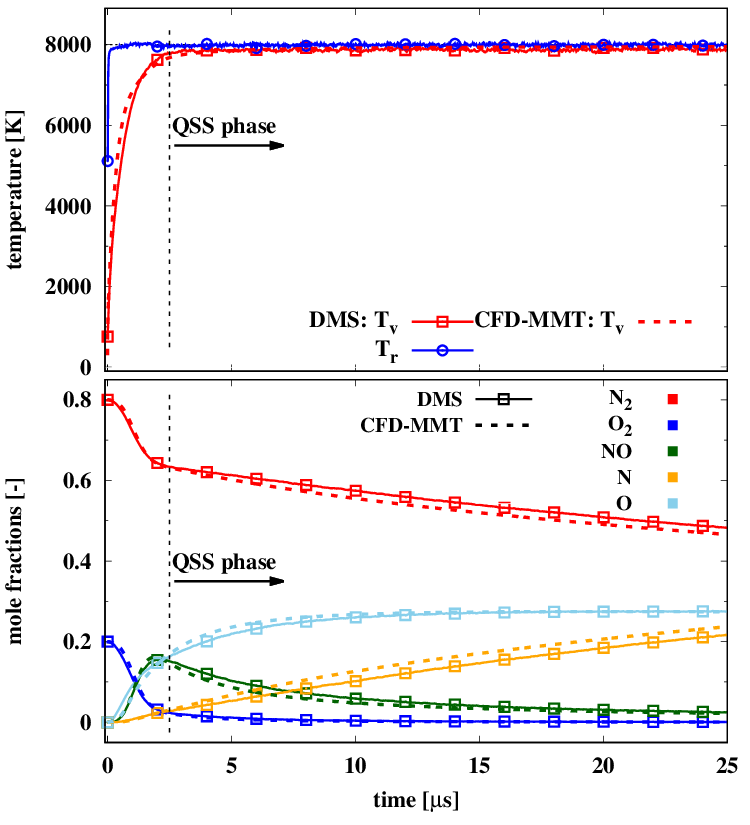}}~
 \subfloat[Isothermal heat bath at $T = 10\,000 \, \mathrm{K}$]{\label{fig:iso_10k_dms_vs_cfd}
 \includegraphics[width=0.5\textwidth]{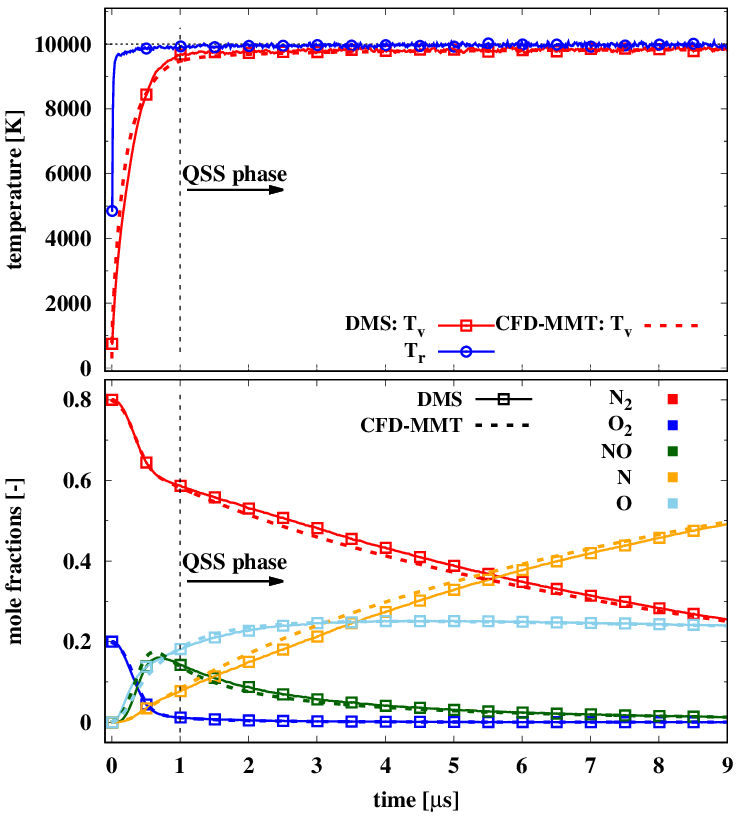}}
 
 \subfloat[Isothermal heat bath at $T = 15\,000 \, \mathrm{K}$]{\label{fig:iso_15k_dms_vs_cfd}
 \includegraphics[width=0.5\textwidth]{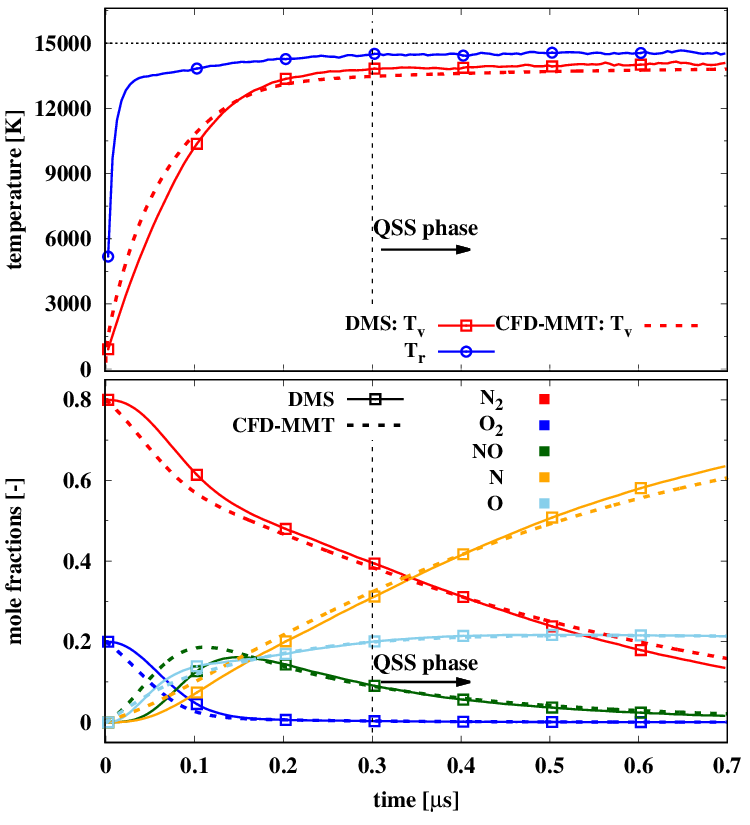}}~
 \subfloat[Isothermal heat bath at $T = 20\,000 \, \mathrm{K}$]{\label{fig:iso_20k_dms_vs_cfd}
 \includegraphics[width=0.5\textwidth]{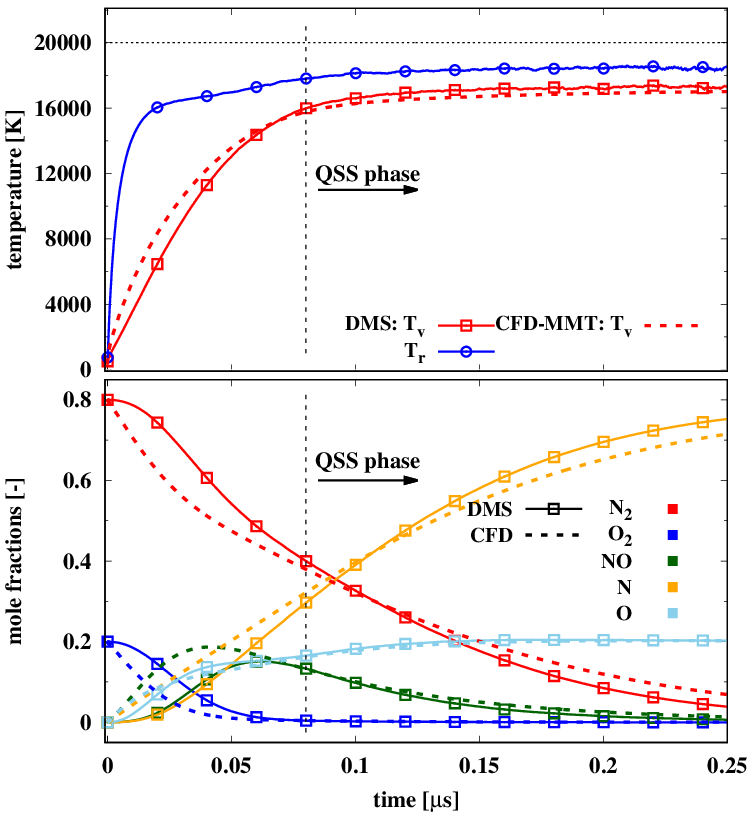}}
 
 \caption{Comparison of CFD-MMT model against DMS: 5-species air in isothermal heat bath}
 \label{fig:iso_combined_vs_cfd}
\end{figure}

In Figs.~\ref{fig:iso_8k_dms_vs_cfd}-\ref{fig:iso_20k_dms_vs_cfd} we compare predictions with the CFD-MMT model to DMS reference solutions at $T = 8\,000\, \mathrm{K}$-$20\,000\, \mathrm{K}$ respectively. All figures follow the pattern established in Sec.~VII.A of Ref.~\cite{chaudhry25a}, with time-dependent vibrational and rotational temperatures plotted in the upper half and corresponding species mole fraction profiles in the lower half. The CFD-MMT profiles are plotted with dotted lines and corresponding DMS results with unfilled symbols over continuous ones of the same color. Some of the DMS solutions used for reference were originally presented in Ref.~\cite{torres24a} and additional details may be found therein.

Note that both for the CFD-MMT and DMS calculations the temperature profiles shown represent mixture-weighted values. Since these air mixtures may contain up to three diatomic species simultaneously, in the CFD-MMT calculations the mixture vibrational temperature $T_\mathrm{v}$ at every time step is calculated as implicit solution to the equation:
\begin{equation}
 E_\mathrm{v} = \sum_{s \in D} \rho_s \, e_{\mathrm{v}, s} (T_\mathrm{v}) \label{eq:rhoevib_mix}
\end{equation}
where the instantaneous mixture vibrational energy per unit volume $E_\mathrm{v}$ is one of the state variables in the governing equations, see Eq.~(\ref{eq:vib_energy_balance_0d}), and each diatom's specific vibrational energy is calculated from Eq.~(S32), or alternatively Eq.~(S41) of the supplemental information. As discussed there, for these comparisons against DMS we employ the custom PES-derived fit parameters from Table~S2 to evaluate $h_s^\mathrm{fit} |_{T_\mathrm{ref}} (T_\mathrm{v})$. Therefore, by construction we deliberately exclude electronic excited state contributions of any of the 5 species.

For the $T = 8\,000\, \mathrm{K}$-case the upper half of Fig.~\ref{fig:iso_8k_dms_vs_cfd} shows that the CFD model does an excellent job of replicating the vibrational excitation behavior predicted by DMS (dotted red vs. solid $T_\mathrm{v}$ lines). The rotational temperature profile for the DMS solution (blue line) reaches the heat bath temperature almost instantaneously and remains there for the rest of the simulation. Thus, the assumption of constant trans-rotational temperature $T$ inherent in the two-temperature CFD model is satisfied for practically the entire simulated time. Furthermore, just as was observed in Sec.~VII.A of Ref.~\cite{chaudhry25a}, in Fig.~\ref{fig:iso_8k_dms_vs_cfd} we see that both the CFD- and DMS-derived vibrational temperature profiles level off at a value slightly below $T$ and do not fully equilibrate with the heat bath temperature for the remainder of simulated time. Thus, just as was seen for the simpler $\mathrm{N_2/N}$ and $\mathrm{O_2/O}$ mixtures in Ref.~\cite{chaudhry25a}, the DMS reference calculations for 5-species air point to the existence of a QSS-dissociation phase responsible for this temperature gap. Note that, as discussed in Ref.~\cite{torres24a}, the presence of additional reactions in 5-species air (e.g. Zeldovich exchange and other reactions) does not prevent this QSS dissociation phase from being established.

In the lower half of Fig.~\ref{fig:iso_8k_dms_vs_cfd} we compare mole fraction profiles for all 5 species ($\mathrm{N_2}$ in red, $\mathrm{O_2}$ in dark blue, $\mathrm{NO}$ in green, $\mathrm{N}$ in orange and $\mathrm{O}$ in light blue) and observe very close agreement between CFD (dotted lines) and DMS (solid lines) results. The CFD-MMT results reproduce all major features of the reference calculations, such as the early drop in molecular oxygen from $x_\mathrm{O_2} = 0.2$ to nearly zero (primarily due to rapid $\mathrm{O_2-O_2}$, $\mathrm{O_2-N_2}$ and $\mathrm{O_2-O}$ dissociation), the early decrease  from $x_\mathrm{N_2} = 0.8$ to about $0.65$ and near-simultaneous rise in nitric oxide mole fraction from zero to its peak of $x_\mathrm{NO} \approx 0.18$ (both caused by the first Zeldovich reaction $\mathrm{N_2 + O} \rightarrow \mathrm{NO + N}$). Beyond this point CFD-MMT closely matches DMS in predicting the subsequent slow decline in $x_\mathrm{N_2}$ and $x_\mathrm{NO}$ due to dissociation, accompanied by a simultaneous rise in atomic nitrogen and oxygen mole fractions. The only notable discrepancy is a small time lag building up at later times between the CFD- and DMS-derived $x_\mathrm{N_2}$-profiles in red, mirrored by the corresponding $x_\mathrm{N}$-profiles in orange. However, these relative discrepancies in mixture composition never exceed a few percent.

In Figs.~\ref{fig:iso_10k_dms_vs_cfd}-\ref{fig:iso_20k_dms_vs_cfd} we compare CFD-MMT and DMS predictions for the same 5-species air mixture, but at increasingly higher heat bath temperatures. Overall agreement between both methods' predicted $T_\mathrm{v}$ profiles (dotted vs. continuous red lines) is good, despite stronger coupling between vibrational and rotational modes in the DMS results. This effect is particularly noticeable for cases $T = 15\,000 \, \mathrm{K}$ and $20\,000 \, \mathrm{K}$ where the DMS calculations predict that rotational and translational modes remain slightly out of equilibrium for the duration of the QSS phase. Furthermore, mixture composition profiles in Figs.~\ref{fig:iso_10k_dms_vs_cfd}-\ref{fig:iso_20k_dms_vs_cfd} reveal that, with increasing $T$ a greater portion of chemical activity takes place before the gas even enters the QSS dissociation regime. Differences between predicted mixture compositions are most noticeable early on, during the vibrational (and rotational in DMS) excitation phase. The dissociation process occurring during the rotational and vibrational energy excitation phases, including non-Boltzmann effects, has been studied previously for nitrogen using a three-temperature continuum model~\cite{singh20b, singh20c}. However, recall that in the two-temperature formulation used here with the MMT model a common translation-rotational temperature $T = T_\mathrm{t} = T_\mathrm{r}$ is being imposed. Separate rotational relaxation is not modeled in these CFD calculations and any rotation-vibration coupling is ignored. Since the rotational mode starts out fully excited at $t = 0$, the initial gas in the CFD calculation possesses a higher thermal energy than in the DMS solution. This allows for faster initial dissociation of $\mathrm{O_2}$, in turn causing faster production of atomic oxygen and subsequent faster $\mathrm{N_2}$-consumption coupled with faster $\mathrm{NO}$-formation through the first Zeldovich reaction. Despite these built-in simplifications the MMT model does a remarkably good job of reproducing all major features of the DMS reference solutions at these higher temperatures.

\subsection{Vibrational excitation and dissociation of 5-species air mixture in adiabatic heat baths} \label{sec:nitro_oxy_adia_again}

After the isothermal heat bath comparisons of Sec.~\ref{sec:isothermal_air5_mmt_vs_dms}, we perform a similar analysis for adiabatic conditions. In this scenario the heat bath's total energy density $E$ remains constant and translational temperature freely adjusts in response to internal energy redistribution processes and chemical reactions, which gradually drive the gas closer to equilibrium. These conditions allow us to better assess the MMT model's behavior over a wide temperature range within a single calculation and can be considered more representative of post-shock conditions.

Two separate cases for 5-species air are simulated using the MMT model and compared to corresponding DMS reference solutions. Both cases are distinguished by their heat bath's respective specific thermal energy $e$ (constant throughout the simulation\footnote{Gas velocity is set to $\vec{u} = \vec{0}$ and $\rho$ in the reactor remains constant. Therefore $e = E/\rho - \tfrac{1}{2} |\vec{u}|^2$ is also constant.}). Initial conditions are summarized in Table~\ref{tab:air5_adia_conditions} and, as was the case for the isothermal heat baths of Sec.~\ref{sec:isothermal_air5_mmt_vs_dms}, they differ slightly between CFD and the equivalent DMS calculations. In the DMS calculations the translational temperature $T_\mathrm{t}^\mathrm{DMS}(t=0)$ (same for $\mathrm{N_2}$ and $\mathrm{O_2}$) is initialized at a higher value than the common rotational and vibrational temperatures $T_\mathrm{r}^\mathrm{DMS}(t=0) = T_\mathrm{v}^\mathrm{DMS}(t=0)$, causing both molecular oxygen and nitrogen to undergo rapid early translation-rotational relaxation, in parallel with a slower vibrational excitation phase. By construction in the two-temperature CFD model, mixture translational and rotational modes remain in equilibrium at all times. Thus, we initialize the combined translation-rotational temperature to the weighted average $T^\mathrm{CFD}(t=0) = \tfrac{3}{5} T_\mathrm{t}^\mathrm{DMS}(t=0) + \tfrac{2}{5} T_\mathrm{r}^\mathrm{DMS}(t=0)$ in order to match the specific energy of the DMS case. Recall that the weights $c_\mathrm{v}^\mathrm{t} / (c_\mathrm{v}^\mathrm{t} + c_\mathrm{v}^\mathrm{r}) = 3/5$ and $c_\mathrm{v}^\mathrm{r} / (c_\mathrm{v}^\mathrm{t} + c_\mathrm{v}^\mathrm{r}) = 2/5$ correspond to the relative energy contributions to heat capacity at constant volume of the translational and rotational modes in a fully excited diatomic gas. In both cases the initial translation-rotational temperature of the CFD calculations will be lower than the DMS's initial translational one, because the two modes in CFD begin ``pre-equilibrated''. As seen in Table~\ref{tab:air5_adia_conditions}, this also has repercussions for the initial pressures in the CFD and DMS calculations, because $p^\mathrm{DMS} = \rho \, ( \bar{R} / \sum_{s \in S} \{ x_s \, M_s \} ) \, T_\mathrm{t}^\mathrm{DMS}$ vs. $p^\mathrm{CFD} = \rho \, ( \bar{R} / \sum_{s \in S} \{ x_s \, M_s \} ) \, T^\mathrm{CFD}$.


\begin{table}
 \centering
 \caption{DMS and CFD initial conditions for adiabatic heat bath calculations in 5-species air}
 \label{tab:air5_adia_conditions}
 
 \begin{tabular}{c c c c c c c c c}
                   &        & initial & \multicolumn{3}{c}{DMS initial conditions} & \multicolumn{3}{c}{CFD initial conditions} \\
               $e^{[*]}$ & $\rho \times 10^3$ & moles & $T_\mathrm{t}$ & $T_\mathrm{r} = T_\mathrm{v}$ & $p$ & $T$ & $T_\mathrm{v}$      & $p$ \\
  $\mathrm{[MJ/kg]}$ & $\mathrm{[kg/m^3]}$ & $\mathrm{[-]}$ & $\mathrm{[K]}$ & $\mathrm{[K]}$ & $\mathrm{[kPa]}$ & $\mathrm{[K]}$ & $\mathrm{[K]}$      & $\mathrm{[kPa]}$ \\ \hline
  $8.409$ & $1.961$ & $80\% \mathrm{N_2}, \, 20\%\mathrm{O_2}$ & $19\,920$ & $300$ & $11.27$ & $12\,070$ & $300$ & $6.832$ \\
  $15.61$ & $2.524$ & $80\% \mathrm{N_2}, \, 20\%\mathrm{O_2}$ & $36\,570$ & $300$ & $26.64$ & $22\,050$ & $300$ & $16.06$ \\
  \\
  \multicolumn{9}{l}{[*] Thermodynamic properties for benchmarking against DMS \emph{exclude} electronic energy} \\
  \multicolumn{9}{l}{contributions and are computed using fit coefficients from Table~S2 of the} \\
  \multicolumn{9}{l}{supplemental information.}
 \end{tabular}
 
\end{table}

\begin{figure}
 \centering
 
 \subfloat[Adiabatic heat bath at {$e = 8.409 \, \mathrm{MJ/kg}$}]{\label{fig:adia_h10p1_combined_vs_dms}
 \includegraphics[width=0.5\textwidth]{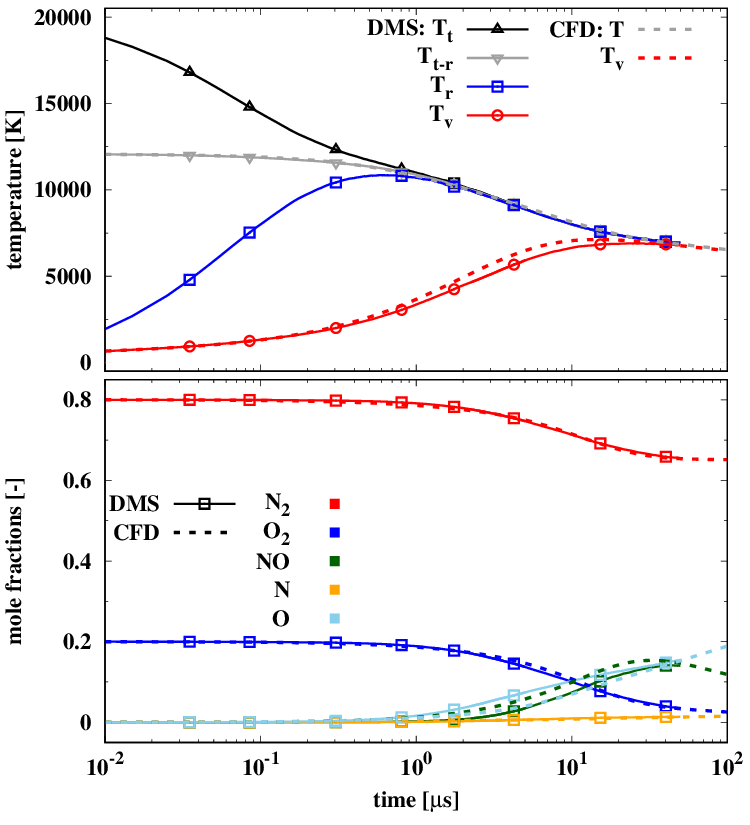}}~
 \subfloat[Adiabatic heat bath at {$e = 15.61 \, \mathrm{MJ/kg}$}]{\label{fig:adia_h17p9_combined_vs_dms}
 \includegraphics[width=0.5\textwidth]{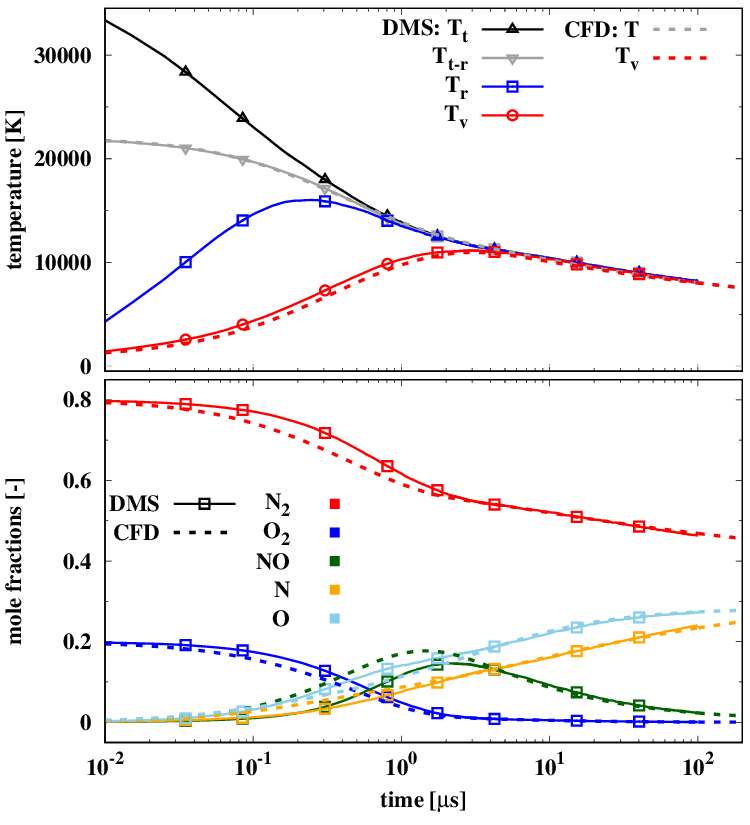}}
 
 \caption{Comparison of CFD-MMT model against DMS: 5-species air in adiabatic heat bath}
 \label{fig:adia_combined_vs_dms}
\end{figure}

Focusing on the lower-energy case first, the upper half Fig.~\ref{fig:adia_h10p1_combined_vs_dms}, shows DMS profiles of mixture translational, rotational and vibrational temperature as continuous black, blue and red lines respectively. The mixture rotational and vibrational temperatures are computed in the same manner as in Sec.~\ref{sec:isothermal_air5_mmt_vs_dms}, taking into account composition-weighted contributions of all three diatomic species. The mixture translational temperature is calculated directly as $T_\mathrm{t}^\mathrm{DMS} = p / (n \, \mathrm{k_B})$, where $n$ is the instantaneous mixture number density and the instantaneous mixture static pressure in turn is calculated as the trace of the mixture kinetic pressure tensor $p = \frac{1}{3} \left( p_{xx} + p_{yy} + p_{zz} \right)$, as defined in App.~C of Ref.~\cite{torres22b}. A combined mixture trans-rotational temperature for DMS (continuous gray line) is now computed as $T_\mathrm{t-r}^\mathrm{DMS} = 3/5 \, T_\mathrm{t}^\mathrm{DMS} + 2/5 \, T_\mathrm{r}^\mathrm{DMS}$. Corresponding CFD profiles for mixture trans-rotational and vibrational temperatures (dotted gray and red lines) are shown in the upper half of Fig.~\ref{fig:adia_h10p1_combined_vs_dms}. Near-exact overlap between the $T$ and $T_\mathrm{t-r}$-profiles obtained with both methods can be observed, whereas the CFD vibrational temperature runs slightly ahead the DMS counterpart during the initial relaxation phase. However, this temperature gap never exceeds a few percent and narrows toward the end of the simulated time. The bottom half of Fig.~\ref{fig:adia_h10p1_combined_vs_dms} shows mole fraction profiles of all five mixture components using the same color scheme and line patterns established in Sec.~\ref{sec:isothermal_air5_mmt_vs_dms}. Over the course of the first $50$ microseconds one observes the initial $20\%$ of $\mathrm{O_2}$ (dark blue lines) being almost completely consumed, while the mole fraction of $\mathrm{N_2}$ decreases from $0.8$ to roughly $0.65$ in the same time span. The decrease in $\mathrm{O_2}$ mole fraction can be attributed almost exclusively to dissociation (reactions 6-7 and 9-10 in Table~\ref{tab:mmt_2024_reactions}). On the other hand, the main cause for the early drop in $\mathrm{N_2}$ concentration is the first Zeldovich reaction (reaction 16 of Table~\ref{tab:mmt_2024_reactions}) rather than $\mathrm{N_2}$-dissociation itself. The exchange reaction activates as soon as the first oxygen atoms become available, producing nitric oxide and atomic nitrogen in the process. Very close agreement between the CFD- and DMS-predicted mole fraction profiles for $\mathrm{N_2}$ (dotted vs. continuous red lines), $\mathrm{O_2}$ (dark blue) and $\mathrm{N}$ (orange) can be observed for the entire $100$ microseconds plotted. For nitric oxide (dark green lines) and atomic oxygen (light blue) differences become most noticeable in the time window spanning $1-50 \, \mathrm{\mu s}$. During this interval the CFD model predicts simultaneously higher $\mathrm{NO}$ and lower $\mathrm{O}$ mole fractions than the DMS calculation.

Results for the higher-energy case are shown in Fig.~\ref{fig:adia_h17p9_combined_vs_dms}. As seen in the upper half of the figure, vibration-translational relaxation is effectively complete after the first $5$ microseconds, nearly ten times quicker than at the lower-energy conditions. Again, very close agreement between the CFD-predicted profiles for $T$ and $T_\mathrm{v}$ (dotted gray and red lines) and the equivalent DMS solution (continuous lines) can be observed. These two adiabatic cases further confirm that the MMT model does a remarkably good job of reproducing all major features observed in the benchmark DMS solutions, insofar as this can be expected given the simplifying assumptions of the two-temperature CFD implementation.


\section{MMT model behavior when approaching chemical equilibrium} \label{sec:mmt_nb_vs_vnb}

Section~VI of Ref.~\cite{chaudhry25a} discusses limitations inherent to the use of constant non-Boltzmann correction factors when evaluating dissociation rate coefficients and associated vibrational energy removed by dissociation, i.e. Eqs.~(\ref{eq:mmt_rate_coefficient_again}) and (\ref{eq:devib_mmt}), in the MMT model. These limitations were previously investigated by Singh and Schwartzentruber~\cite{singh22a}, who proposed varying the non-Boltzmann factors as the gas approaches chemical equilibrium. Specifically, they derived an expression where the factor was a function of the ratio between the instantaneous chemical composition and the local equilibrium composition, such that the non-Boltzmann factor smoothly approaches unity at equilibrium. Their original expression was slightly modified in Sec.~VI of Ref.~\cite{chaudhry25a} to yield the currently used functional form, i.e. Eq.~(31) in Ref.~\cite{chaudhry25a}. When these variable (i.e. composition-dependent) non-Boltzmann factors are employed, they enable the model to reproduce the correct forward (dissociation) and backward (recombination) reaction rates, as well as vibrational energy removal and replenishment rates as the gas mixture reaches thermo-chemical equilibrium.

\begin{figure}
 \centering
 
 \begin{minipage}{0.5\textwidth}
  \subfloat[Isothermal heat bath at $T = 8\,000 \, \mathrm{K}$]{\label{fig:iso_8k_combined_mmt_nb_vnb}
  \includegraphics[width=\textwidth]{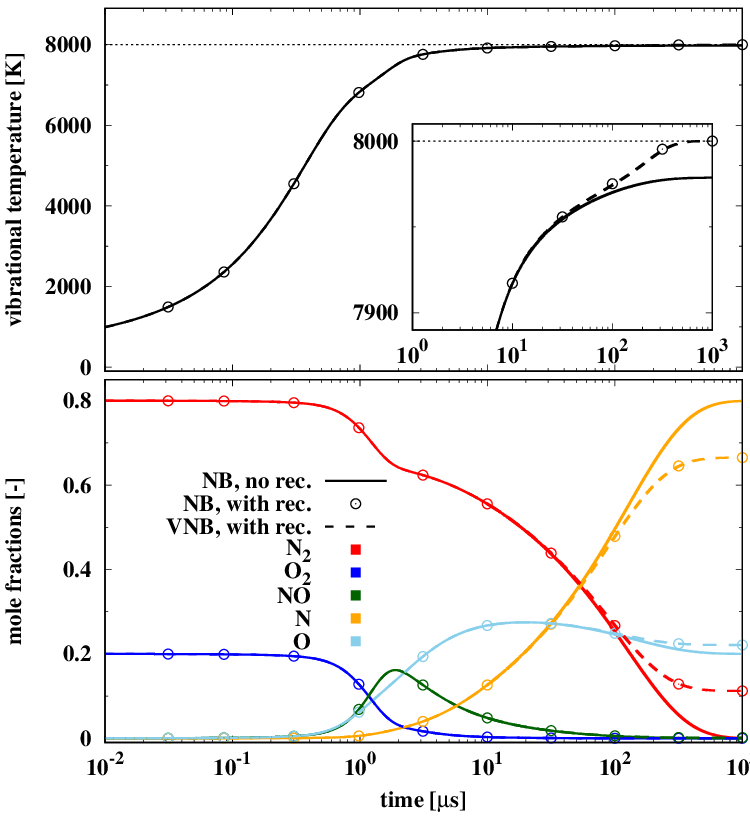}}
 \end{minipage}~
 \begin{minipage}{0.5\textwidth}
  \subfloat[Isothermal heat bath at $T = 15\,000 \, \mathrm{K}$]{\label{fig:iso_15k_combined_mmt_nb_vnb}
  \includegraphics[width=\textwidth]{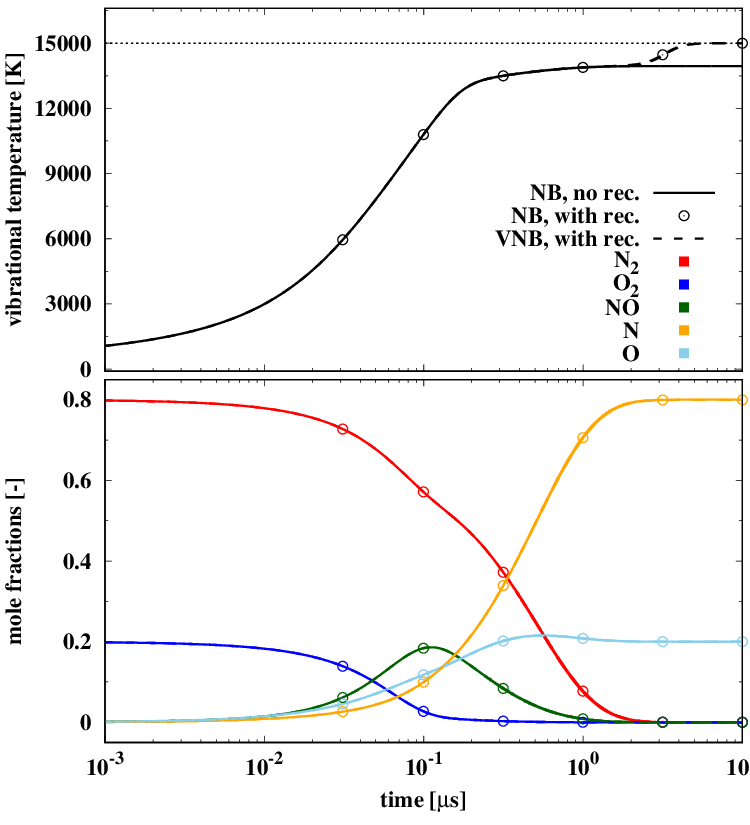}}
 \end{minipage}
 
 \caption{Isothermal heat bath: MMT solutions obtained using constant (NB) vs. variable (VNB) non-Boltzmann factors}
 \label{fig:iso_combined_mmt_nb_vnb}
\end{figure}

In this section we examine the aggregate effect of switching from constant to variable non-Boltzmann factors in actual CFD simulations. Three sets of results, all using the MMT model, are shown in Fig.~\ref{fig:iso_combined_mmt_nb_vnb} at the two lower heat bath temperatures from Sec.~\ref{sec:isothermal_air5_mmt_vs_dms}. All use the exact same reaction rate and vibrational relaxation parameters as in the model benchmarking calculations of Sec.~\ref{sec:isothermal_air5_mmt_vs_dms}. The only difference is that the first one employs constant values $f_k^\mathrm{NB} = 0.5$ and $f_\varepsilon^\mathrm{NB} = 0.85$ and has all recombination reactions disabled (continuous lines, labeled as ``NB, no rec.'' in Figs.~\ref{fig:iso_8k_combined_mmt_nb_vnb}  and \ref{fig:iso_15k_combined_mmt_nb_vnb}). Thus, it is identical with the MMT solutions previously presented in Figs.~\ref{fig:iso_8k_dms_vs_cfd} and \ref{fig:iso_15k_dms_vs_cfd}, but integrated further in time (up to a total of $1\,000$ microseconds for the lower-temperature case). The much longer integration times in Fig.~\ref{fig:iso_combined_mmt_nb_vnb} relative to Fig.~\ref{fig:iso_combined_vs_cfd} become necessary to observe the gas mixture's eventual approach to chemical equilibrium. The second set (open circles labeled as ``NB, with rec.'' in Figs.~\ref{fig:iso_8k_combined_mmt_nb_vnb} and \ref{fig:iso_15k_combined_mmt_nb_vnb}) uses the same constant factors, but has recombination reactions enabled. Finally, the third set of results (dashed lines labeled as ``VNB, with rec.'' in Figs.~\ref{fig:iso_8k_combined_mmt_nb_vnb} and \ref{fig:iso_15k_combined_mmt_nb_vnb}) employs the variable factors as given by Eq.~(31) of Ref.~\cite{chaudhry25a} and also has recombination reactions enabled.

Focusing on the lower-temperature case first, Fig.~\ref{fig:iso_8k_combined_mmt_nb_vnb} reveals several commonalities. First, during the early stages, when the gas consists primarily of molecular species $\mathrm{N_2}$ and $\mathrm{O_2}$, all three solutions are identical. This is to be expected, given that during this period the nonequilibrium concentration ratios $\zeta_r$ (Eq.~(30) in Ref.~\cite{chaudhry25a}) of all major dissociation reactions are close to zero (thus $f_k^\mathrm{VNB}(\zeta_r\rightarrow0) \approx 0.5$ and $f_\varepsilon^\mathrm{VNB}(\zeta_r \rightarrow 0) \approx 0.85$) and recombination rates are still negligible. Later on, by the time $T_\mathrm{v}$ has almost reached the heat bath temperature and sufficient atomic nitrogen and oxygen have built up in the mixture, differences become more noticeable. In the solution with constant non-Boltzmann factors and recombination disabled, $T_\mathrm{v}$ eventually attains its QSS-dissociation-phase value and remains unchanged for the rest of the calculation. Meanwhile, dissociation of $\mathrm{N_2}$, $\mathrm{O_2}$ and $\mathrm{NO}$ continue unopposed until nothing but atomic nitrogen and oxygen remain in the mixture and true thermodynamic equilibrium is never reached. In the two solutions with recombination enabled, $T_\mathrm{v}$ also first approaches the QSS-phase plateau, but eventually continues climbing to fully equalize with the heat bath temperature (see close-ups in temperature plots). This occurs in tandem with the species mole fraction profiles leveling off and approaching the thermodynamic equilibrium composition at temperature $T$ and final equilibrium pressure in the heat bath.  

Any differences in final mixture composition between the three calculations are most noticeable at this lower-temperature condition, because the respective equilibrium compositions comprise small, but still noticeable amounts of molecular nitrogen. By contrast, at the higher-temperature condition shown in Fig.~\ref{fig:iso_15k_combined_mmt_nb_vnb}, true thermodynamic equilibrium compositions are already skewed so far toward the atomic species that any differences in final mole fractions between the three calculations become negligible and the only straightforward way to distinguish them is through the temperature profiles. For this higher-temperature heat bath the gaps between $T$ and $T_\mathrm{v}$ during QSS are much more noticeable and all three solutions reach this plateau together. As before, the solution using constant non-Boltzmann factors and neglecting recombination settles in at this QSS value, whereas in the other two solutions recombination reactions end up balancing with dissociation and prevent the entirety of diatoms from disappearing. Vibrational relaxation of this small remaining amount of diatomic species (mostly $\mathrm{N_2}$) ultimately drives $T_\mathrm{v}$ toward equilibrium with the heat bath temperature.

At both conditions examined, the two solutions that include recombination reactions, regardless of whether they employ constant or variable non-Boltzmann factors (unfilled circles vs. dashed lines), exhibit nearly identical behavior. Indeed, any differences are so small that they become almost indistinguishable at the scale of the plots. Thus, paradoxically the use of the arguably more accurate variable non-Boltzmann factors does not alter the overall results significantly. This apparent contradiction has a rather simple explanation: as discussed in Sec.~VI of Ref.~\cite{chaudhry25a}, the whole point of making the non-Boltzmann factors dependent on nonequilibrium concentration ratio $\zeta_r$ is that this dynamically re-scales dissociation and recombination rate coefficients (and related vibrational energy removal/replenishment rates) so that each, on their own, attain their expected \emph{thermal} rates as the mixture approaches equilibrium. By construction, when $\zeta_r \rightarrow 1$ close to the mixture's chemical equilibrium composition, differences between reaction rates computed using constant vs. variable non-Boltzmann factors are greatest. However, these differences are only evident when comparing one-way forward or backward rates. The \emph{net} rates of reaction and vibrational energy removal (e.g. Eqs.~(28) and (29) in Ref.~\cite{chaudhry25a}) on the other hand, are barely affected by this difference, since the same factors (constant or variable) are applied simultaneously to the forward and backward contributions to these terms.

In summary, the result of the comparisons in Fig.~\ref{fig:iso_combined_mmt_nb_vnb} is that at low- and high-temperature conditions the main distinguishing factor between the three sets of calculations is not whether constant or variable non-Boltzmann factors were used, but rather whether recombination reactions were included or neglected. Despite this, we recommend the MMT model to always be used with recombination enabled and to employ variable non-Boltzmann factors.

\section{Final additions to internal energy and MMT reaction models} \label{sec:dms_vs_real}

In Sec.~\ref{sec:testing_vs_dms_again} we compared the MMT model for 5-species air against benchmark solutions and all efforts concentrated on ensuring a consistent CFD vs. DMS comparison. It meant employing a set of model parameters that accounted for the limitations inherent to the DMS reference solutions. However, the MMT model is ultimately intended for studying real-world high-enthalpy flows generated in ground test facilities, or for simulating actual hypersonic flight conditions, which involves completing several components of the model.

First, we add electronic-excited-state energy contributions into the calculation of thermodynamic properties by reverting from the ground-electronic-state-only (PES-derived) curve fits to the original NASA Lewis ones (see discussion in Sec.~S1.A of the supplemental information). This implies a shift from viewing $T_\mathrm{v}$ as a purely vibrational temperature to one representing the energy content of a combined vibrational-electronic mode. It also implies that, by construction, vibrational and electronic energy relaxation rates will now be tightly coupled. As described in Sec.~\ref{sec:tau_vt_fits_again}, we also modify the manner in which we calculate the pair-wise vibrational relaxation times. We switch from the slower $\tau^\mathrm{v}$-parameters for some of the relaxation pairs (marked with an asterisk superscript in Table~\ref{tab:tau_v_fit_combined}) to the faster ones (marked with a double-asterisk superscript). Furthermore, we replace the parameters yielding near-infinite relaxation times for $\mathrm{NO - N_2}$,$\mathrm{NO - O_2}$, $\mathrm{N_2 - NO}$ and $\mathrm{O_2 - NO}$ with available parameters of what we consider the most similar interaction pairs (see Table~\ref{tab:tau_v_fit_combined}). We now also include all reactions absent from the ``benchmarking'' set (see footnote [a] in Tables~\ref{tab:mmt_2024_reactions} and \ref{tab:mmt_2024_parameters_complete}), again by using data from what we consider the most similar available reactions to act as stand-ins for the unavailable ones. Finally, as mentioned in Sec.~\ref{sec:mmt_params_for_real}, it may become necessary to modify the oxygen dissociation rate coefficients to include a multi-surface factor $\eta_\mathrm{O_2} = 16/3$ (see footnote [b] in Tables~\ref{tab:mmt_2024_reactions} and \ref{tab:mmt_2024_parameters_complete} and related discussion in Sec.~\ref{sec:mmt_params_for_real}) in order to account for enhanced dissociation rates from electronically excited $\mathrm{O_2}$. 

In Table~\ref{tab:air5_models_summary} we summarize the aforementioned changes, which result in three distinct versions of the MMT model for 5-species air: the MMT-benchmark version, the MMT version without enhancement factor $\eta_\mathrm{O_2}$ and the MMT model with the enhancement factor applied equally to all 5 oxygen dissociation reactions. In the remainder of this section we compare predictions with these three versions against each other to assess the effect that these changes have. 


\begin{table}
 \centering
 \caption{Choice of MMT model parameters used to simulate model-benchmarking-only vs. complete air mixtures}
 \label{tab:air5_models_summary}
 
 \begin{tabular}{p{1.4in} p{1.4in} p{1.4in} p{1.4in}}
                         & MMT-benchmark & MMT (no $\eta_\mathrm{O_2}$ factor) & MMT (full $\eta_\mathrm{O_2}$ factor) \\ \hline
  thermodynamics fits    & PES-derived                   & NASA Lewis                     & NASA Lewis        \\
  vibrational relaxation & parameter set [*]             & parameter set [**]             & parameter set [**] \\
  kinetic rate data      & 17 reactions                  & 21 reactions                   & 21 reactions  \\
  multi-surface factor: $\eta_\mathrm{O_2}$        & no                             & no & applied to reactions 6-10 in Table~\ref{tab:mmt_2024_parameters_complete} \\
 \end{tabular}

\end{table}

\begin{figure}
 \centering
 
 \begin{minipage}{0.5\textwidth}
  \subfloat[Isothermal heat bath at $T = 8\,000 \, \mathrm{K}$]{\label{fig:iso_8k_combined_mmt_vs_real}
  \includegraphics[width=\textwidth]{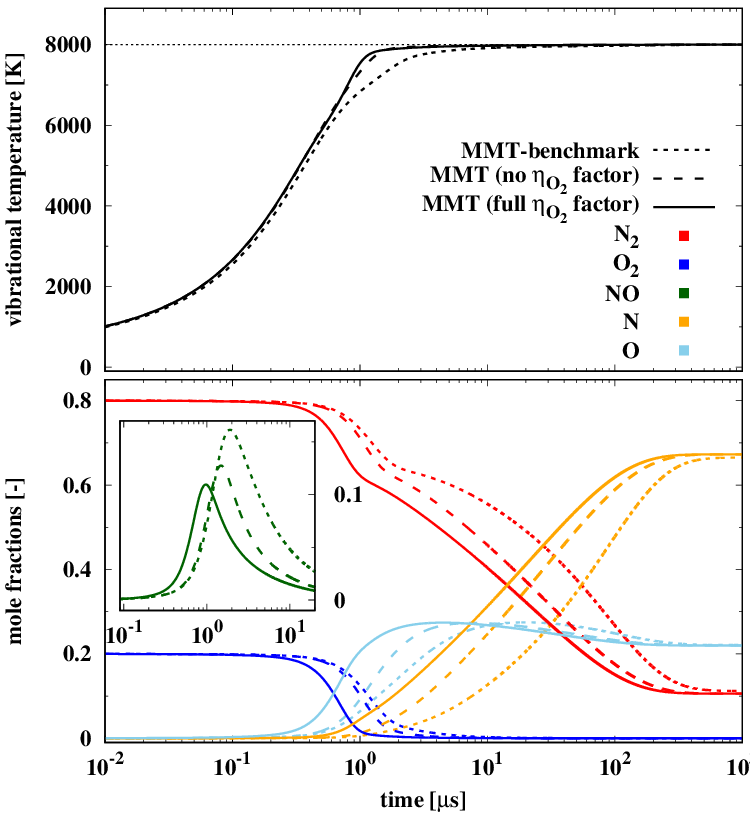}}
 \end{minipage}~
 \begin{minipage}{0.5\textwidth}
  \subfloat[Isothermal heat bath at $T = 15\,000 \, \mathrm{K}$]{\label{fig:iso_15k_combined_mmt_vs_real}
  \includegraphics[width=\textwidth]{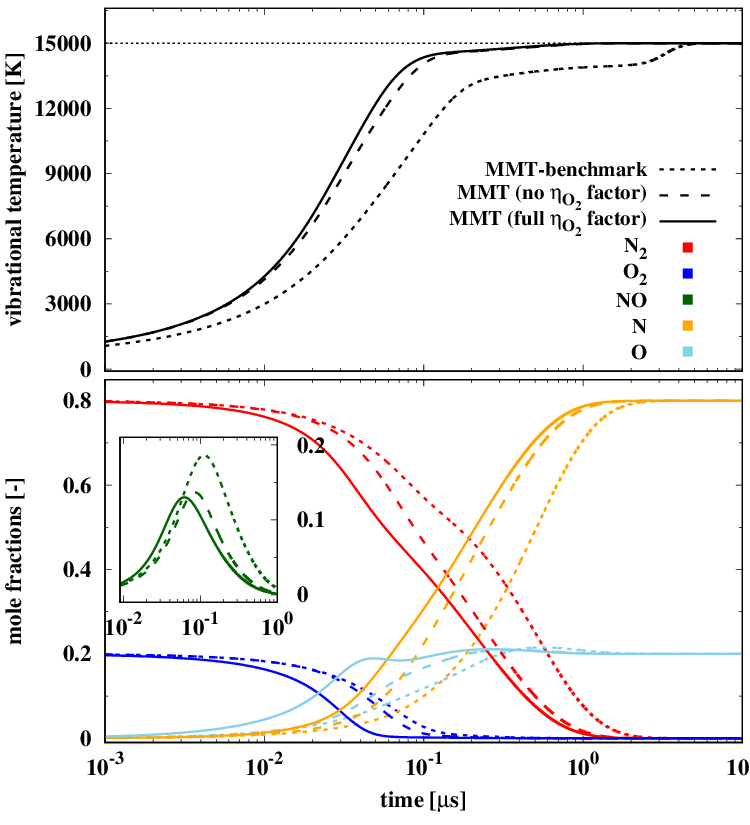}}
 \end{minipage}
 
 \caption{Isothermal heat bath results for the tree MMT parameter sets of Table~\ref{tab:air5_models_summary}}
 \label{fig:iso_combined_vs_real}
\end{figure}

In Fig.~\ref{fig:iso_combined_vs_real}, comparisons are made for heat bath temperatures $T = 8\,000 \, \mathrm{K}$ and $15\,000 \, \mathrm{K}$ and three sets of calculations are shown in each sub-figure: the first using the MMT-benchmark parameter set (dotted lines), the second using the MMT parameter set without the multi-surface enhancement factor for oxygen dissociation (dashed lines) and the third applying said factor equally to the 5 oxygen dissociation reactions (reactions 6-10 in Table~\ref{tab:mmt_2024_parameters_complete}). All three sets of calculations employ the variable non-Boltzmann correction factors discussed in Sec.~\ref{sec:mmt_nb_vs_vnb} and treat all reactions in Table~\ref{tab:mmt_2024_parameters_complete} as reversible. Thus, the MMT-benchmark solutions in Fig.~\ref{fig:iso_combined_vs_real} are identical with the ``VNB, with rec.'' solutions from Sec.~\ref{sec:mmt_nb_vs_vnb} (dashed lines in Fig.~\ref{fig:iso_combined_mmt_nb_vnb}).

As expected, the MMT-benchmark results are significantly different compared to the other two solutions, as this parameter set uses slower vibrational relaxation times for some species and is missing internal energy relaxation pairs and reactions for some species. Clear differences seen in Fig.~\ref{fig:iso_combined_vs_real} include that the rise in $T_\mathrm{v}$ is noticeably slower in the MMT-benchmark result and that the QSS temperature plateau is established at a much lower value compared to the other two solutions.

The important differences in Fig.~\ref{fig:iso_combined_vs_real} to focus on are those between the MMT results with and without the electronic excitation factor. The addition of the multi-surface enhancement factor to the 5 oxygen dissociation reactions has a significant effect in speeding up the rate at which $\mathrm{O_2}$ is removed from the mixture. Any uncertainty in this parameter will thus have a major effect on the predicted overall $\mathrm{O_2}$ consumption and associated $\mathrm{O}$ production rate. Furthermore, the rate at which $\mathrm{O}$-atoms become available early on also controls the rate at which the first Zeldovich reaction is able to begin producing the first $\mathrm{NO}$-molecules. Both figures also highlight the crucial role that the two exchange reactions play in shaping the overall evolution of the mixture composition, as the first Zeldovich reaction is responsible for producing much of the early $\mathrm{N}$ atoms, which may then be partially consumed in the second Zeldovich reaction running in its exothermic sense. This further increases the $\mathrm{O_2}$ consumption rate in the process. Beyond this, the first Zeldovich reaction remains the greatest contributor to the net rate of $\mathrm{N_2}$ destruction taking place over the entire simulated time, whereas the effect of the 5 nitrogen dissociation reactions is small by comparison. Although not shown here, these findings are broadly replicated at the higher-temperature heat bath conditions.

%
\section{Comparison of MMT and Park+M\&W models} \label{sec:mmt_vs_park}

In this section we compare the MMT model's predictions against those of the well-established Park $TT_\mathrm{v}$ model for high-temperature air~\cite{park90a}. The Park model is usually employed in combination with vibrational relaxation rates from the Millikan and White correlation~\cite{millikan63a} and Park's own high-temperature correction~\cite{park84a}. In the Park simulations we include only the reactions listed in Table~2 of Ref.~\cite{park93a} under the heading ``Dissociation Reactions'' which involve the five neutral air species $\mathrm{N_2}, \mathrm{O_2}, \mathrm{NO}, \mathrm{N}$ and $\mathrm{O}$, in addition to the two Zeldovich reactions listed in the same table under the heading ``NO Exchange reactions''. The M\&W and high-temperature correction parameters are taken from the section entitled ``Vibrational Relaxation Parameters'', also in Ref.~\cite{park93a}.

For the MMT model in this section we report two separate results using the parameter set either with, or without applying the multi-electronic-surface rate enhancement factor to all oxygen dissociation reactions (see two rightmost columns in Table~\ref{tab:air5_models_summary}). As a reminder, in the MMT model we evaluate the vibrational energy loss per dissociation using Eq.~(\ref{eq:devib_mmt}) for most dissociation reactions (reactions 1-10 and 14-15 and in Table~\ref{tab:mmt_2024_reactions}), whereas the remaining reactions assume non-preferential (i.e. $\langle \varepsilon_{v,s} \rangle_r = e_{\mathrm{v},s} (T_\mathrm{v}) / M_s$) vibrational energy-chemistry coupling. By contrast, in the Park model simulations we treat the vibrational energy change per reaction as non-preferential for all processes, regardless of whether they represent dissociation, or exchange reactions. To our knowledge, this currently represents the default choice in the CFD community for the Park model. Finally, note that in both the MMT and Park+M\&W simulations we now compute the thermodynamic properties (heat capacities, vibrational-electronic energies, Gibbs free energies) of all five species taking into account the contributions of excited electronic states (from the original NASA Lewis-fit~\cite{mcbride93a} parameters as listed in Table~S1 of the supplemental information). Thus, for both models in this section the temperature labeled as $T_\mathrm{v}$ will represent a combined vibrational-electronic one for the gas mixture and we now evaluate the relaxation term in the modified form as described in Sec.~S3.C of the supplemental information.

\begin{figure}
 \centering
 
 \begin{minipage}{0.5\textwidth}
  \subfloat[MMT (full $\eta_\mathrm{O_2}$ factor) vs. Park+M\&W]{\label{fig:iso_8k_combined_mmt_vs_park}
  \includegraphics[width=\textwidth]{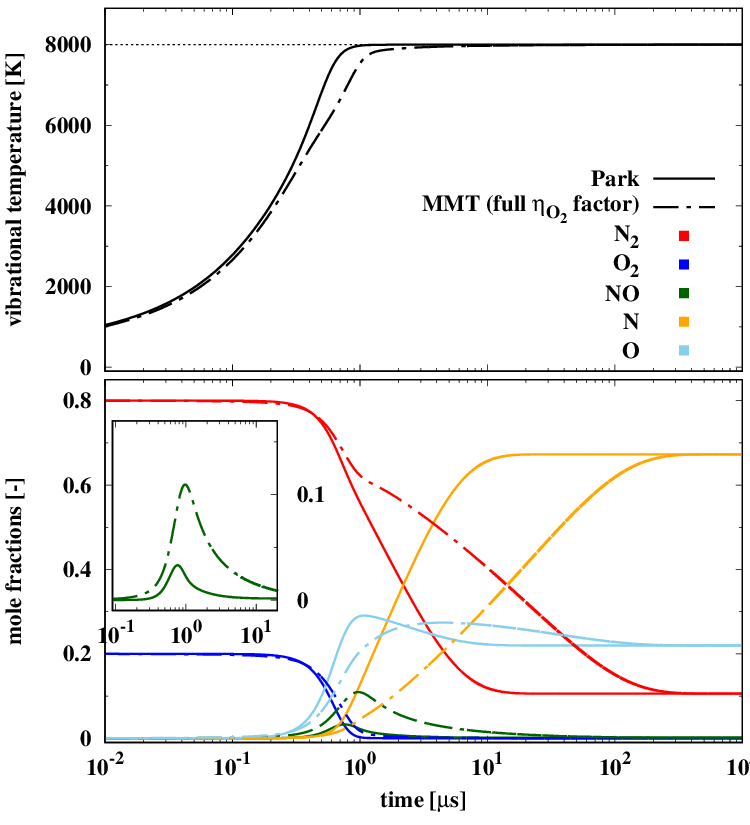}}
 \end{minipage}~
 \begin{minipage}{0.5\textwidth}
  \subfloat[MMT (no $\eta_\mathrm{O_2}$ factor) vs. Park+M\&W]{\label{fig:iso_8k_combined_mmt_nofac_vs_park}
  \includegraphics[width=\textwidth]{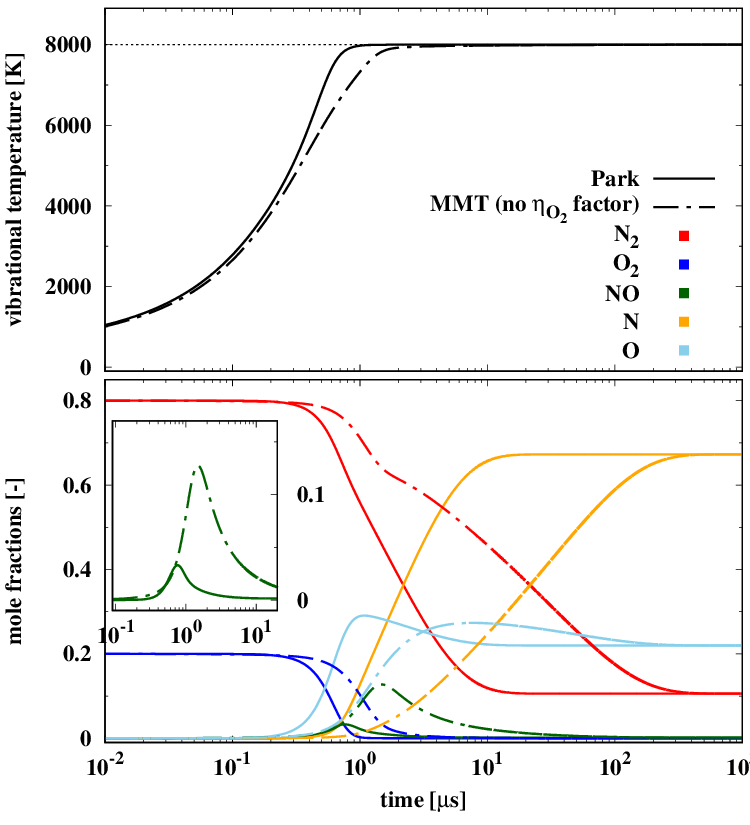}}
 \end{minipage}
 
 \caption{Isothermal heat bath $T = 8\,000 \, \mathrm{K}$, complete MMT vs. Park+M\&W}
 \label{fig:iso_8k_mmt_vs_park}
\end{figure}

We first compare the predictions of MMT (solid lines) and Park+M\&W (dash-dotted lines) in isothermal heat bath simulations at different temperatures. Results for $T = 8\,000 \, \mathrm{K}$ are shown in Fig.~\ref{fig:iso_8k_mmt_vs_park}, with vibrational temperatures in the upper half and species mole fractions in the lower half. Two sub-plots are shown. In Fig.~\ref{fig:iso_8k_combined_mmt_vs_park} we compare Park against MMT results including the multi-surface enhancement factors for $\mathrm{O_2}$-dissociation, whereas in Fig.~\ref{fig:iso_8k_combined_mmt_nofac_vs_park} this factor is absent. The vibrational temperature profiles for MMT and Park show close agreement throughout most of the simulated time, although both MMT curves begin to lag slightly behind Park+M\&W the more $T_\mathrm{v}$ approaches the heat bath temperature. The gap forming between MMT and Park roughly coincides with the onset of chemical reactions, as can be seen from the changing mole fractions in the lower half of both sub-figures. Whereas the $T_\mathrm{v}$-profile in the MMT results seems to develop a QSS-like plateau roughly between $t = 1 \, \mathrm{\mu s}$ and $5 \, \mathrm{\mu s}$, the vibrational temperature in the Park+M\&W case does not and relaxes toward $T$ without interruption. 

Differences between Park+M\&W and MMT become more apparent when comparing mixture composition histories. In Fig.~\ref{fig:iso_8k_combined_mmt_vs_park}, both models predict a similar early decrease in $\mathrm{O_2}$ mole fraction (dark blue lines), but the differences for $\mathrm{N_2}$, $\mathrm{N}$ and $\mathrm{NO}$ are significant. Focusing on nitric oxide first, a small inset in the lower half of Fig.~\ref{fig:iso_8k_combined_mmt_vs_park} shows a close-up around the time when the $\mathrm{NO}$ mole fraction profiles with both models attain their respective peaks. Although there exists little difference in the time when this maximum is reached, the Park+M\&W profile (solid green line) peaks at a value only about 25\% that of the MMT model (dash-dotted green line). Furthermore, the rate of decline in the $\mathrm{NO}$ mole fraction over time is noticeably slower in the MMT model than for Park+M\&W.

\begin{figure}
 \centering
 
 \begin{minipage}{0.5\textwidth}
  \subfloat[Nitrogen dissociation]{\label{fig:N2_dissociation_rate_coeff_ratios_Park_MMT}
  \includegraphics[width=\textwidth]{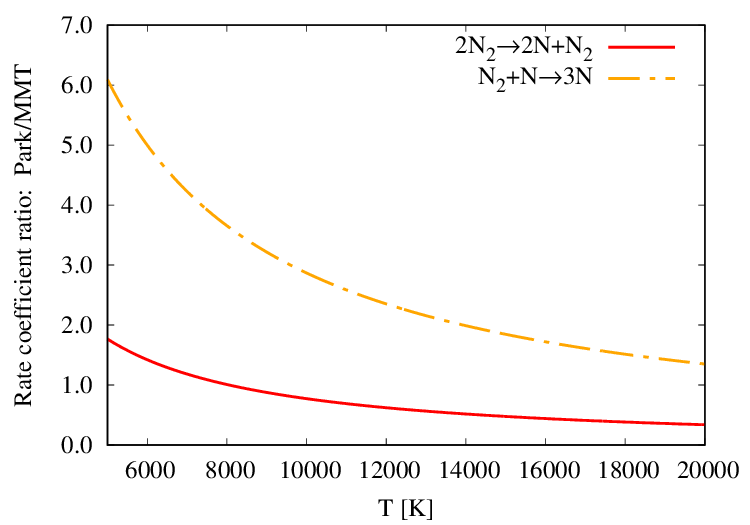}}
 \end{minipage}~
 \begin{minipage}{0.5\textwidth}
  \subfloat[Zeldovich reactions]{\label{fig:Zeldovich_rate_coeff_ratios_Park_MMT}
  \includegraphics[width=\textwidth]{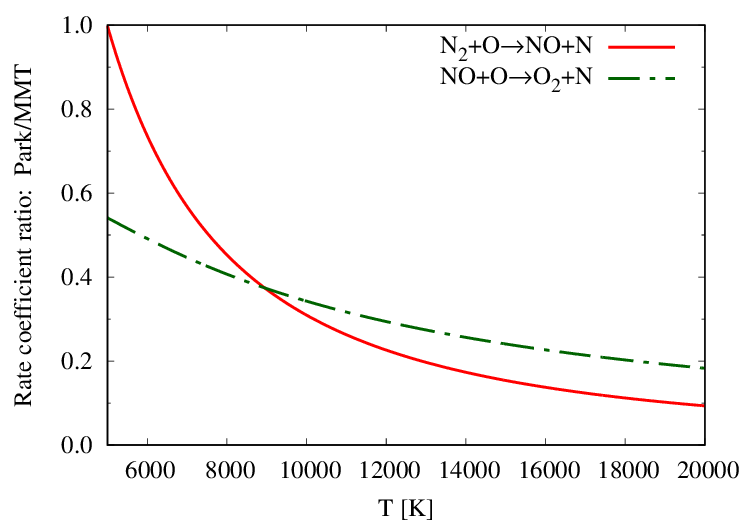}}
 \end{minipage}
 
 \caption{Ratio of thermal rate coefficients $k_\mathrm{Park} (\sqrt{T\,T}) / k_\mathrm{MMT}^\mathrm{Arrh.} (T)$}
\end{figure}

At this heat bath temperature the Park model also predicts a significantly faster decrease in molecular nitrogen mole fraction (solid red line) than the MMT model (dash-dotted red line). This drop in $\mathrm{N_2}$ is almost exactly mirrored by a faster rise of the atomic nitrogen mole fraction for Park+M\&W when compared to MMT (solid vs. dash-dotted orange line). Due to their high activation energies, nitrogen dissociation reactions proceed at rates considerably slower than most other processes in the mixture and most of the $\mathrm{N_2}$ is consumed only after the early significant chemical reactions ($\mathrm{O_2}$-dissociation and Zeldovich reactions) have already subsided. Thus, differences between the Park and MMT nitrogen profiles at later stages can primarily be traced back to the $\mathrm{N_2}$-dissociation rates predicted by both models. From the parameters in Table~\ref{tab:mmt_2024_parameters_complete} for MMT and those from Table~2 of Ref.~\cite{park93a} for Park, we can estimate that at $8\,000 \, \mathrm{K}$ Park's $\mathrm{N_2-N_2}$ thermal dissociation rate coefficient is about the same in magnitude as for MMT (see red line in Fig.~\ref{fig:N2_dissociation_rate_coeff_ratios_Park_MMT}). However, as shown by the orange line in Fig.~\ref{fig:N2_dissociation_rate_coeff_ratios_Park_MMT}, for $\mathrm{N_2-N}$ dissociation Park's thermal dissociation rate coefficient is roughly $3.5$ times the one predicted for MMT. Therefore, the noticeably faster $\mathrm{N_2}$ dissociation predicted by the Park model relative to MMT can be attributed primarily to the $\mathrm{N_2-N}$ reaction.

Figure~\ref{fig:iso_8k_combined_mmt_nofac_vs_park} shows that removing the multi-surface factor from the MMT rates has the overall effect of shifting most of the early reaction dynamics backward by roughly one microsecond. Whereas in Fig.~\ref{fig:iso_8k_combined_mmt_vs_park} the $\mathrm{O_2}$ mole fraction profiles for Park and MMT were almost identical, without the enhancement factor early $\mathrm{O_2}$-dissociation is slower, which causes the MMT $\mathrm{O_2}$ mole fraction profile in Fig.~\ref{fig:iso_8k_combined_mmt_nofac_vs_park} to lag behind that of the Park model. This delay has a knock-on effect on the dynamics of all the other species in the MMT result. The decline in $\mathrm{N_2}$ mole fraction is delayed by a similar amount, as is the location of the $\mathrm{NO}$ mole fraction peak and the initial rise in both atomic species' mole fractions.

\begin{figure}
 \centering
 
  \begin{minipage}{0.5\textwidth}
  \subfloat[MMT (full $\eta_\mathrm{O_2}$ factor) vs. Park+M\&W]{\label{fig:iso_15k_combined_mmt_vs_park}
  \includegraphics[width=\textwidth]{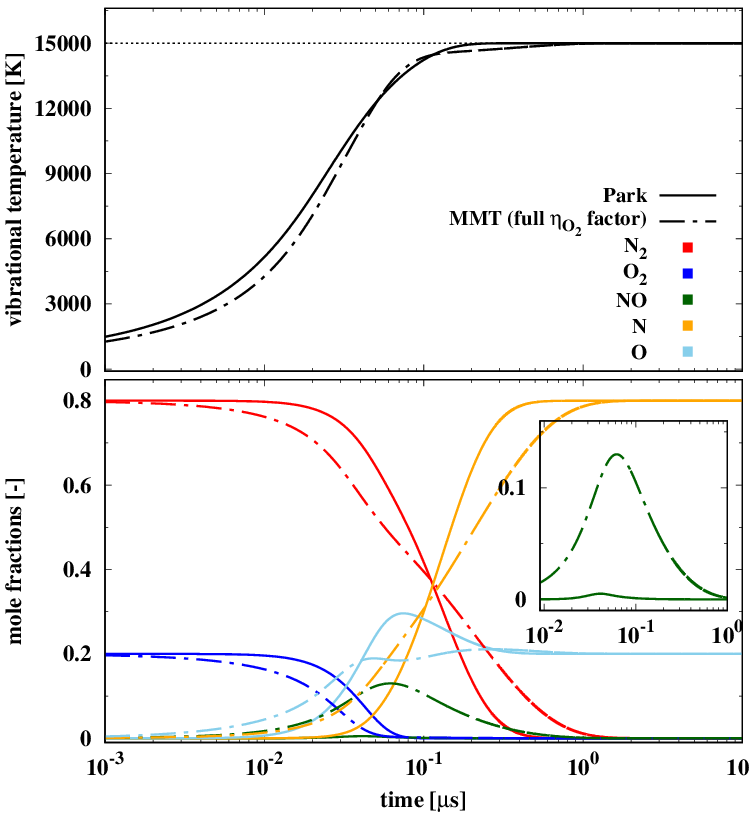}}
 \end{minipage}~
 \begin{minipage}{0.5\textwidth}
  \subfloat[MMT (no $\eta_\mathrm{O_2}$ factor) vs. Park+M\&W]{\label{fig:iso_15k_combined_mmt_nofac_vs_park}
  \includegraphics[width=\textwidth]{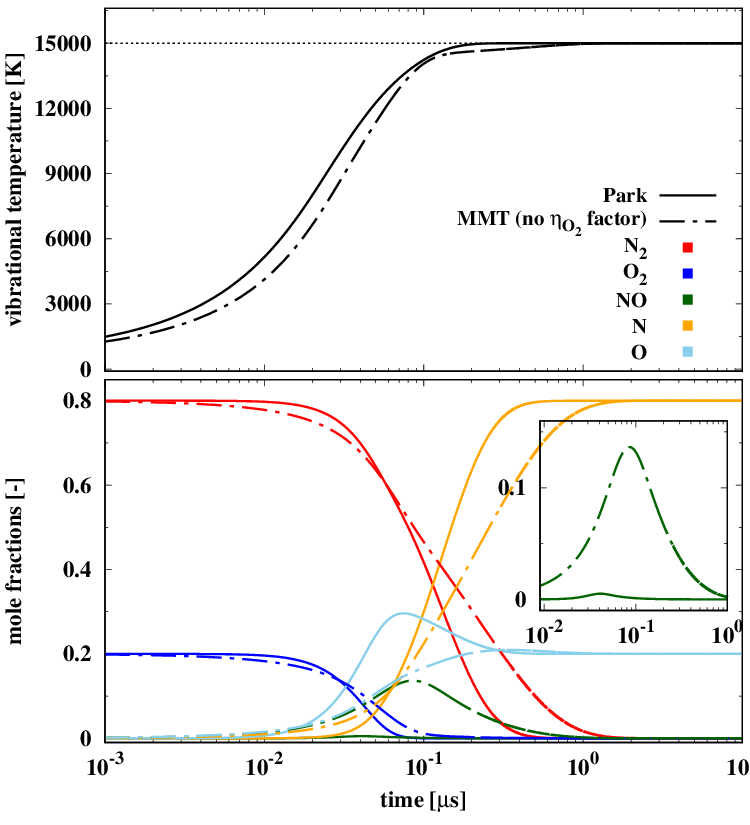}}
 \end{minipage}
 
 \caption{Isothermal heat bath $T = 15\,000 \, \mathrm{K}$, complete MMT vs. Park+M\&W}
 \label{fig:iso_15k_mmt_vs_park}
\end{figure}

Equivalent results for the $T = 15\,000 \, \mathrm{K}$ case are shown in Figs.~\ref{fig:iso_15k_combined_mmt_vs_park} and \ref{fig:iso_15k_combined_mmt_nofac_vs_park}. By comparing the Park and MMT $T_\mathrm{v}$-profiles in the upper half of both sub-plots, one sees that vibrational relaxation again occurs at nearly the same rate in either of the two MMT and the Park simulations. As before, the MMT profiles exhibit a mild tendency to form a QSS-like plateau, whereas this feature is absent from the Park+M\&W profiles. With regard to the evolving mixture composition, differences between both models are clearly visible. Focusing on Fig.~\ref{fig:iso_15k_combined_mmt_vs_park} first, very early on both the $\mathrm{O_2}$ and $\mathrm{N_2}$ mole fraction profiles in the MMT model (dash-dotted dark blue and red lines) run ahead of their counterparts from the Park model (solid dark blue and red lines). The decrease in molecular nitrogen is mirrored by a rise in atomic nitrogen (orange lines) in both models. Coincidentally, the intersection of $\mathrm{N_2}$ and $\mathrm{N}$ mole fraction profiles happens at roughly the same time for MMT (full $\eta_\mathrm{O_2}$ factor) and Park. By this time the rate at which $\mathrm{N_2}$ is being consumed is noticeably faster in the Park simulation. Again, this can be attributed mainly to differences in the rate coefficients for nitrogen dissociation in both models. As time progresses and more $\mathrm{N}$-atoms appear in the mixture, the relative importance to overall $\mathrm{N_2}$ decline gradually shifts from $\mathrm{N_2-N_2}$ dissociation to $\mathrm{N_2-N}$ dissociation. At $T = 15\,000 \, \mathrm{K}$ the Park $\mathrm{N_2-N_2}$ thermal dissociation rate coefficient is about half that in the MMT model, but its $\mathrm{N_2-N}$ dissociation rate coefficient is roughly twice that of MMT (see red and orange lines in Fig.~\ref{fig:N2_dissociation_rate_coeff_ratios_Park_MMT}). Thus, it makes sense that the Park simulation exhibits a faster rate of $\mathrm{N_2}$-consumption at the later stages. 

Figure~\ref{fig:iso_15k_combined_mmt_nofac_vs_park} shows that removing the multi-surface enhancement factor from the MMT rates delays the early decrease in $\mathrm{O_2}$ mole fraction. Contrary to what could be observed in the comparisons at $T = 8\,000 \, \mathrm{K}$, at this higher heat bath temperature the MMT and Park profiles now show closer agreement without the $\eta_\mathrm{O_2}$ factor. As before, the slower $\mathrm{O_2}$ dissociation without the enhancement factor causes most other changes in the mixture composition to be slightly delayed when compared to Fig.~\ref{fig:iso_15k_combined_mmt_vs_park}. Nevertheless, differences between Park and MMT are significant with, or without the enhancement factor applied.

Figures~\ref{fig:iso_15k_combined_mmt_vs_park} and \ref{fig:iso_15k_combined_mmt_nofac_vs_park} also show clear differences in the amount of nitric oxide being produced between models. Whereas MMT predicts peak $\mathrm{NO}$ mole fractions of roughly $0.13$, barely any nitric oxide appears in the mixture with the Park model. At this higher heat bath temperature the differences between Park and MMT are even more pronounced than those observed in Figs.~\ref{fig:iso_8k_combined_mmt_vs_park} and \ref{fig:iso_8k_combined_mmt_nofac_vs_park}. As before, they can be attributed to significant differences between both models' rate coefficients for the Zeldovich reactions. In Fig.~\ref{fig:Zeldovich_rate_coeff_ratios_Park_MMT} we plot the ratio $k_\mathrm{Park} / k_\mathrm{MMT}$ for the first Zeldovich reaction $\mathrm{N_2 + O \rightarrow NO + N}$ as the red line. Over the interval $8\,000 \, \mathrm{K} < T < 15\,000 \, \mathrm{K}$ this ratio gradually decreases from roughly $0.45$ to just above $0.15$, which means that at both heat bath temperatures in this comparison the Park model predicts noticeably lower rates for the first Zeldovich reaction than MMT and the amount of nitric oxide produced is correspondingly lower. Simultaneously, the ratio $k_\mathrm{Park} / k_\mathrm{MMT}$ for the second Zeldovich reaction $\mathrm{NO + O \rightarrow O_2 + N}$ ranges from $0.4$ to $0.25$ over the same temperature interval (dash-dotted green line in Fig.~\ref{fig:Zeldovich_rate_coeff_ratios_Park_MMT}). This reaction, as written in its endothermic sense, consumes the $\mathrm{NO}$ produced via the first Zeldovich reaction. Thus, both $\mathrm{NO}$ production and destruction rates via the first and second Zeldovich reactions respectively are significantly lower with the Park rates compared to MMT. Nevertheless, the net effect is that with the Park rates the second reaction remains more active at higher temperatures, rapidly consuming almost all of the $\mathrm{NO}$ being produced by the first one. This likely results in the much smaller peaks observed of nitric oxide mole fraction in the Park simulations compared to MMT.

Finally, we compare predictions of the MMT and Park models for the two adiabatic heat bath cases listed in Table~\ref{tab:air5_adia_conditions}. Initial values of $T$, $T_\mathrm{v}$ and $\rho$ specified for the CFD calculations, as well as the initial mole fractions are carried over directly from Table~\ref{tab:air5_adia_conditions} for both cases. The thermodynamic properties in these calculations now include contributions of electronic energy, which in principle slightly alters the specific energy $e$ in both cases. However, since the electronic energy contributions at the initial value of $T_\mathrm{v} = 300\, \mathrm{K}$ are negligible in both cases, the specific thermal energies remain practically unchanged from those listed in Table~\ref{tab:air5_adia_conditions}.

\begin{figure}
 \centering
 
 \begin{minipage}{0.5\textwidth}
  \subfloat[MMT (full $\eta_\mathrm{O_2}$ factor) vs. Park+M\&W]{\label{fig:adia_h10p1_combined_mmt_vs_park}
  \includegraphics[width=\textwidth]{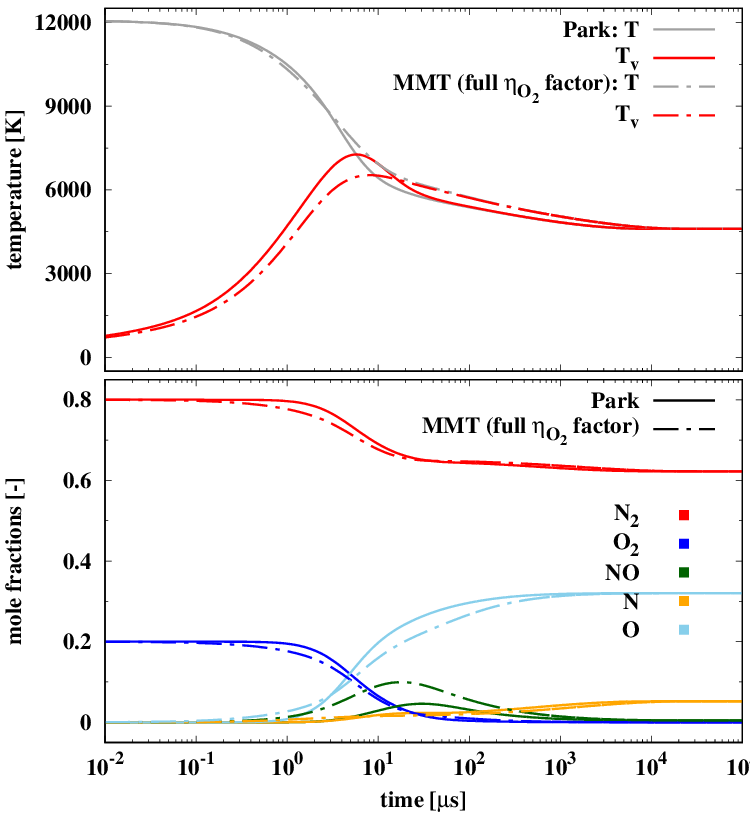}}
 \end{minipage}~
 \begin{minipage}{0.5\textwidth}
  \subfloat[MMT (no $\eta_\mathrm{O_2}$ factor) vs. Park+M\&W]{\label{fig:adia_h10p1_combined_mmt_nofac_vs_park}
  \includegraphics[width=\textwidth]{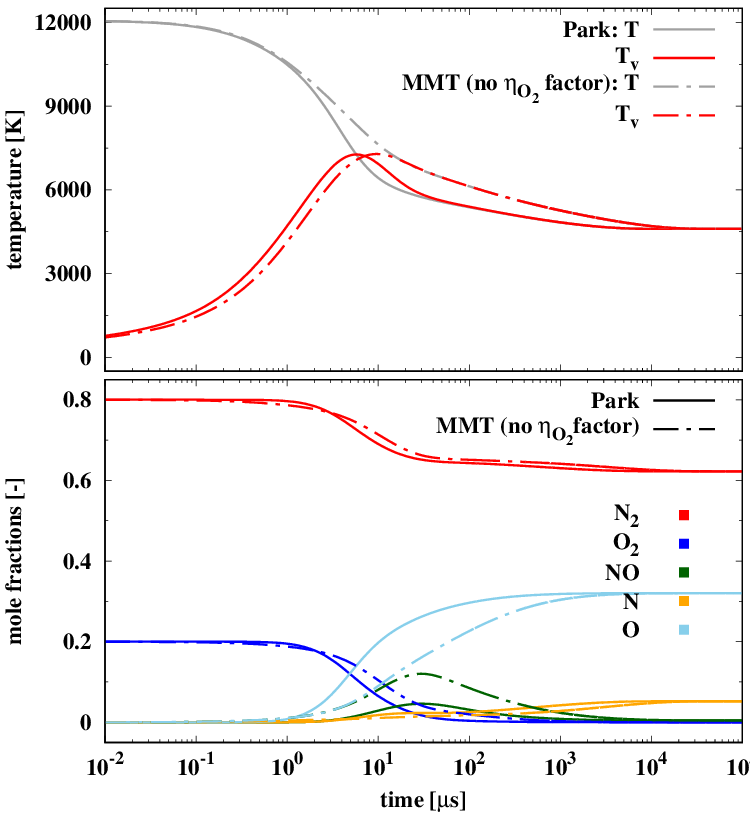}}
 \end{minipage}
  
 \caption{Adiabatic heat bath {$e = 8.409 \, \mathrm{MJ/kg}$}, complete MMT vs. Park+M\&W}
 \label{fig:adia_h10p1_mmt_vs_park}
\end{figure}

Results for the lower-energy case are shown in Figs.~\ref{fig:adia_h10p1_combined_mmt_vs_park} and \ref{fig:adia_h10p1_combined_mmt_nofac_vs_park} (MMT rates with and without multi-surface enhancement factor respectively), with temperature profiles plotted in the upper half and mixture composition below. Focusing on Fig.~\ref{fig:adia_h10p1_combined_mmt_vs_park} first, both for Park and MMT we report vibrational (red lines) and trans-rotational (gray lines) temperatures. As was the case in the preceding isothermal cases, here the rates of vibrational relaxation for Park+M\&W and MMT are roughly the same. With both models this process is complete after about $20$ microseconds. However, one major difference is that the Park+M\&W model produces a clearly visible crossing of the  $T_\mathrm{v}$ and $T$ profiles (solid red vs. gray lines) before relaxing to a common temperature. This feature is absent from the MMT profiles (dash-dotted red vs. gray lines). This overshoot, or lack thereof, may be attributed to differences in how the two source terms $w_\mathrm{v}^\mathrm{relax}$ and $w_\mathrm{v}^\mathrm{chem}$ in Eq.~(\ref{eq:vib_energy_balance}) are computed with both models. In the MMT approach, vibrational relaxation and energy removal rates of a dissociating species have both been calibrated to ab inito data from a common source (see Refs.~\cite{torres24a, torres24b}). Furthermore, Knab's expression for average vibrational energy removed per dissociation, employed in the MMT model, Eq.~(\ref{eq:devib_mmt}), explicitly includes a dependence on both $T$ and $T_\mathrm{v}$. By contrast, the Park+M\&W simulations use non-preferential vibrational energy-chemistry coupling for all species and reactions involved. As Eq.~(\ref{eq:devib_nonpref}) shows, with this modeling assumption the average vibrational energy removed in dissociation depends only on $T_\mathrm{v}$ and is insensitive to the local trans-rotational temperature. 

In the lower half of Fig.~\ref{fig:adia_h10p1_combined_mmt_vs_park}, relative differences between the mole fraction profiles predicted by both models are not too significant, except for $\mathrm{NO}$. Early on, MMT predicts a slightly faster reduction in $\mathrm{O_2}$ and $\mathrm{N_2}$ mole fractions compared to Park, while atomic oxygen production with MMT lags slightly behind the Park profile. The only major discrepancy is again nitric oxide. As seen in the preceding cases, the MMT model predicts a noticeably higher peak in $\mathrm{NO}$ mole fraction compared to Park (dash-dotted vs. solid green lines). For comparison, Fig.~\ref{fig:adia_h10p1_combined_mmt_nofac_vs_park} shows again that, when the MMT model is employed without the multi-surface enhancement factors, all reaction dynamics are slightly delayed. The $\mathrm{O_2}$ mole fraction profile for MMT (no $\eta_\mathrm{O_2}$ factor) now lags slightly behind the Park curve, while that for atomic oxygen is shifted to later times as well. 

\begin{figure}
 \centering
  
 \begin{minipage}{0.5\textwidth}
  \subfloat[MMT (full $\eta_\mathrm{O_2}$ factor) vs. Park+M\&W]{\label{fig:adia_h17p9_combined_mmt_vs_park}
  \includegraphics[width=\textwidth]{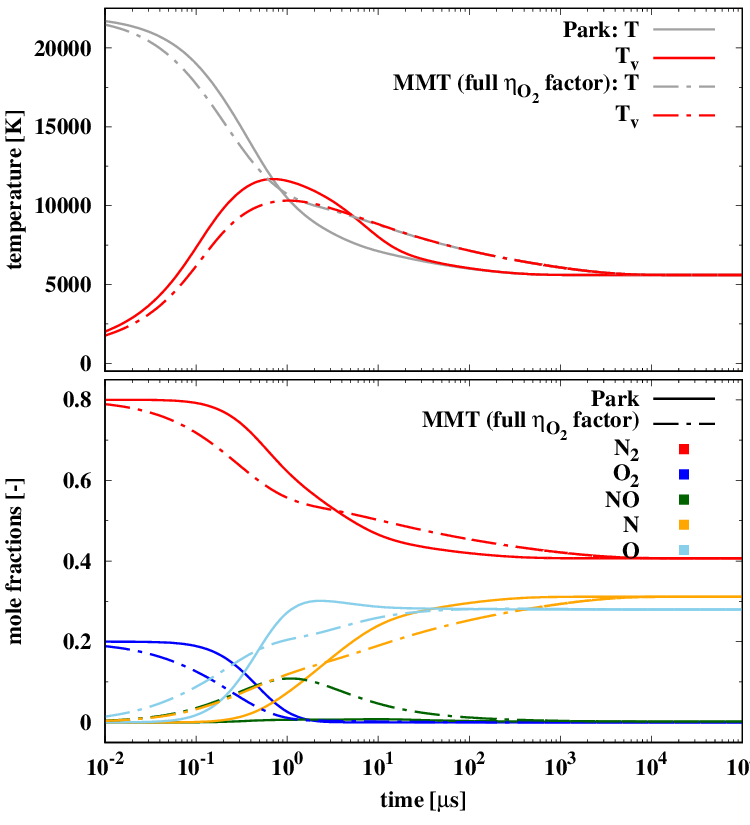}}
 \end{minipage}~ 
 \begin{minipage}{0.5\textwidth}
  \subfloat[MMT (no $\eta_\mathrm{O_2}$ factor) vs. Park+M\&W]{\label{fig:adia_h17p9_combined_mmt_nofac_vs_park}
  \includegraphics[width=\textwidth]{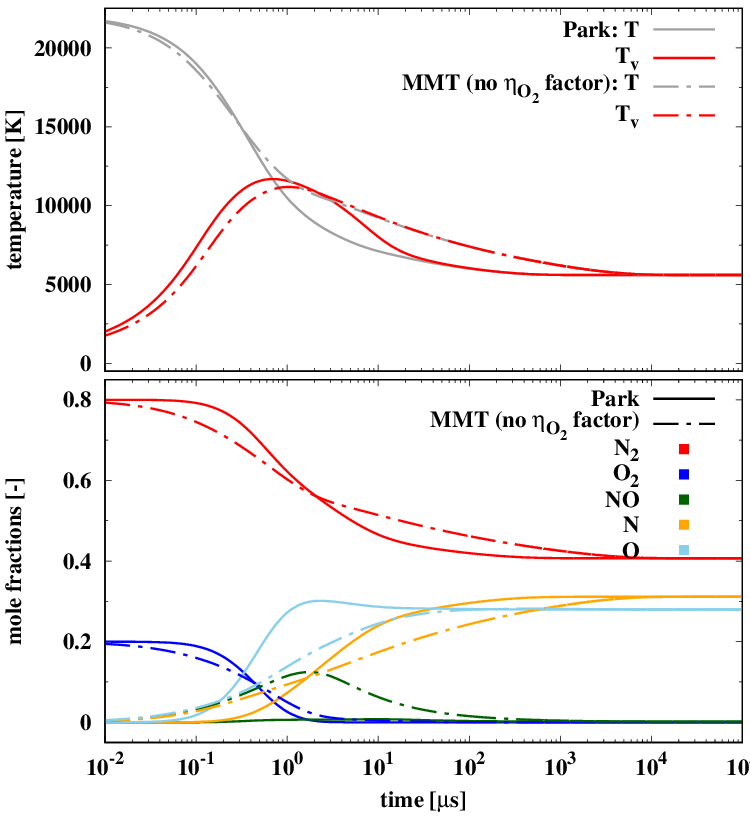}}
 \end{minipage}
 
 \caption{Adiabatic heat bath {$e = 15.61 \, \mathrm{MJ/kg}$}, complete MMT vs. Park+M\&W}
 \label{fig:adia_h17p9_mmt_vs_park}
\end{figure}

Results for the higher-energy case are shown in Figs.~\ref{fig:adia_h17p9_combined_mmt_vs_park} and \ref{fig:adia_h17p9_combined_mmt_nofac_vs_park}. Most features observed at the lower-energy conditions reappear, but are exacerbated here: The crossing of $T_\mathrm{v}$ and $T$ profiles in the Park+M\&W profiles (solid red and gray lines) is more severe, while no $T_\mathrm{v}$-overshoot is observed in the MMT result (dash-totted lines). Differences between MMT and Park+M\&W in the mixture composition time histories are more noticeable now for all species. At these conditions the MMT model predicts a peak $\mathrm{NO}$ mole fraction close to $0.1$, whereas with the Park model, practically no nitric oxide is produced.


%
\section{Conclusions and future work} \label{sec:conclusions}

In this paper we conclude our work detailing all aspects of the newly developed Modified Marrone-Treanor nonequilibrium chemistry model. It follows the model formulation discussed in Ref.~\cite{chaudhry25a} by first listing the most up-to-date  parameters for calculating reaction rates and vibrational relaxation times for 5-species air mixtures ($\mathrm{N_2}$, $\mathrm{O_2}$, $\mathrm{NO}$, $\mathrm{N}$ and $\mathrm{O}$), all of them derived from QCT and DMS calculations on ab initio potential energy surfaces. The new parameter set allows for CFD simulations of high-temperature 5-species air in thermochemical nonequilibrium. The paper presented comparisons of CFD-MMT model predictions against DMS reference solutions in space-homogeneous isothermal and adiabatic heat baths representative of post-shock conditions in hypersonic air flows. The MMT model's mutually consistent expressions for nonequilibrium dissociation rate coefficient and associated vibrational energy removal term ensure that the correct amount of energy is transferred between the vibrational and trans-rotational modes at all times. This feature is particularly important for achieving close agreement between CFD-MMT and DMS under adiabatic conditions, where $T$ and $T_\mathrm{v}$ both vary in response to chemical reactions altering the gas state. Despite the simplicity of the two-temperature MMT model, remarkably close agreement with pure ab-initio-based solutions was observed.

We further investigated the implications of employing constant vs. variable non-Boltzmann correction factors in the MMT model. Such variable (i.e. composition-dependent) factors were introduced in Ref.~\cite{chaudhry25a} to ensure that the MMT model approaches the expected \emph{thermal} dissociation and recombination rates once the gas mixture reaches thermochemical equilibrium. It is found that the overall effect of switching from constant to variable non-Boltzmann factors on the evolution of the gas state is small, as long as reactions are treated as reversible. Since the same factors are applied to forward and backward reaction pairs, changes on the net rates remain minimal.

Results using three MMT parameter sets were presented. The MMT-benchmark parameter set should only be used when comparing to the DMS results used here. The two complete MMT parameter sets are intended for hypersonic CFD simulations, where they represent two limiting assumptions of 1) no oxygen dissociation enhancement due to electronic excitation and 2) full enhancement, described in Sec.~\ref{sec:mmt_params_for_real}. Of particular interest was to assess the influence of the theorized multi-surface enhancement factor for oxygen dissociation on the model predictions. As expected, inclusion of the factor significantly speeds up early oxygen dissociation rates and shifts all other dependent reaction dynamics forward in time. Dissociation of $\mathrm{O_2}$ with collision partner $\mathrm{N_2}$ is the most important reaction for producing atomic oxygen early on. As more $\mathrm{O}$-atoms become available, $\mathrm{O_2 + O}$-dissociation also gains in importance. From there on, the two Zeldovich reactions play a central role in  shaping the production and consumption rates of all five species. Nitrogen dissociation reactions only become dominant at the very late stages, when most other reactions have subsided.

Finally, a comparison of MMT results against equivalent ones obtained with Park's $TT_\mathrm{v}$ model for 5-species air was carried out. At all conditions investigated, the MMT model predicts significantly higher peak mole fractions for nitric oxide than Park. At heat bath temperatures of $10\,000 \, \mathrm{K}$ and below, MMT also predicts noticeably slower conversion of molecular into atomic nitrogen. Both effects can be traced back to the rate parameters for the two Zeldovich reactions ($\mathrm{N_2+O \rightleftharpoons NO+N}$ and $\mathrm{NO+O \rightleftharpoons O_2+N}$). Our new QCT-derived values for these reactions generate noticeably higher rates than those in Park's original 1990 model. The net effect is that far less $\mathrm{NO}$ is produced in all Park simulations. 

In summary, this work demonstrates that the MMT model accurately reproduces all major features present in the DMS reference solutions based on first-principles data. Its computational cost is comparable to Park's model, i.e. orders of magnitude smaller than DMS. This makes it suitable for full-scale CFD simulations of chemically reactive air flows in the hypersonic regime.

Future work to remove some of the MMT model's remaining ad hoc assumptions should include QCT/DMS studies to determine the rates at which electronically excited states of $\mathrm{O_2}$ become populated in shock-heated air and what kind of enhancing effect, if any, this has on the oxygen dissociation rates. Such a study would help narrow down the value of the multi-surface enhancement factor $\eta_\mathrm{O_2}$ and whether such a factor should be applied equally to all $\mathrm{O_2}$-dissociation reactions, or not at all. Computational chemists have already begun generating some of the PESs required for simulating electronically nonadiabatic trajectories in oxygen~\cite{varga22a, shu22a}. Beyond this, additional first-principles studies of non-Boltzmann recombination~\cite{geistfeld23a, pahlani23a, macdonald24a} will be required to better justify the limits placed on the variable non-Boltzmann factors in the MMT model.

%
\section*{Acknowledgments}

This material is based upon work supported by AFOSR grants FA9550-19-1-0219 and FA9550-23-1-0446, and by NASA grant No. 80NSSC20K1061.



 \clearpage
 
 \setcounter{table}{0}
 \setcounter{equation}{0}
 \setcounter{figure}{0}
 \setcounter{section}{0}
 \renewcommand{\thesection}{S\arabic{section}}
 \renewcommand{\thetable}{S\arabic{table}}
 \renewcommand{\thefigure}{S\arabic{figure}}
 \renewcommand{\theequation}{S\arabic{equation}}

\title{Supplemental information to ``Modified Marrone-Treanor model: parameterization and benchmarking for five-species air''}

\maketitle

\begin{abstract} 
 This document contains useful supplemental information. It is not absolutely required for implementing the MMT model, but will be helpful in replicating the simulation results discussed in the main article. Some notation may differ slightly from that used in the main article.
\end{abstract}

\section{Thermodynamic properties in NASA Lewis fit format} \label{sec:thermo_props_lewis}

Any CFD code used to simulate chemically reacting hypersonic flows must accurately describe a mixture's thermodynamic properties, such as species heat capacities, enthalpies and chemical equilibrium constants. A straightforward source for such data is the NASA Lewis database, which forms the basis of the ``Chemical Equilibrium with Applications'' (CEA) software~\cite{mcbride96a}. This database allows one to compute thermodynamic properties for every chemical species following the template of Eqs.~(1)-(3) in McBride et al.~\cite{mcbride93a}. Specifically, the 9-coefficient Lewis curve fits for heat capacity, assigned enthalpy and entropy per unit mole of species $s$ at standard state pressure $p_\standardstate = 1 \, \mathrm{bar}$ may be evaluated as:
\begin{align}
 \hat{c}_{\mathrm{p},s}^\standardstate (T) & = \bar{R} \left( a_{1,s} \, T^{-2} + a_{2,s} \, T^{-1} + a_{3,s} + a_{4,s} \, T + a_{5,s} \, T^2 + a_{6,s} \, T^3 + a_{7,s} \, T^4 \right) \label{eq:cp_lewis} \\
 \hat{h}_s^\standardstate (T) & = \bar{R} \left( - \frac{a_{1,s}}{T} + a_{2,s} \ln (T) + a_{3,s} \, T + \frac{a_{4,s}}{2} \, T^2 + \frac{a_{5,s}}{3} \, T^3 + \frac{a_{6,s}}{4} \, T^4 + \frac{a_{7,s}}{5} \, T^5 + b_{1,s} \right) \label{eq:h_lewis} \\
 \hat{s}_s^\standardstate (T) & = \bar{R} \left( - \frac{a_{1,s}}{2} \, T^{-2} - a_{2,s} \, T^{-1} + a_{3,s} \ln (T) + a_{4,s} \, T + \frac{a_{5,s}}{2} \, T^2 + \frac{a_{6,s}}{3} \, T^3  + \frac{a_{7,s}}{4} \, T^4 + b_{2,s} \right), \label{eq:s_lewis}
\end{align}
where coefficients $a_{1,s}$-$a_{7,s}$ and $b_{1,s}$-$b_{2,s}$ are 9 tabulated fit parameters for a given species and $\bar{R}$ is the universal gas constant. For each of the five species relevant in this work, i.e. $\mathrm{N_2}$, $\mathrm{O_2}$, $\mathrm{NO}$, $\mathrm{N}$ and $\mathrm{O}$, the database provides three separate coefficient sets covering temperature ranges $200\, \mathrm{K} \le T \le 1\,000 \, \mathrm{K}$, $1\,000\, \mathrm{K} \le T \le 6\,000 \, \mathrm{K}$ and $6\,000\, \mathrm{K} \le T \le 20\,000 \, \mathrm{K}$ respectively. Table~\ref{tab:nasa_thermo} lists the numerical values directly extracted from the database. For the monatomic species these coefficients include contributions of translational and electronic modes, whereas for the diatomic species additional contributions from the molecules' rotational and vibrational modes are also included. These coefficients are employed to compute all thermodynamic properties in the complete MMT model (see Sec.~V of the main article). 

For the thermally perfect gases relevant in this work, heat capacity and enthalpy only depend on temperature, which means that Eqs.~(\ref{eq:cp_lewis}) and (\ref{eq:h_lewis}) are valid at any pressure. Entropy on the other hand depends both on pressure and temperature, even for a thermally perfect gas. This means that Eq.~(\ref{eq:s_lewis}) when evaluated with the Lewis fit parameters is only valid at standard pressure $p_\standardstate$. It can however be used to find species entropy at partial pressure $p_s$ via the relation $\hat{s}_s (T, p_s) = \hat{s}_s^\standardstate (T) - \bar{R} \ln \left( p_s / p_\standardstate \right)$. The equilibrium constant for reaction $r$, expressed in terms of equilibrium partial pressures $p_s^\star$, may be directly calculated by combining Eqs.~(\ref{eq:h_lewis}) and (\ref{eq:s_lewis}):
\begin{equation}
 K_{\mathrm{p}, r}^\mathrm{eq} (T) = \prod\limits_{s \in S} \left( p_s^\star \right)^{\nu_{s,r}^b - \nu_{s,r}^f} = \left( p_\standardstate \right)^{\nu_{r}^\mathrm{T}} \exp \left( - \sum\limits_{s \in S} \left\lbrace \left( \nu_{s,r}^b - \nu_{s,r}^f \right) \left( \frac{\hat{h}_s^\standardstate (T) - T \, \hat{s}_s^\standardstate (T)}{\bar{R} T} \right) \right\rbrace \right), \label{eq:kp_eq_lewis}
\end{equation}
where $\nu_{s,r}^b$ and $\nu_{s,r}^f$ represent the stoichiometric coefficients of species $s$ on the product and reactant sides respectively of generic reaction $r$, as written in Eq.~(7) of the main article and the total stoichiometric coefficient difference for reaction $r$ is $\nu_r^\mathrm{T} = \sum_{s \in S} \{ \nu_{s,r}^b - \nu_{s,r}^f \}$. It is sometimes convenient to express a given reaction's equilibrium constant in terms of molar concentrations $C_s = \rho_s / M_s$. This alternative form is linked to Eq.~(\ref{eq:kp_eq_lewis}) as:
\begin{equation}
 K_{\mathrm{c}, r}^\mathrm{eq} = \prod\limits_{s \in S} \left( C_s^\star \right)^{\nu_{s,r}^b - \nu_{s,r}^f} = K_{\mathrm{p}, r}^\mathrm{eq} \left( \frac{1}{\bar{R} T} \right)^{\nu_{r}^\mathrm{T}}. \label{eq:kc_eq}
\end{equation}

For the reactions included in the 5-species air MMT model, Eqs.~(\ref{eq:kp_eq_lewis}) and (\ref{eq:kc_eq}) simplify considerably. Reactions 1-15 in Table~1 of the main article all represent simple dissociation reactions. When these reactions are written in the same left-to-right sense of the table, the corresponding equilibrium constants reduce to: 
\begin{align}
 \mathrm{N_2} + \mathrm{M} \rightleftharpoons \mathrm{N} + \mathrm{N} + \mathrm{M}: \qquad & \Rightarrow \qquad K_{\mathrm{c}, \mathrm{N_2 \rightleftharpoons 2N} }^\mathrm{eq} = ( C_\mathrm{N}^\star )^2 \, ( C_\mathrm{N_2}^\star )^{-1} \label{eq:keq_n2_diss} \\
 \mathrm{O_2} + \mathrm{M} \rightleftharpoons \mathrm{O} + \mathrm{O} + \mathrm{M}: \qquad & \Rightarrow \qquad K_{\mathrm{c}, \mathrm{O_2 \rightleftharpoons 2O} }^\mathrm{eq} = ( C_\mathrm{O}^\star )^2 \, ( C_\mathrm{O_2}^\star )^{-1} \label{eq:keq_o2_diss} \\
 \text{and} & \nonumber \\
 \mathrm{NO} + \mathrm{M} \rightleftharpoons \mathrm{N} + \mathrm{O} + \mathrm{M}: \qquad & \Rightarrow \qquad K_{\mathrm{c}, \mathrm{NO \rightleftharpoons N+O} }^\mathrm{eq} = ( C_\mathrm{N}^\star ) \, ( C_\mathrm{O}^\star ) \, ( C_\mathrm{NO}^\star )^{-1} \label{eq:keq_no_diss}
\end{align}
respectively, with $\mathrm{M}$ a placeholder for the collision partner species. Since this collision partner appears unchanged on both sides of the reaction, it cancels out from the equilibrium concentration ratios in Eqs.~(\ref{eq:keq_n2_diss})-(\ref{eq:keq_no_diss}).

For the three exchange-type reactions, i.e. reactions 16-18 in Table~1 of the main article, the equilibrium constants become:
\begin{align}
 \mathrm{N_2} + \mathrm{O} \rightleftharpoons \mathrm{NO} + \mathrm{N}: \qquad & \Rightarrow \qquad K_{\mathrm{c}, \mathrm{N_2+O \rightleftharpoons NO+N} }^\mathrm{eq} = ( C_\mathrm{NO}^\star ) \, ( C_\mathrm{N}^\star ) \, ( C_\mathrm{N_2}^\star )^{-1} \, ( C_\mathrm{O}^\star )^{-1} \label{eq:keq_exch_1} \\
 \mathrm{NO} + \mathrm{O} \rightleftharpoons \mathrm{O_2} + \mathrm{N}: \qquad & \Rightarrow \qquad K_{\mathrm{c}, \mathrm{NO+O \rightleftharpoons O_2+N} }^\mathrm{eq} = ( C_\mathrm{O_2}^\star ) \, ( C_\mathrm{N}^\star ) \, ( C_\mathrm{NO}^\star )^{-1} \, ( C_\mathrm{O}^\star )^{-1} \label{eq:keq_exch_2} \\
 \text{and} & \nonumber \\
 \mathrm{N_2} + \mathrm{O_2} \rightleftharpoons \mathrm{NO} + \mathrm{NO}: \qquad & \Rightarrow \qquad K_{\mathrm{c}, \mathrm{N_2+O_2 \rightleftharpoons 2NO} }^\mathrm{eq} = ( C_\mathrm{NO}^\star )^2 \, ( C_\mathrm{N_2}^\star )^{-1} \, ( C_\mathrm{O_2}^\star )^{-1}. \label{eq:keq_exch_3}
\end{align}

Finally, the equilibrium constants for the remaining mixed-type reactions, i.e. reactions 19-21 in Table~1 of the main article, are defined as:
\begin{align}
 \mathrm{NO} + \mathrm{NO} \rightleftharpoons \mathrm{O} + \mathrm{O} + \mathrm{N_2}: \qquad & \Rightarrow \qquad K_{\mathrm{c}, \mathrm{2NO \rightleftharpoons 2O+N_2} }^\mathrm{eq} = ( C_\mathrm{O}^\star )^2 \, ( C_\mathrm{N_2}^\star ) \, ( C_\mathrm{NO}^\star )^{-2} \label{eq:keq_mixed_1} \\
 \mathrm{NO} + \mathrm{NO} \rightleftharpoons \mathrm{N} + \mathrm{N} + \mathrm{O_2}: \qquad & \Rightarrow \qquad K_{\mathrm{c}, \mathrm{2NO \rightleftharpoons O_2+2N} }^\mathrm{eq} = ( C_\mathrm{O_2}^\star ) \, ( C_\mathrm{N}^\star )^2 \, ( C_\mathrm{NO}^\star )^{-2} \label{eq:keq_mixed_2} \\
 \text{and} & \nonumber \\
 \mathrm{N_2} + \mathrm{O_2} \rightleftharpoons \mathrm{NO + N + O}: \qquad & \Rightarrow \qquad K_{\mathrm{c}, \mathrm{N_2+O_2 \rightleftharpoons NO+N+O} }^\mathrm{eq} = ( C_\mathrm{NO}^\star ) \, ( C_\mathrm{N}^\star ) \, ( C_\mathrm{O}^\star ) \, ( C_\mathrm{N_2}^\star )^{-1} \, ( C_\mathrm{O_2}^\star )^{-1}. \label{eq:keq_mixed_3}
\end{align}

Note that Eqs.~(\ref{eq:keq_n2_diss})-(\ref{eq:keq_no_diss}) and (\ref{eq:keq_mixed_1})-(\ref{eq:keq_mixed_3}) all involve two separate particles (molecules or atoms) on the reactant (left), but three on the product (right) side of the reaction. Simultaneously the total stoichiometric coefficient difference in all these reactions is $\nu_r^\mathrm{T} = 1$, which in turn implies that the corresponding equilibrium constants have units of molar concentration (or of pressure if defined by Eq.~(\ref{eq:kp_eq_lewis}) instead). By contrast, Eqs.~(\ref{eq:keq_exch_1})-(\ref{eq:keq_exch_3}) involve two colliding particles on both sides, which also means that $\nu_r^\mathrm{T} = 0$. Therefore, for these reactions the equilibrium constants are dimensionless and $K_{\mathrm{p}, r}^\mathrm{eq}$ and $K_{\mathrm{c}, r}^\mathrm{eq}$ are numerically identical.


\begin{table}[htb]
 \centering
 \caption{Parameters for computing thermodynamic properties extracted from NASA Lewis database (including contributions of electronic excited states)} \label{tab:nasa_thermo}
 
 \begin{tabular}{c c c c c c}
  cf. &  $\mathrm{N_2}$  & $\mathrm{O_2}$ & $\mathrm{NO}$ & $\mathrm{N}$ & $\mathrm{O}$ \\ \hline \hline
  \multicolumn{6}{c}{$200 \, \mathrm{K} \le T \le 1\,000 \, \mathrm{K}$} \\ \hline
  $a_1$ &  2.210371497E+04 & -3.425563420E+04 & -1.143916503E+04 &  0.000000000E+00 & -7.953611300E+03 \\
  $a_2$ & -3.818461820E+02 &  4.847000970E+02 &  1.536467592E+02 &  0.000000000E+00 &  1.607177787E+02 \\
  $a_3$ &  6.082738360E+00 &  1.119010961E+00 &  3.431468730E+00 &  2.500000000E+00 &  1.966226438E+00 \\
  $a_4$ & -8.530914410E-03 &  4.293889240E-03 & -2.668592368E-03 &  0.000000000E+00 &  1.013670310E-03 \\
  $a_5$ &  1.384646189E-05 & -6.836300520E-07 &  8.481399120E-06 &  0.000000000E+00 & -1.110415423E-06 \\
  $a_6$ & -9.625793620E-09 & -2.023372700E-09 & -7.685111050E-09 &  0.000000000E+00 &  6.517507500E-10 \\
  $a_7$ &  2.519705809E-12 &  1.039040018E-12 &  2.386797655E-12 &  0.000000000E+00 & -1.584779251E-13 \\
  $b_1$ &  7.108460860E+02 & -3.391454870E+03 &  9.098214410E+03 &  5.610463780E+04 &  2.840362437E+04 \\
  $b_2$ & -1.076003744E+01 &  1.849699470E+01 &  6.728725490E+00 &  4.193905036E+00 &  8.404241820E+00 \\ \hline
  \multicolumn{6}{c}{$1\,000 \, \mathrm{K} \le T \le 6\,000 \, \mathrm{K}$} \\ \hline
  $a_1$ &  5.877124060E+05 & -1.037939022E+06 &  2.239018716E+05 &  8.876501380E+04 &  2.619020262E+05 \\
  $a_2$ & -2.239249073E+03 &  2.344830282E+03 & -1.289651623E+03 & -1.071231500E+02 & -7.298722030E+02 \\
  $a_3$ &  6.066949220E+00 &  1.819732036E+00 &  5.433936030E+00 &  2.362188287E+00 &  3.317177270E+00 \\
  $a_4$ & -6.139685500E-04 &  1.267847582E-03 & -3.656034900E-04 &  2.916720081E-04 & -4.281334360E-04 \\
  $a_5$ &  1.491806679E-07 & -2.188067988E-07 &  9.880966450E-08 & -1.729515100E-07 &  1.036104594E-07 \\
  $a_6$ & -1.923105485E-11 &  2.053719572E-11 & -1.416076856E-11 &  4.012657880E-11 & -9.438304330E-12 \\
  $a_7$ &  1.061954386E-15 & -8.193467050E-16 &  9.380184620E-16 & -2.677227571E-15 &  2.725038297E-16 \\
  $b_1$ &  1.283210415E+04 & -1.689010929E+04 &  1.750317656E+04 &  5.697351330E+04 &  3.392428060E+04 \\
  $b_2$ & -1.586640027E+01 &  1.738716506E+01 & -8.501669090E+00 &  4.865231506E+00 & -6.679585350E-01 \\ \hline
  \multicolumn{6}{c}{$6\,000 \, \mathrm{K} \le T \le 20\,000 \, \mathrm{K}$} \\ \hline
  $a_1$ &  8.310139160E+08 &  4.975294300E+08 & -9.575303540E+08 &  5.475181050E+08 &  1.779004264E+08 \\
  $a_2$ & -6.420733540E+05 & -2.866106874E+05 &  5.912434480E+05 & -3.107574980E+05 & -1.082328257E+05 \\
  $a_3$ &  2.020264635E+02 &  6.690352250E+01 & -1.384566826E+02 &  6.916782740E+01 &  2.810778365E+01 \\
  $a_4$ & -3.065092046E-02 & -6.169959020E-03 &  1.694339403E-02 & -6.847988130E-03 & -2.975232262E-03 \\
  $a_5$ &  2.486903333E-06 &  3.016396027E-07 & -1.007351096E-06 &  3.827572400E-07 &  1.854997534E-07 \\
  $a_6$ & -9.705954110E-11 & -7.421416600E-12 &  2.912584076E-11 & -1.098367709E-11 & -5.796231540E-12 \\
  $a_7$ &  1.437538881E-15 &  7.278175770E-17 & -3.295109350E-16 &  1.277986024E-16 &  7.191720164E-17 \\
  $b_1$ &  4.938707040E+06 &  2.293554027E+06 & -4.677501240E+06 &  2.550585618E+06 &  8.890942630E+05 \\
  $b_2$ & -1.672099740E+03 & -5.530621610E+02 &  1.242081216E+03 & -5.848769753E+02 & -2.181728151E+02
 \end{tabular}
 
\end{table}

\subsection{Custom curve fits for PES-consistent thermodynamic properties} \label{sec:thermo_comparison}

The default curve-fit parameters from the Lewis database (see Table~\ref{tab:nasa_thermo}) yield accurate thermodynamic data over the entire temperature range $200 \, \mathrm{K} \le T \le 20\,000 \, \mathrm{K}$ for the set of five species relevant in this work ($S = \{ \mathrm{N_2}, \mathrm{O_2}, \mathrm{NO}, \mathrm{N}, \mathrm{O} \}$) and should be employed whenever CFD results are being compared to experiments, or when simulating realistic flight conditions. However, the inclusion of electronic energy mode contributions to the thermodynamic properties is not appropriate when we attempt to benchmark our new MMT model predictions against DMS in Sec.~III of the main article. Under these circumstances it is important to ensure that thermodynamic properties in the CFD implementation are as consistent as possible with the ab initio potential energy surfaces used in the original DMS calculations. Currently these PESs only describe electronically adiabatic trajectories between ground states of the colliding species. No transitions to, or from electronically excited states are included in the reference calculations and no excited states of any species are assumed to be populated in the benchmarking runs.

Section~\ref{sec:thermo_props_from_pes} provides more details on how these ``DMS-consistent'' ground-state-only thermodynamic properties were derived and Table~\ref{tab:pes_thermo} lists the equivalent ground-state-only 9-coefficient parameters in NASA-Lewis fit format employed in the model benchmarking runs of Sec.~III of the main article.


\begin{table}[htb]
 \centering
 \caption{Ground-electronic-state, PES-derived curve fit parameters for computing thermodynamic properties in 9-coefficient format} \label{tab:pes_thermo}
 
 \begin{tabular}{c c c c c c}
  cf. &  $\mathrm{N_2}$  & $\mathrm{O_2}$ & $\mathrm{NO}$ & $\mathrm{N}$ & $\mathrm{O}$ \\ \hline \hline
  \multicolumn{6}{c}{$200 \, \mathrm{K} \le T \le 1\,000 \, \mathrm{K}$} \\ \hline
  $a_1$ &  2.326052032e+04 & -3.881838306e+04 & -2.421964397e+03 &  0.000000000e+00 &  0.000000000e+00 \\
  $a_2$ & -3.959537642e+02 &  5.490702438e+02 & -4.207286817e+01 &  0.000000000e+00 &  0.000000000e+00 \\
  $a_3$ &  6.136874298e+00 &  7.644655543e-01 &  4.400469183e+00 &  2.500000000e+00 &  2.500000000e+00 \\
  $a_4$ & -8.571113320e-03 &  5.273260868e-03 & -5.117857455e-03 &  0.000000000e+00 &  0.000000000e+00 \\
  $a_5$ &  1.370135972e-05 & -2.155525019e-06 &  1.171301899e-05 &  0.000000000e+00 &  0.000000000e+00 \\
  $a_6$ & -9.421499287e-09 & -9.036227826e-10 & -9.860269507e-09 &  0.000000000e+00 &  0.000000000e+00 \\
  $a_7$ &  2.446930284e-12 &  7.033475202e-13 &  2.979579868e-12 &  0.000000000e+00 &  0.000000000e+00 \\
  $b_1$ &  7.816667395e+02 & -3.700387574e+03 &  9.970366243e+03 &  5.585984812e+04 &  2.892114779e+04 \\
  $b_2$ & -1.109221108e+01 &  2.047126993e+01 &  1.249707038e+00 &  4.193533513e+00 &  5.203964057e+00 \\ \hline
  \multicolumn{6}{c}{$1\,000 \, \mathrm{K} \le T \le 6\,000 \, \mathrm{K}$} \\ \hline
  $a_1$ &  5.454004664e+05 & -4.014143683e+04 &  1.940280236e+05 &  0.000000000e+00 &  0.000000000e+00 \\
  $a_2$ & -2.086277240e+03 & -5.035795069e+02 & -1.216718047e+03 &  0.000000000e+00 &  0.000000000e+00 \\
  $a_3$ &  5.833819866e+00 &  4.740919616e+00 &  5.306332593e+00 &  2.500000000e+00 &  2.500000000e+00 \\
  $a_4$ & -4.613787447e-04 &  7.381501944e-07 & -2.681005265e-04 &  0.000000000e+00 &  0.000000000e+00 \\
  $a_5$ &  9.786643869e-08 & -1.095155626e-08 &  5.941165696e-08 &  0.000000000e+00 &  0.000000000e+00 \\
  $a_6$ & -1.047202046e-11 &  4.351743287e-12 & -6.096049390e-12 &  0.000000000e+00 &  0.000000000e+00 \\
  $a_7$ &  4.660049516e-16 & -3.542774272e-16 &  2.758636876e-16 &  0.000000000e+00 &  0.000000000e+00 \\
  $b_1$ &  1.189936810e+04 &  1.428431833e+03 &  1.696708979e+04 &  5.585984812e+04 &  2.892114779e+04 \\
  $b_2$ & -1.426142011e+01 & -3.976910243e+00 & -7.647046816e+00 &  4.193533513e+00 &  5.203964057e+00 \\ \hline
  \multicolumn{6}{c}{$6\,000 \, \mathrm{K} \le T \le 20\,000 \, \mathrm{K}$} \\ \hline
  $a_1$ & -1.678129830e+07 &  1.826900078e+08 & -1.365586424e+08 &  0.000000000e+00 &  0.000000000e+00 \\
  $a_2$ &  3.311043178e+03 & -9.391194781e+04 &  9.859364412e+04 &  0.000000000e+00 &  0.000000000e+00 \\
  $a_3$ &  5.644407531e+00 &  2.147513208e+01 & -2.364619885e+01 &  2.500000000e+00 &  2.500000000e+00 \\
  $a_4$ & -4.358560705e-04 & -1.145626885e-03 &  3.984968128e-03 &  0.000000000e+00 &  0.000000000e+00 \\
  $a_5$ &  5.620799501e-08 &  1.651628592e-08 & -2.796926055e-07 &  0.000000000e+00 &  0.000000000e+00 \\
  $a_6$ & -2.780311114e-12 &  7.289307242e-13 &  9.262427905e-12 &  0.000000000e+00 &  0.000000000e+00 \\
  $a_7$ &  4.563924616e-17 & -2.071417179e-17 & -1.188448333e-16 &  0.000000000e+00 &  0.000000000e+00 \\
  $b_1$ & -3.610437453e+04 &  7.633975930e+05 & -7.569135711e+05 &  5.585984812e+04 &  2.892114779e+04 \\
  $b_2$ & -1.177560915e+01 & -1.560500417e+02 &  2.385691860e+02 &  4.193533513e+00 &  5.203964057e+00
 \end{tabular}

\end{table}


\subsection{Comparison of Lewis-database vs. PES-consistent properties} \label{sec:comparison_lewis_pes}

\begin{figure}
 \centering
 
 \includegraphics[width=0.85\textwidth]{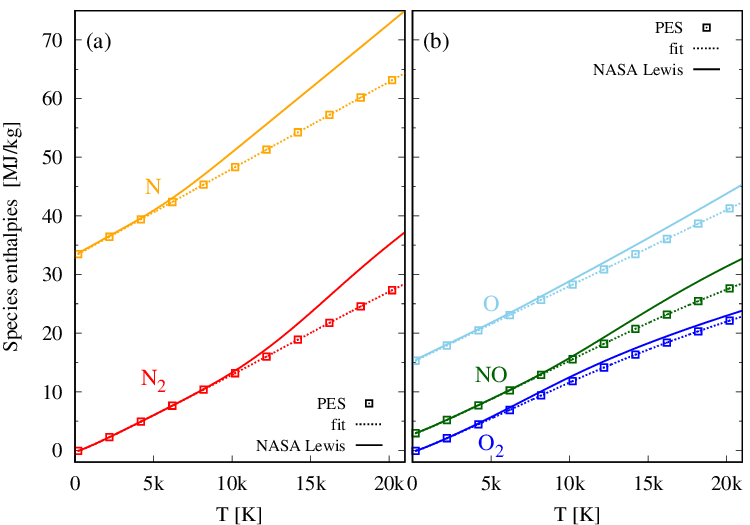}
 \caption{Species enthalpies per unit mass for 5-species air, derived from diatomic PES energy levels (symbols) vs. PES curve fit (dotted) vs. NASA Lewis fits (solid)}
 \label{fig:enthalpy_combined_air5_comparison}
\end{figure}

\begin{figure}
 \centering
 
 \includegraphics[width=\textwidth]{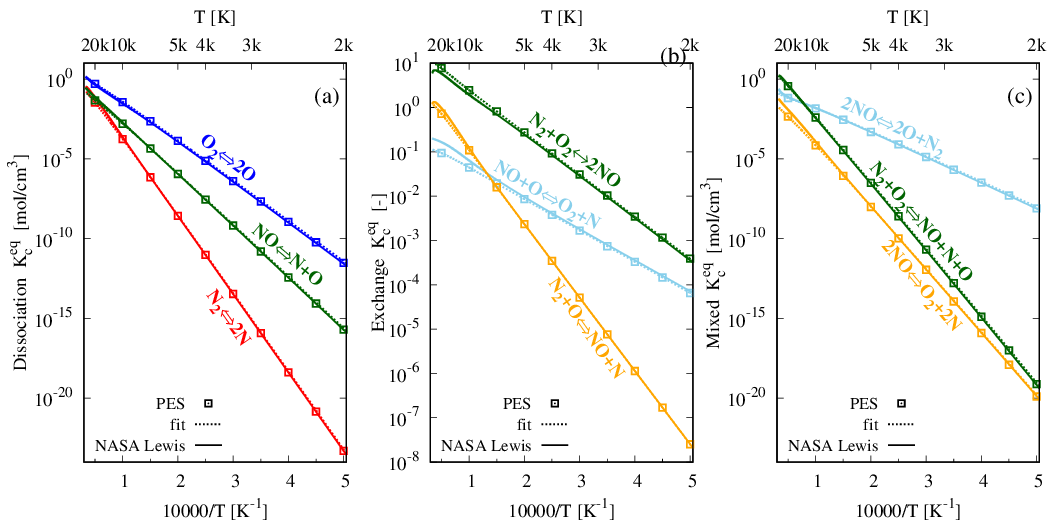}
 \caption{Equilibrium constants for 5-species air, derived from diatomic PES energy levels (symbols) vs. PES curve fit (dotted) vs. NASA Lewis fits (solid)}
 \label{fig:Keq_combined_air5_comparison}
\end{figure}

The degree to which thermodynamic properties derived from the ground-state-only PES fit parameters differ from those using the NASA Lewis ones is illustrated in Figs.~\ref{fig:enthalpy_combined_air5_comparison} and \ref{fig:Keq_combined_air5_comparison}. Enthalpies per unit mass, $h_s (T) = \hat{h}_s (T) / M_s$, for molecular and atomic nitrogen ($\mathrm{N_2}$: red, $\mathrm{N}$: orange) are shown in Fig.~\ref{fig:enthalpy_combined_air5_comparison}(a) and for the remaining species ($\mathrm{O_2}$: dark blue, $\mathrm{NO}$: green, $\mathrm{O}$: light blue) in Fig.~\ref{fig:enthalpy_combined_air5_comparison}(b). Enthalpies calculated directly from PES-derived partition functions through Eq.~(\ref{eq:enthalpy_direct}) are labeled as ``PES'' (symbols), whereas the enthalpies using the corresponding curve fit coefficients from Table~\ref{tab:pes_thermo} are plotted using dotted lines. Enthalpies obtained with the default Lewis database parameters from Table~\ref{tab:nasa_thermo} are labeled as ``NASA Lewis'' (solid lines). For each of the five species the PES-derived and Lewis database enthalpies yield essentially the same low-temperature behavior. This is to be expected, since for every species only the ground electronic state is significantly populated in the low-temperature limit. As temperature increases, electronic excited state contributions gradually become more significant, causing the default NASA Lewis predictions (solid lines) to diverge from those derived from the ground-state PESs (symbols and dotted lines) alone. The departure becomes most noticeable for $\mathrm{N_2}$ above $10\,000 \, \mathrm{K}$, whereas for $\mathrm{O_2}$ it already begins at $3\,000 \, \mathrm{K}$, but remains much less severe from there on.


In Fig.~\ref{fig:Keq_combined_air5_comparison} we plot the equilibrium constants expressed in terms of equilibrium concentration ratios. We again compare the prediction obtained from the PES-derived partition functions, i.e. Eq.~(\ref{eq:kc_eq_direct}) (symbols), the corresponding PES-derived curve fits (dotted lines, evaluated according to Eq.~(\ref{eq:kc_eq}) with fit parameters from Table~\ref{tab:pes_thermo}) and those from the Lewis fits (solid lines, fit parameters from Table~\ref{tab:nasa_thermo}). Figure~\ref{fig:Keq_combined_air5_comparison}(a) shows the equilibrium constants for the three dissociation-type reactions as defined in Eqs.~(\ref{eq:keq_n2_diss})-(\ref{eq:keq_no_diss}): $K_\mathrm{c}^{\mathrm{N_2 \rightleftharpoons 2N}}$ (red), $K_\mathrm{c}^{\mathrm{O_2 \rightleftharpoons 2O}}$ (blue) and $K_\mathrm{c}^{\mathrm{NO \rightleftharpoons N+O}}$ (green) respectively. Figure~\ref{fig:Keq_combined_air5_comparison}(b) shows the equilibrium constants for the three exchange exchange-type reactions as defined in Eqs.~(\ref{eq:keq_exch_1})-(\ref{eq:keq_exch_3}): $K_\mathrm{c}^{\mathrm{NO+O \rightleftharpoons O_2+N}}$ (light blue), $K_\mathrm{c}^{\mathrm{N_2+O \rightleftharpoons NO+N}}$ (orange) and $K_\mathrm{c}^{\mathrm{N_2+O_2 \rightleftharpoons 2NO}}$ (green). Finally, Fig.~\ref{fig:Keq_combined_air5_comparison}(c) shows the equilibrium constants for the three exchange mixed-type reactions as defined in Eqs.~(\ref{eq:keq_mixed_1})-(\ref{eq:keq_mixed_3}): $K_\mathrm{c}^{\mathrm{2NO \rightleftharpoons 2O+N_2}}$ (light blue), $K_\mathrm{c}^{\mathrm{2NO \rightleftharpoons 2N+O_2}}$ (orange) and $K_\mathrm{c}^{\mathrm{N_2+O_2 \rightleftharpoons NO+N+O}}$ (green). As was the case for species enthalpies, differences between the ground-state-only PES-derived and NASA Lewis equilibrium constants become more prominent at higher temperatures, where electronic excited states of all species would increasingly become populated.


\section{Calculating thermodynamic properties consistent with the Minnesota ab initio potentials} \label{sec:thermo_props_from_pes}

In this section we summarize the procedure followed to derive thermodynamic property tables for a 5-species air mixture consistent with the behavior of the Minnesota potential energy surfaces. Curve fits to these custom-built properties in the format of Eqs.(\ref{eq:cp_lewis}-\ref{eq:s_lewis}) yield the parameters in Table~\ref{tab:pes_thermo}.

Dilute gas-phase thermodynamic properties for any diatomic, or atomic species can be calculated if one is able to construct the respective overall partition function and temperature derivatives thereof. Each species' overall partition function is computed as the product of translational, electronic and, for diatomic species, rovibrational mode contributions. We write the translational mode partition function as:
\begin{equation}
 Q_s^\mathrm{t} (T, V) = \left( \frac{2 \pi \, M_s \bar{R} \, T}{h_\mathrm{P}^2} \right)^{3/2} V,
\end{equation}
where $M_s$ is the species molar mass, $\bar{R}$ the universal gas constant, $h_\mathrm{P}$ Planck's constant and $V$ the volume. In the present case the combined rovibrational partition function for each of the three diatomic species is calculated as the ``state sum'':
\begin{equation}
 Q_s^\mathrm{r-v} (T) = \frac{1}{\sigma_s} \sum\limits_{v \in \mathcal{V}_s} \sum\limits_{J \in \mathcal{J}_v} (2J + 1) \exp \left( - \frac{\hat{\varepsilon}_{s(v,J)}}{\bar{R} T} \right), \quad \text{for} \quad s = \mathrm{N_2}, \mathrm{O_2}, \mathrm{NO}  \label{eq:rovib_partition}
\end{equation}
where the $\hat{\varepsilon}_{s(v,J)}$ are the energies per unit mole of rovibrational level $(v,J)$ of the given diatom $s$ in its ground electronic state. The full set of rovibrational energies was derived from the diatomic potential curves for $\mathrm{N_2} = \mathrm{N_2}(X\, ^1\Sigma_g^+)$, $\mathrm{O_2} = \mathrm{\mathrm{O_2}(X\, ^3\Sigma_g^-)}$ and $\mathrm{NO} = \mathrm{NO}(X \, ^2\Pi_r)$, which are included with the Minnesota ab initio PESs employed in this work. Major parameters relevant to these diatomic potentials and the derived rovibrational levels are listed in Table~A3 of Ref.~\cite{torres24a} and a complete listing of all three species' rovibrational energy levels can be found in the supplemental information to Ref.~\cite{torres24b}. The symmetry factor $\sigma_s$ in Eq.~(\ref{eq:rovib_partition}) is equal to 2 for homonuclear diatoms, i.e. $\mathrm{N_2}$ and $\mathrm{O_2}$, and equal to 1 for the heteronuclear one, i.e. $\mathrm{NO}$.

Contributions from the electronic mode partition function are trivial in this work, since only ground-electronic-state (i.e. electronic energy level $i=0$) contributions are included in our calculation. In our convention the ground-state electronic energy $\varepsilon_{s(i=0)}^\mathrm{el}$, (which by itself is zero) is added on top of the ground-state formation energy that species $\hat{\varepsilon}_{s}^\mathrm{f}$, which means the state sum to compute the electronic partition function reduces to:
\begin{equation}
 Q_s^\mathrm{el} (T) = g_{s(i=0)}^\mathrm{el} \exp \left( - \frac{\hat{\varepsilon}_{s}^\mathrm{f} }{\bar{R} T} \right)
\end{equation}
with $g_{s(i=0)}^\mathrm{el}$ as the ground-electronic-state degeneracy of species $s$. The final expression for the overall partition function of species $s$ thus becomes:
\begin{equation}
 Q_s (T, V) = g_{s(i=0)}^{\mathrm{el}} \, \exp \left( - \frac{\hat{\varepsilon}_{s}^\mathrm{f} }{\bar{R} T} \right) Q_s^\mathrm{t} (T,V) \times \left\lbrace \begin{array}{l l}
                         Q_s^\mathrm{r-v} (T) & \text{for} \quad s = \mathrm{N_2}, \mathrm{O_2}, \mathrm{NO} \\
                         1 &  \text{for} \quad s = \mathrm{N}, \mathrm{O}
                        \end{array} \right. \label{eq:overall_partition_function}
\end{equation}


The specific values for $\sigma_s$, $g_{s(i=0)}^\mathrm{el}$ and formation energy $\hat{\varepsilon}_{s}^\mathrm{f}$ employed in the current work are summarized in Table~A4 of Ref.~\cite{torres24a}. With partition functions for each of the 5 species defined in Eq.~(\ref{eq:overall_partition_function}), what remains is to derive the necessary thermodynamic properties.

Following statistical mechanics relations the thermal energy\footnote{We deliberately replace the commonly used term ``internal energy'' with ``thermal energy'' when referring to $\hat{e}_s$ so as to prevent confusion with internal (i.e. rotational, vibrational and electronic) energy mode contributions.} per unit mole of species $s$ is found as $\hat{e}_{s} (T) = \bar{R} T^2 \, ( \partial \ln Q_s / \partial T) = \bar{R} T^2 / Q_s (\partial Q_s / \partial T)$. The corresponding species enthalpy $\hat{h}_s = \hat{e}_s + \bar{R} T$ becomes:
\begin{equation}
 \hat{h}_{s} (T) = \frac{5}{2} \bar{R} T + \hat{\varepsilon}_{s}^\mathrm{f} + \left\lbrace \begin{array}{l l}
                                                                                                         \dfrac{A_s^\mathrm{r-v} (T)}{Q_s^\mathrm{r-v} (T)} & \text{for} \quad s = \mathrm{N_2}, \mathrm{O_2}, \mathrm{NO} \\
                                                                                                         0 & \text{for} \quad s = \mathrm{N}, \mathrm{O},
                                                                                                        \end{array} \right. \label{eq:enthalpy_direct}
\end{equation}
while the associated molar heat capacity at constant pressure $\hat{c}_{\mathrm{p}, s} = \partial \hat{h}_s / \partial T$ is:
\begin{equation}
 \hat{c}_{\mathrm{p}, s} (T) = \frac{5}{2} \bar{R} + \left\lbrace \begin{array}{l l}
                                                                                                           \dfrac{1}{\bar{R} T} \left[ \left( \dfrac{A_s^\mathrm{r-v} (T)}{Q_s^\mathrm{r-v} (T)} \right)^2 - \dfrac{B_s^\mathrm{r-v} (T)}{Q_s^\mathrm{r-v} (T)} \right] & \quad s = \mathrm{N_2}, \mathrm{O_2}, \mathrm{NO} \\
                                                                                                           0 & \text{for} \quad s = \mathrm{N}, \mathrm{O}.
                                                                                                          \end{array} \right. \label{eq:cp_direct}
\end{equation}

Equations~(\ref{eq:enthalpy_direct}) and (\ref{eq:cp_direct}) involve first and second temperature derivatives of the partition function, which make the two rovibrational ``state sums'':
\begin{equation}
 A_s^\mathrm{r-v}(T) = \frac{1}{\sigma_s} \sum\limits_{v \in \mathcal{V}_s} \sum\limits_{J \in \mathcal{J}_v} (2J + 1) \, \hat{\varepsilon}_{s(v,J)} \exp \left( - \frac{\hat{\varepsilon}_{s(v,J)}}{\bar{R} T} \right), \label{eq:a_rovib}
\end{equation}
and:
\begin{equation}
 B_s^\mathrm{r-v}(T) = \frac{1}{\sigma_s} \sum\limits_{v \in \mathcal{V}_s} \sum\limits_{J \in \mathcal{J}_v} (2J + 1) \, \hat{\varepsilon}_{s(v,J)}^2 \exp \left( - \frac{\hat{\varepsilon}_{s(v,J)}}{\bar{R} T} \right), \label{eq:b_rovib}
\end{equation}
appear.

Finally, species entropy can be derived via the expression $\hat{s}_s = \bar{R} \left[ \ln \left( Q_s (T,V) / N_s \right) + 1 + (\partial \ln Q / \partial T) \right]$, where $N_s$ is the number of molecules, or atoms of species $s$ occupying the volume $V$. This can be re-expressed in a more convenient manner in terms of species partial pressure $p_s$ and enthalpy: 
\begin{equation}
 \hat{s}_s (T,p_s) = \bar{R} \ln \left( \frac{\bar{R} T \, \check{Q}_s (T)}{p_s \mathcal{N}_A} \right) + \frac{\hat{h}_s (T)}{T}, \label{eq:species_entropy}
\end{equation}
where we have introduced the species partition function per unit volume $\check{Q}_s (T) = Q_s (T, V) / V$ and the substitution $N_s = \mathcal{N}_A p_s V / \bar{R} T$ with $\mathcal{N}_A$ being Avogadro's number.

Enthalpies defined according to Eq.~(\ref{eq:enthalpy_direct}), implicitly place the common reference temperature for all species at $T_\mathrm{ref} = 0 \, \mathrm{K}$. This is a perfectly valid \emph{choice} and, as long as thermodynamic properties for all species are consistently defined in this way, is completely adequate for calculations. However, values for thermal energy and associated enthalpy calculated in this manner will be offset from the ``assigned'' enthalpies provided by the Lewis database, which uses a reference temperature of $T_\mathrm{ref} = 298.15 \, \mathrm{K}$ instead. For consistency with the convention followed by the Lewis database, we thus adjust our definition for enthalpy by correcting for this temperature offset. The ``assigned'' enthalpy per unit mole of species $s$ equivalent to the Lewis curve fits ends up being:
\begin{equation}
 \hat{h}_{s} (T) = \frac{5}{2} \bar{R} \left( T - T_\mathrm{ref} \right) + \hat{h}_{s}^\mathrm{f}(T_\mathrm{ref}) + \left\lbrace \begin{array}{l l}
                                                                                                           \dfrac{A_s^\mathrm{r-v} (T)}{Q_s^\mathrm{r-v} (T)} - \dfrac{A_s^\mathrm{r-v} (T_\mathrm{ref})}{Q_s^\mathrm{r-v} (T_\mathrm{ref})} & \text{for} \quad s = \mathrm{N_2}, \mathrm{O_2}, \mathrm{NO} \\
                                                                                                           0 & \text{for} \quad s = \mathrm{N}, \mathrm{O},
                                                                                                          \end{array} \right. \label{eq:enthalpy_assigned}
\end{equation}
where $\hat{h}_{s}^\mathrm{f}(T_\mathrm{ref}) = \hat{\varepsilon}_s^\mathrm{f} + \bar{R} T_\mathrm{ref}$ is now the species enthalpy of formation at reference temperature $T_\mathrm{ref}$. Note that the heat capacity in Eq.~(\ref{eq:cp_direct}) does not require any temperature offset and therefore remains unaffected by this redefinition. The expression for entropy also remains valid, only that the enthalpy term in Eq.~(\ref{eq:species_entropy}) now introduces a corresponding temperature offset to the calculated values of $\hat{s}_s$. However, since the same offset is also present in the ``assigned'' entropies of the Lewis database, Eq.~(\ref{eq:species_entropy}) automatically remains consistent. In summary, Eqs.~(\ref{eq:cp_direct}), (\ref{eq:enthalpy_assigned}) and (\ref{eq:species_entropy}) evaluated with $T_\mathrm{ref} = 298.15\, \mathrm{K}$ will produce the PES-derived thermodynamic properties using the same reference state as the ``assigned'' values of the Lewis database. 

By further combining Eqs.~(\ref{eq:enthalpy_assigned}) and Eq.~(\ref{eq:species_entropy}), one is able to compute the Gibbs free energy per unit mole (or chemical potential) of each species $\hat{\mu}_s = \hat{h}_s - T \, \hat{s}_s (T,p_s)$ directly from the partition functions. Minimization of the Gibbs free energy then makes it possible to derive the condition for obtaining partial pressures at equilibrium for a given reaction ``$r$'':
\begin{equation}
 \sum\limits_{s \in S} ( \nu_{s,r}^b - \nu_{s,r}^f ) \left( \frac{\hat{h}_s (T) - T \, \hat{s}_s (T, p_s^\star)}{\bar{R} T} \right) = 0,
\end{equation}
which itself can be re-written as the more familiar equilibrium constant in terms of partial pressures:
\begin{equation}
 K_\mathrm{p}^r (T) = \prod\limits_{s \in S} ( p_s^\star )^{\nu_{s,r}^b - \nu_{s,r}^f} =  \prod\limits_{s \in S} \left( \frac{\bar{R} T \, \check{Q}_s (T)}{\mathcal{N}_A} \right)^{\nu_{s,r}^b - \nu_{s,r}^f}. \label{eq:kp_eq_direct}
\end{equation}
or the equivalent form in terms of molar concentrations:
\begin{equation}
 K_\mathrm{c}^r (T) = \prod\limits_{s \in S} ( C_s^\star )^{\nu_{s,r}^b - \nu_{s,r}^f} =  \prod\limits_{s \in S} \left( \frac{\check{Q}_s (T)}{\mathcal{N}_A} \right)^{\nu_{s,r}^b - \nu_{s,r}^f}. \label{eq:kc_eq_direct}
\end{equation}

By design, Eqs.~(\ref{eq:kp_eq_direct}) and (\ref{eq:kc_eq_direct}) produce PES-derived values for the equilibrium constant consistent with the ones derived from the Lewis fits expressed in Eq.~(\ref{eq:kp_eq_lewis}) and (\ref{eq:kc_eq}) respectively. As shown in Sec.~\ref{sec:thermo_comparison}, just as is the case for enthalpy both sets of equilibrium constants will be in close agreement at low to moderate temperatures, but will diverge at temperatures where electronic excited states become increasingly populated. 

In Table~\ref{tab:pes_thermo} we list curve fit parameters to these PES-derived properties in the 9-coefficient format of the Lewis database. The CFD benchmarking calculations in Sec.~III of the main article employ Eqs.~(\ref{eq:cp_lewis}), (\ref{eq:h_lewis}) and (\ref{eq:s_lewis}) together with the values of Table~\ref{tab:pes_thermo} to replicate the behavior of the DMS calculations as closely as possible.

\section{Evaluating thermodynamic properties in a two-temperature \texorpdfstring{$(T,T_\mathrm{v})$}{(T,Tv)} fluid model} \label{sec:multitemp_thermo}

In multi-temperature fluid models thermodynamic nonequilibrium manifests itself in local departure of individual energy mode temperatures (translational, vibrational, etc.) from a common equilibrium value. When numerically simulating such flows it becomes necessary to evaluate the species enthalpies at conditions where they depend on more than one temperature simultaneously. Direct evaluation of the Lewis fit expressions Eqs.~(\ref{eq:cp_lewis}) and (\ref{eq:h_lewis}) however only yields enthalpy and heat capacity of species $s$ at the \emph{equilibrium} temperature. Thus, a few adjustments must be made to make the Lewis fits applicable to nonequilibrium situations. Consider the expression for enthalpy:
\begin{equation}
 h_s |_{T_\mathrm{ref}} (T) = h_{s}^\mathrm{f, \standardstate} (T_\mathrm{ref}) + \int_{T_\mathrm{ref}}^{T} c_{\mathrm{p}, s} (\bar{T}) \, \mathrm{d} \bar{T}, \qquad s \in S \label{eq:enthalpy_tref}
\end{equation}
with $h_{s}^\mathrm{f, \standardstate} (T_\mathrm{ref})$ as the formation enthalpy of species $s$ at standard pressure $p_\standardstate$ and reference temperature $T_\mathrm{ref}$. For the thermally perfect gas species we consider in this work, the heat capacity at constant pressure is a function of temperature alone. 
When coupling between translational and different internal energy modes is ignored, it is possible to split the heat capacity into contributions due to a species' translational (t) and relevant internal (rotation (r), vibration (v) and electronic (el)) energy:
\begin{equation}
 h_s |_{T_\mathrm{ref}} (T) = h_{s}^\mathrm{f, \standardstate} (T_\mathrm{ref}) + \left\lbrace \begin{array}{l l}
            \int_{T_\mathrm{ref}}^{T} \left( c_{\mathrm{p}, s}^\mathrm{t}(\bar{T}) + c_{\mathrm{p}, s}^\mathrm{r}(\bar{T}) + c_{\mathrm{p}, s}^\mathrm{v}(\bar{T}) + c_{\mathrm{p}, s}^\mathrm{el}(\bar{T}) \right) \mathrm{d} \bar{T}, & \text{for} \quad s = \mathrm{N_2}, \mathrm{O_2}, \mathrm{NO}, \\
            \int_{T_\mathrm{ref}}^{T} \left( c_{\mathrm{p}, s}^\mathrm{t}(\bar{T}) + c_{\mathrm{p}, s}^\mathrm{el}(\bar{T}) \right) \mathrm{d} \bar{T}. & \text{for} \quad s = \mathrm{N}, \mathrm{O}.
           \end{array} \right. \label{eq:enthalpy_decomposed_tref}
\end{equation}



In general every species' up to four energy modes may relax at separate characteristic time scales and be out of equilibrium with each other, resulting in a local dependence on multiple mode temperatures. In practice, several simplifying assumptions are often made to reduce the dependence down to two. First, since the molar masses of all five mixture components are comparable in order of magnitude, we implicitly assume that their translational modes remain thermalized at a common mixture translational temperature $T_\mathrm{t} = T_{\mathrm{t}, s}$ with $s = \mathrm{N_2}, \mathrm{O_2}, \mathrm{NO} , \mathrm{N}$ and $\mathrm{O}$ at all times\footnote{We ignore the possibility of a separate translational temperature for free electrons here, since we do not deal with ionized flows.}. Second, since rotational relaxation times of all major diatomic species are similar in magnitude (for instance, see Ref.~\cite{torres24b}) they may also be assumed to remain in equilibrium at a common mixture rotational temperature $T_\mathrm{r} = T_{\mathrm{r}, s}$ with $s = \mathrm{N_2}, \mathrm{O_2}, \mathrm{NO}$. Furthermore, it is usually assumed that translational and rotational modes remain in equilibrium at a common trans-rotational temperature $T = T_\mathrm{t} = T_\mathrm{r}$ for most portions of a hypersonic flow field. This assumption may not be strictly valid immediately downstream of strong shock waves, or within boundary layers developing on sharp leading edges of slender bodies at high-altitudes~\cite{bhide21a}. However, at flow conditions relevant to multi-temperature CFD models the extent of this relaxation region is often relatively small and may be neglected. Within the temperature range relevant to this work any diatomic species' translational and rotational degrees of freedom always remain fully excited. Thus, for $s = \mathrm{N_2}, \mathrm{O_2}, \mathrm{NO}$ their contributions to the heat capacity at constant pressure become $c_{\mathrm{p}, s}^\mathrm{t} + c_{\mathrm{p},s}^\mathrm{r} = \tfrac{5}{2} R_s + R_s= \tfrac{7}{2} R_s$. For monatomic species the rotational mode obviously does not contribute to heat capacity, thus for them only the term $c_{\mathrm{p}, s}^\mathrm{t} = \tfrac{5}{2} R_s$ remains.

A less well-founded approximation is that vibrational modes of all diatomic species relax at a common rate, resulting in $T_\mathrm{v} = T_{\mathrm{v}, s}$ for $s = \mathrm{N_2}, \mathrm{O_2}, \mathrm{NO}$. Indeed, experiments~\cite{streicher20c} and molecular-scale simulations of vibrational relaxation in $\mathrm{O_2}/\mathrm{N_2}$ mixtures~\cite{torres24b} contradict this assumption. Nevertheless, this simplification is often made in production-level CFD codes as a reasonable trade-off between higher physical fidelity and the added computational cost associated with solving separate vibrational energy equations for each diatomic species. The last major simplification is to assume that, regardless of chemical species, the gas' vibrational and electronic modes relax at a common time scale and remain in equilibrium with each other at a common vibration-electronic temperature $T_\mathrm{v} = T_\mathrm{el}$.

When all these assumptions are incorporated into Eq.~(\ref{eq:enthalpy_decomposed_tref}), the two-temperature expression for enthalpy may be written as:
\begin{equation}
 h_s |_{T_\mathrm{ref}} (T, T_\mathrm{v}) =  h_{s}^\mathrm{f, \standardstate} (T_\mathrm{ref}) + \left\lbrace \begin{array}{l l}
            \tfrac{7}{2} R_s \left( T - T_\mathrm{ref} \right) + \Delta e_{\mathrm{v-el}, s} (T_\mathrm{v}, T_\mathrm{ref}) & \text{for} \quad s = \mathrm{N_2}, \mathrm{O_2}, \mathrm{NO} \\
            \tfrac{5}{2} R_s \left( T - T_\mathrm{ref} \right) + \Delta e_{\mathrm{el}, s} (T_\mathrm{v}, T_\mathrm{ref}) & \text{for} \quad s = \mathrm{N}, \mathrm{O}.
           \end{array} \right. \label{eq:enthalpy_decomposed_tref_ttv}
\end{equation}
where we have introduced a shorthand notation for vibrational + electronic mode contributions:
\begin{equation}
 \begin{array}{l l}
  \Delta e_{\mathrm{v-el}, s} (T_\mathrm{v},T_\mathrm{ref}) = \int_{T_\mathrm{ref}}^{T_\mathrm{v}} \left( c_{\mathrm{p}, s}^\mathrm{v}(\bar{T}) + c_{\mathrm{p}, s}^\mathrm{el}(\bar{T}) \right) \mathrm{d} \bar{T} & \text{for} \quad s = \mathrm{N_2}, \mathrm{O_2}, \mathrm{NO} \\
  \Delta e_{\mathrm{el}, s} (T_\mathrm{v},T_\mathrm{ref}) = \int_{T_\mathrm{ref}}^{T_\mathrm{v}} c_{\mathrm{p}, s}^\mathrm{el}(\bar{T}) \, \mathrm{d} \bar{T} & \text{for} \quad s = \mathrm{N}, \mathrm{O}.
 \end{array} \label{eq:devel_species}
\end{equation}

Unlike translation-rotation, vibrational and electronic energy modes never become fully excited at the temperatures of interest in this work and their thermal energy contributions remain nonlinear, non-trivial functions of temperature. When thermalized at temperature $T_\mathrm{v}$ they may be computed as Boltzmann-weighted ``state sums'' over all discrete vibrational and electronic energy levels. This pre-supposes knowledge of all discrete vibrational level energies, electronic level energies and electronic level degeneracies of each species $s$. 

In CFD practice, calculating these state sums at every iteration and in every cell usually is not desirable, because it adds significant computational overhead. An alternative approach is to outsource this expense to a pre-computed thermodynamic database containing all necessary properties as function of temperature and to evaluate curve fits to these quantities during the CFD calculations instead. Thus, we may back out the vibrational and electronic contributions to the two-temperature enthalpies in Eq.~(\ref{eq:enthalpy_decomposed_tref_ttv}) from a curve-fit expression $h_s^\mathrm{fit} |_{T_\mathrm{ref}} (T_\mathrm{v})$. We do this by evaluating the left hand side of Eq.~(\ref{eq:enthalpy_decomposed_tref_ttv}) for $T\!=\!T_\mathrm{v}$, i.e. $h_s |_{T_\mathrm{ref}} (T_\mathrm{v}, T_\mathrm{v}) = h_s^\mathrm{fit} |_{T_\mathrm{ref}} (T_\mathrm{v})$ and re-arranging terms: 
\begin{equation}
 \begin{array}{r l}
  \Delta e_{\mathrm{v-el}, s} (T_\mathrm{v}, T_\mathrm{ref}) = h_s^\mathrm{fit} |_{T_\mathrm{ref}} (T_\mathrm{v}) - \tfrac{7}{2} R_s \left(T_\mathrm{v} - T_\mathrm{ref} \right) - h_{s}^\mathrm{f, \standardstate} (T_\mathrm{ref}) & \text{for} \quad s= \mathrm{N_2}, \mathrm{O_2}, \mathrm{NO} \\
  \Delta e_{\mathrm{el}, s} (T_\mathrm{v}, T_\mathrm{ref}) = h_s^\mathrm{fit} |_{T_\mathrm{ref}} (T_\mathrm{v}) - \tfrac{5}{2} R_s \left( T_\mathrm{v} - T_\mathrm{ref} \right) - h_{s}^\mathrm{f, \standardstate} (T_\mathrm{ref}) & \text{for} \quad s = \mathrm{N}, \mathrm{O}.
 \end{array} \label{eq:evel_species_lewis}
\end{equation}

When we evaluate $h_s^\mathrm{fit} |_{T_\mathrm{ref}} (T_\mathrm{v}) = \hat{h}_s^\standardstate (T_\mathrm{v}) / M_s$ using Eq.~(\ref{eq:h_lewis}) and the original NASA Lewis coefficients of Table~\ref{tab:nasa_thermo}, Eq.~(\ref{eq:evel_species_lewis}) will automatically yield the combined contributions from vibrational and/or electronic energy for each species. When Eq.~(\ref{eq:evel_species_lewis}) is then substituted back into Eq.~(\ref{eq:enthalpy_decomposed_tref_ttv}), one ends up with an expression for the two-temperature enthalpy that requires one only to have the curve fits $h_s^\mathrm{fit} |_{T_\mathrm{ref}} (T_\mathrm{v})$ available.
\begin{equation}
 h_s |_{T_\mathrm{ref}} ( T, T_\mathrm{v} ) = h_s^\mathrm{fit} |_{T_\mathrm{ref}} (T_\mathrm{v}) + \left\lbrace
 \begin{array}{r l}
  \tfrac{7}{2} R_s \left( T - T_\mathrm{v} \right) & \text{for} \quad s = \mathrm{N_2}, \mathrm{O_2}, \mathrm{NO} \\
  \tfrac{5}{2} R_s \left( T - T_\mathrm{v} \right) & \text{for} \quad s = \mathrm{N}, \mathrm{O}.
 \end{array} \right. \label{eq:two_temp_enthalpy_fit}
\end{equation}

The corresponding expressions for the translational-rotational / translational heat capacities at constant pressure become:
\begin{equation} 
 \frac{\partial h_s}{\partial T} \biggr|_{p, T_\mathrm{v}} = \left\lbrace
 \begin{array}{r l}
  c_{\mathrm{p}, s}^\mathrm{t-r} = \tfrac{7}{2} R_s & \text{for} \,\, s = \mathrm{N_2}, \mathrm{O_2}, \mathrm{NO} \\
  c_{\mathrm{p}, s}^\mathrm{t} = \tfrac{5}{2} R_s & \text{for} \,\, s = \mathrm{N}, \mathrm{O},
 \end{array} \right. \label{eq:cp_tra_rot}
\end{equation}
while the vibrational-electronic / electronic ones using the Lewis fits from Eq.~(\ref{eq:cp_lewis}), $c_{\mathrm{p}, s}^\mathrm{fit} (T_\mathrm{v}) = \hat{c}_{\mathrm{p}, s}^\standardstate (T_\mathrm{v}) / M_s$, end up being:
\begin{equation} 
 \frac{\partial h_s}{\partial T_\mathrm{v}} \biggr|_{p, T} = \left\lbrace 
 \begin{array}{r l}
  c_{\mathrm{p}, s}^\mathrm{v-el} (T_\mathrm{v}) = c_{\mathrm{p}, s}^\mathrm{fit} (T_\mathrm{v}) - \tfrac{7}{2} R_s & \text{for} \,\, s = \mathrm{N_2}, \mathrm{O_2}, \mathrm{NO} \\
  c_{\mathrm{p}, s}^\mathrm{el} (T_\mathrm{v}) = c_{\mathrm{p}, s}^\mathrm{fit} (T_\mathrm{v}) - \tfrac{5}{2} R_s & \text{for} \,\, s = \mathrm{N}, \mathrm{O}.
 \end{array} \right. \label{eq:cp_vib_el}
\end{equation}

The corresponding expression for thermal energy $e_s |_{T_\mathrm{ref}} = h_s |_{T_\mathrm{ref}} - R_s \, T$ becomes:
\begin{equation}
 e_s |_{T_\mathrm{ref}} ( T, T_\mathrm{v} ) = h_s^\mathrm{fit} |_{T_\mathrm{ref}} (T_\mathrm{v}) + \left\lbrace
 \begin{array}{r l}
  \tfrac{5}{2} R_s \, T - \tfrac{7}{2} R_s \, T_\mathrm{v} & \text{for} \quad s = \mathrm{N_2}, \mathrm{O_2}, \mathrm{NO} \\
  \tfrac{3}{2} R_s \, T - \tfrac{5}{2} R_s \, T_\mathrm{v} & \text{for} \quad s = \mathrm{N}, \mathrm{O}.
 \end{array} \right. \label{eq:two_temp_energy_fit}
\end{equation}
and those for the translational-rotational / translational heat capacities at constant volume become:
\begin{equation}
 \frac{\partial e_s}{\partial T} \biggr|_{v, T_\mathrm{v}} = \left\lbrace
 \begin{array}{r l}
  c_{\mathrm{v}, s}^\mathrm{t-r} = \tfrac{5}{2} R_s & \text{for} \,\, s = \mathrm{N_2}, \mathrm{O_2}, \mathrm{NO} \\
  c_{\mathrm{v}, s}^\mathrm{t} = \tfrac{3}{2} R_s & \text{for} \,\, s = \mathrm{N}, \mathrm{O},
 \end{array} \right.
\end{equation}
whereas $c_{\mathrm{v}, s}^\mathrm{v-el} (T_\mathrm{v}) = ( \partial e_s / \partial T_\mathrm{v} )|_{v, T} \equiv c_{\mathrm{p}, s}^\mathrm{v-el} (T_\mathrm{v})$ as given by Eq.~(\ref{eq:cp_vib_el}).

Alternatively, if one employs the PES-derived fit parameters of Table~\ref{tab:pes_thermo} in evaluating Eq.~(\ref{eq:evel_species_lewis}), the expression for diatoms now represents only the vibrational energy at vibrational temperature $T_\mathrm{v}$, since the PES-derived thermodynamic fits do not account for any electronic excited state contributions. In other words, $\Delta e_{\mathrm{v-el}, s} (T_\mathrm{v}, T_\mathrm{ref})$ effectively becomes $\Delta e_{\mathrm{v}, s}(T_\mathrm{v}, T_\mathrm{ref})$. Furthermore, the bottom row for monatomic species in Eq.~(\ref{eq:evel_species_lewis}) becomes trivial, since atoms do not possess vibrational degrees of freedom and their respective PES-derived fit by construction only includes contributions from the translational mode together with their formation enthalpy. Thus, the corresponding expression for the two-temperature enthalpy based on the PES fit parameters is functionally identical to Eq.~(\ref{eq:two_temp_enthalpy_fit}), but excludes any energy contributions of electronic excited states. From that it also follows that the vibrational mode heat capacity is now $c_{\mathrm{p}, s}^\mathrm{v} (T_\mathrm{v}) = c_{\mathrm{p}, s}^\mathrm{fit} (T_\mathrm{v}) - \tfrac{7}{2} R_s$ for diatomic species, whereas for monatomic species $c_{\mathrm{p}, s}^\mathrm{fit} (T_\mathrm{v}) \equiv c_{\mathrm{p}, s}^\mathrm{t} = \frac{5}{2} R_s$, which turns the second row of Eq.~(\ref{eq:cp_vib_el}) into a trivial expression.

\subsection{Notation for thermodynamic properties with reference temperature at absolute zero} \label{sec:alternative}

The NASA Lewis fits were constructed for $T_\mathrm{ref} = 298.15 \, \mathrm{K}$, but a common alternative in compressible CFD modeling is to choose $T_\mathrm{ref} = 0\mathrm{K}$ instead. In that case, the integral in Eq.~(\ref{eq:enthalpy_tref}) is evaluated between absolute zero and $T$ and the species formation enthalpies change to $h_{s}^\mathrm{f, \standardstate} (0\mathrm{K})$. The two enthalpy definitions are related as:
\begin{equation}
 h_s |_{0\mathrm{K}} (T) = h_s |_{T_\mathrm{ref}} (T) + \left[ h_{s}^\mathrm{f, \standardstate} (0\mathrm{K}) - h_{s}^\mathrm{f, \standardstate} (T_\mathrm{ref}) \right] + \int_{0\mathrm{K}}^{T_\mathrm{ref}} \, c_{\mathrm{p}, s} (\bar{T}) \, \mathrm{d} \bar{T}, \qquad s \in S. \label{eq:enthalpy_0k_to_tref}
\end{equation}

The term in brackets represents the difference between formation enthalpies at absolute zero and a different reference temperature $T_\mathrm{ref}$. The standard formation enthalpies are related to their respective formation energies by $h_{s}^\mathrm{f, \standardstate} (T_\mathrm{ref}) = e_{s}^\mathrm{f, \standardstate} + R_s \, T_\mathrm{ref}$, where the formation energies $e_{s}^\mathrm{f, \standardstate}$ are constants for each given species. This allows one to simplify the difference as $[ h_{s}^\mathrm{f, \standardstate} (0\mathrm{K}) - h_{s}^\mathrm{f, \standardstate} (T_\mathrm{ref}) ] =  - R_s \, T_\mathrm{ref}$. With $T_\mathrm{ref} = 0\mathrm{K}$ the equivalent form to Eq.~(\ref{eq:enthalpy_decomposed_tref_ttv}) becomes:
\begin{equation}
 h_s |_{0\mathrm{K}} (T, T_\mathrm{v}) =  h_{s}^\mathrm{f, \standardstate} (0\mathrm{K}) + \left\lbrace \begin{array}{l l}
            \tfrac{7}{2} R_s T + e_{\mathrm{v-el}, s}(T_\mathrm{v}) & \text{for} \quad s = \mathrm{N_2}, \mathrm{O_2}, \mathrm{NO} \\
            \tfrac{5}{2} R_s T + e_{\mathrm{el}, s}(T_\mathrm{v}) & \text{for} \quad s = \mathrm{N}, \mathrm{O},
           \end{array} \right. \label{eq:enthalpy_decomposed_0k_ttv}
\end{equation}
with vibrational-electronic / electronic energies for diatomic /atomic species now defined as:
\begin{equation}
 \begin{array}{l l}
  e_{\mathrm{v-el}, s} (T_\mathrm{v}) = \int_{0\mathrm{K}}^{T_\mathrm{v}} \left( c_{\mathrm{p}, s}^\mathrm{v}(\bar{T}) + c_{\mathrm{p}, s}^\mathrm{el}(\bar{T}) \right) \mathrm{d} \bar{T} & \text{for} \quad s = \mathrm{N_2}, \mathrm{O_2}, \mathrm{NO} \\
  e_{\mathrm{el}, s} (T_\mathrm{v}) = \int_{0\mathrm{K}}^{T_\mathrm{v}} c_{\mathrm{p}, s}^\mathrm{el}(\bar{T}) \, \mathrm{d} \bar{T} & \text{for} \quad s = \mathrm{N}, \mathrm{O}.
 \end{array} \label{eq:evel_species_0k}
\end{equation}

If we wish to employ the same Lewis curve fits as before to evaluate the vibrational-electronic contributions in Eq.~(\ref{eq:evel_species_0k}) as we did previously to arrive at Eq.~(\ref{eq:evel_species_lewis}), we must now account for the fact that these enthalpy fits were constructed for a reference temperature different from absolute zero. Following the same steps as before, but now also incorporating Eq.~(\ref{eq:enthalpy_0k_to_tref}) with $h_s |_{T_\mathrm{ref}} (T_\mathrm{v}) = h_s^\mathrm{fit} |_{T_\mathrm{ref}} (T_\mathrm{v})$ yields and recognizing that $c_{\mathrm{p}, s}^\mathrm{v}(\bar{T}) + c_{\mathrm{p}, s}^\mathrm{el}(\bar{T}) = c_{\mathrm{p}, s}^\mathrm{fit} (\bar{T}) - \frac{7}{2} R_s$ for diatoms and that $c_{\mathrm{p}, s}^\mathrm{el}(\bar{T}) = c_{\mathrm{p}, s}^\mathrm{fit} (\bar{T}) - \frac{5}{2} R_s$ for atoms, yields:
\begin{equation}
 \begin{array}{r l}
  e_{\mathrm{v-el}, s} (T_\mathrm{v}) = h_s^\mathrm{fit} |_{T_\mathrm{ref}} (T_\mathrm{v}) - \tfrac{7}{2} R_s T_\mathrm{v} - h_{s}^\mathrm{f, \standardstate} (T_\mathrm{ref}) + \int_{0\mathrm{K}}^{T_\mathrm{ref}} c_{\mathrm{p}, s}^\mathrm{fit} (\bar{T}) \, \mathrm{d} \bar{T} & \text{for} \quad s= \mathrm{N_2}, \mathrm{O_2}, \mathrm{NO} \\
  e_{\mathrm{el}, s} (T_\mathrm{v}) = h_s^\mathrm{fit} |_{T_\mathrm{ref}} (T_\mathrm{v}) - \tfrac{5}{2} R_s T_\mathrm{v} - h_{s}^\mathrm{f, \standardstate} (T_\mathrm{ref}) + \int_{0\mathrm{K}}^{T_\mathrm{ref}} c_{\mathrm{p}, s}^\mathrm{fit} (\bar{T}) \, \mathrm{d} \bar{T} & \text{for} \quad s = \mathrm{N}, \mathrm{O}.
 \end{array} \label{eq:evel_species_lewis_0k}
\end{equation}

Again, if the $h_s^\mathrm{fit} |_{T_\mathrm{ref}} (T_\mathrm{v})$ and $c_{\mathrm{p}, s}^\mathrm{fit} (\bar{T})$ in Eq.~(\ref{eq:evel_species_lewis_0k}) are evaluated using the PES-derived fit parameters of Table~\ref{tab:pes_thermo}, $e_{\mathrm{v-el}, s} (T_\mathrm{v})$  effectively represents $e_{\mathrm{v}, s} (T_\mathrm{v})$ for diatoms, whereas $e_{\mathrm{el}, s} (T_\mathrm{v})$ becomes zero for the atomic species. Finally, plugging Eq.~(\ref{eq:evel_species_lewis_0k}) back into Eq.~(\ref{eq:enthalpy_decomposed_0k_ttv}) and simplifying terms yields:
\begin{equation}
 h_s |_{0\mathrm{K}} (T, T_\mathrm{v}) = h_s^\mathrm{fit} |_{T_\mathrm{ref}} (T_\mathrm{v}) - R_s T_\mathrm{ref}  + \int_{0\mathrm{K}}^{T_\mathrm{ref}} c_{\mathrm{p}, s}^\mathrm{fit} (\bar{T}) \, \mathrm{d} \bar{T} + \left\lbrace \begin{array}{l l}
            \tfrac{7}{2} R_s (T - T_\mathrm{v}) + e_{\mathrm{v-el}, s}(T_\mathrm{v}) & \text{for} \quad s = \mathrm{N_2}, \mathrm{O_2}, \mathrm{NO} \\
            \tfrac{5}{2} R_s (T - T_\mathrm{v}) + e_{\mathrm{el}, s}(T_\mathrm{v}) & \text{for} \quad s = \mathrm{N}, \mathrm{O},
           \end{array} \right. \label{eq:two_temp_enthalpy_fit_0k}
\end{equation}

The corresponding thermal energies $e_s |_{0\mathrm{K}} = h_s |_{0\mathrm{K}} - R_s \, T$ become:
\begin{equation}
 e_s |_{0\mathrm{K}} (T, T_\mathrm{v}) = h_s^\mathrm{fit} |_{T_\mathrm{ref}} (T_\mathrm{v}) - R_s T_\mathrm{ref}  + \int_{0\mathrm{K}}^{T_\mathrm{ref}} c_{\mathrm{p}, s}^\mathrm{fit} (\bar{T}) \, \mathrm{d} \bar{T} + \left\lbrace \begin{array}{l l}
            \tfrac{5}{2} R_s T - \tfrac{7}{2} T_\mathrm{v} + e_{\mathrm{v-el}, s}(T_\mathrm{v}) & \text{for} \quad s = \mathrm{N_2}, \mathrm{O_2}, \mathrm{NO} \\
            \tfrac{3}{2} R_s T - \tfrac{5}{2} T_\mathrm{v} + e_{\mathrm{el}, s}(T_\mathrm{v}) & \text{for} \quad s = \mathrm{N}, \mathrm{O},
           \end{array} \right. \label{eq:two_temp_energy_fit_0k}
\end{equation}

Equations~(\ref{eq:two_temp_enthalpy_fit_0k}) and (\ref{eq:two_temp_energy_fit_0k}) for the two-temperature enthalpy and thermal energy expressions are equivalent to Eqs.~(\ref{eq:two_temp_enthalpy_fit}) and (\ref{eq:two_temp_energy_fit}), but contain additional terms to compensate for the fact that the Lewis fits were constructed with $T_\mathrm{ref} = 298.15 \, \mathrm{K}$ in mind, instead of absolute zero. Of course, the expressions for heat capacity are not affected by the choice of reference temperature.

\subsection{Note on the vibrational energies in the relaxation source term} \label{sec:small_note}

The vibrational energies $e_{\mathrm{v}, s} (T)$ and $e_{\mathrm{v}, s} (T_\mathrm{v})$ appearing in Eq.~(20) of the main article represent integrals of vibrational heat capacity $c_{\mathrm{p}, s}^\mathrm{v} (\bar{T}) \equiv c_{\mathrm{v}, s}^\mathrm{v} (\bar{T})$ from absolute zero to the given values $T$ and $T_\mathrm{v}$ respectively. This integral can be split into two parts:
\begin{equation}
 e_{\mathrm{v}, s} (T) = \int_{T=0\mathrm{K}}^\mathrm{T} c_{\mathrm{p}, s}^\mathrm{v} (\bar{T}) \, \mathrm{d} \bar{T} = \int_{T=0\mathrm{K}}^\mathrm{T_\mathrm{ref}} c_{\mathrm{p}, s}^\mathrm{v} (\bar{T}) \, \mathrm{d} \bar{T} + \int_\mathrm{T_\mathrm{ref}}^{T} c_{\mathrm{p}, s}^\mathrm{v} (\bar{T}) \, \mathrm{d} \bar{T}
\end{equation}

The rightmost term is just $\Delta e_{\mathrm{v}, s} (T, T_\mathrm{ref})$, which allows us to re-write:
\begin{equation}
 \Delta e_{\mathrm{v}, s} (T, T_\mathrm{ref}) = e_{\mathrm{v}, s} (T) - \int_{T=0\mathrm{K}}^\mathrm{T_\mathrm{ref}} c_{\mathrm{p}, s}^\mathrm{v} (\bar{T}) \, \mathrm{d} \bar{T} = e_{\mathrm{v}, s} (T) - e_{\mathrm{v}, s} (T_\mathrm{ref}).
\end{equation}

Therefore, the vibrational energy differences appearing in the numerators of the relaxation source term are not affected by the choice of reference temperature, i.e.:
\begin{equation}
 e_{\mathrm{v}, s} (T) - e_{\mathrm{v}, s} (T_\mathrm{v}) = \Delta e_{\mathrm{v}, s} (T, T_\mathrm{ref}) - \Delta e_{\mathrm{v}, s} (T_\mathrm{v}, T_\mathrm{ref})
\end{equation}
and Eq.~(20) in the main article remains valid regardless of the chosen value of $T_\mathrm{ref}$.

\subsection{Calculation of combined vibrational-electronic relaxation time in full MMT model simulations} \label{sec:velrelax}

In Sec.~III of the main article we benchmark the MMT model against DMS reference solutions and, since these DMS calculations exclusively simulate electronically adiabatic trajectories between the 5 species' respective ground electronic states, we neglect the effect of electronic excited states and only account for vibrational relaxation. However, later on in Sec.~VI of the main article, we modify the CFD implementation to improve its predictive capabilities in realistic flows and compare to Park's model. Among other things, this means switching to thermodynamic properties based on the Lewis fits Table~\ref{tab:nasa_thermo} that include contributions to internal energy of all 5 species' excited electronic states. In the context of our two-temperature model the mixture vibrational energy transport equation (i.e. Eq.~(4) in the main article) now becomes one for a combined vibrational-electronic mode, which effectively ties the electronic mode to the same time scale as that of vibrational energy relaxation.

Instead of the usual definition for the vibrational relaxation source term given by Eq.~(20) in the main article, we then resort to an alternative definition:
\begin{equation}
 w_\mathrm{v-el}^\mathrm{relax, *} = \frac{E_\mathrm{v-el} (T) - E_\mathrm{v-el} (T_\mathrm{v})}{\tau_v} = \frac{\sum\limits_{s \in S} \rho_s \left( e_{\mathrm{v-el}, s} (T) - e_{\mathrm{v-el}, s} (T_\mathrm{v}) \right)}{\tau_v} \label{eq:q_relax_alt}
\end{equation}
where the numerator now represents the difference between mixture vibrational-electronic energies per unit volume at translational-rotational and vibrational-electronic temperatures as driving force. Note that, contrary to Eq.~(20) in the main article, these combined vibrational-electronic energy terms include contributions from all species, not just the diatomic ones. This change becomes necessary, because the term as defined in Eq.~(\ref{eq:q_relax_alt}) is meant to account for relaxation of the gas mixture's combined vibrational-electronic mode with the translation-rotational one. Since all mixture components, regardless of whether diatomic or monatomic, may be electronically excited to some degree, the driving force in the numerator of $w_\mathrm{v-el}^\mathrm{relax, *}$ must now include these electronic energy contributions.

The second modification to Eq.~(20) in the main article is that the denominator in Eq.~(\ref{eq:q_relax_alt}) now contains a single, mixture-averaged vibrational-electronic relaxation time, defined as:
\begin{equation}
 \tau_v = \left( \sum\limits_{s \in D} \dfrac{x_s}{\langle \tau_s^v \rangle} \right)^{-1} \sum\limits_{q \in D} x_q, \label{eq:tau_v_mix}
\end{equation}
where each diatomic species' $\langle \tau_s^\mathrm{v} \rangle$ is still computed according to Eq.~(21) in the main article, but these individual contributions are then weighted by their respective mole fractions to arrive at the final result. The sums in Eq.~(\ref{eq:tau_v_mix}) only span the set of diatomic species, emphasizing the fact that this characteristic relaxation time $\tau_v$ is ultimately tied to the vibrational mode.

\section{Rate coefficient and vibrational energy change plots for MMT reactions} \label{sec:reactions_plots}

\begin{figure}
 \centering

 \subfloat[Rate coefficient $\mathrm{\, [cm^3 \cdot mol^{-1} \cdot s^{-1}]}$]{\label{fig:N2-N2_diss_arrhenius_fits2}
 \includegraphics[width=0.49\textwidth]{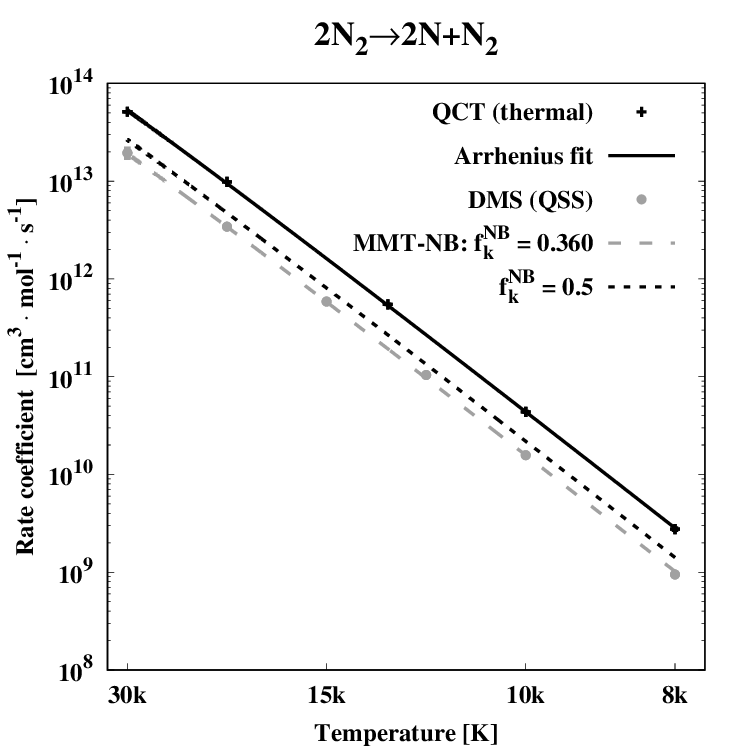}}~
 \subfloat[Vibrational energy change per reaction $\mathrm{\, [eV]}$]{\label{fig:mmt_evib_fit_N2-N2_diss2}
 \includegraphics[width=0.49\textwidth]{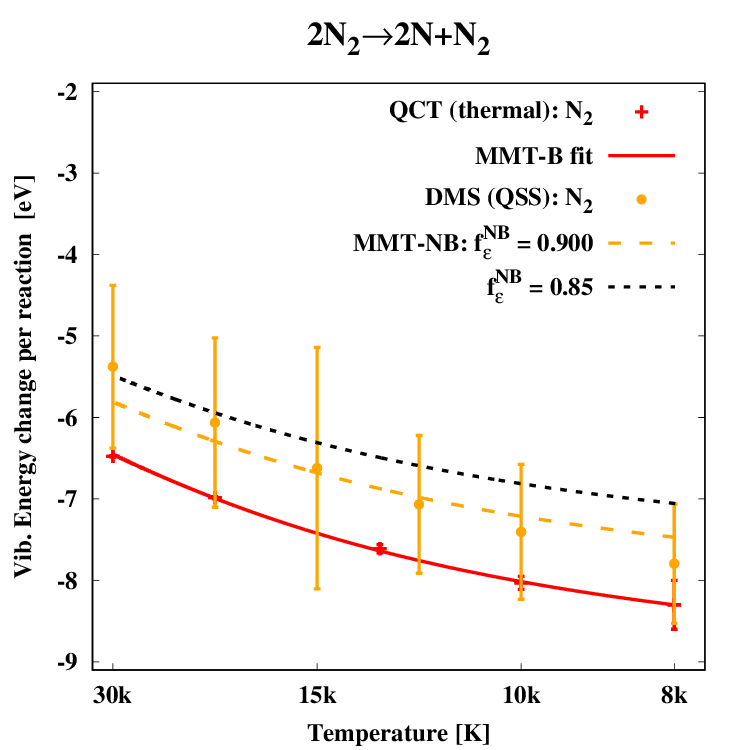}}
 
 \caption{$\boldsymbol{\mathrm{2N_2 \rightleftharpoons 2N + N_2}}$}
 \label{fig:N2-N2_diss}
\end{figure}

\begin{figure}
 \centering

 \subfloat[Rate coefficient $\mathrm{\, [cm^3 \cdot mol^{-1} \cdot s^{-1}]}$]{\label{fig:N2O2_N2_diss_arrhenius_fits}
 \includegraphics[width=0.49\textwidth]{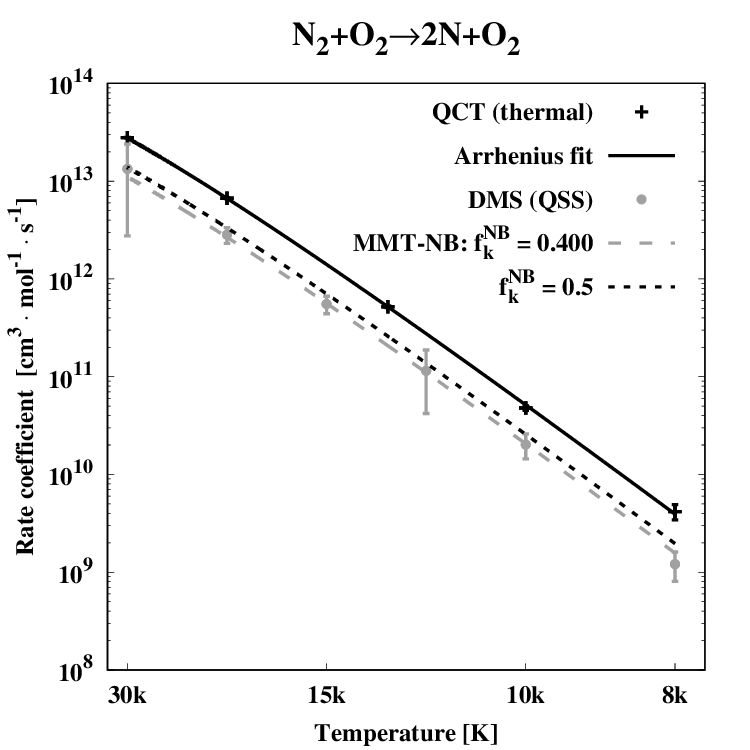}}~
 \subfloat[Vibrational energy change per reaction $\mathrm{\, [eV]}$]{\label{fig:mmt_evib_fit_N2O2_N2_diss}
 \includegraphics[width=0.49\textwidth]{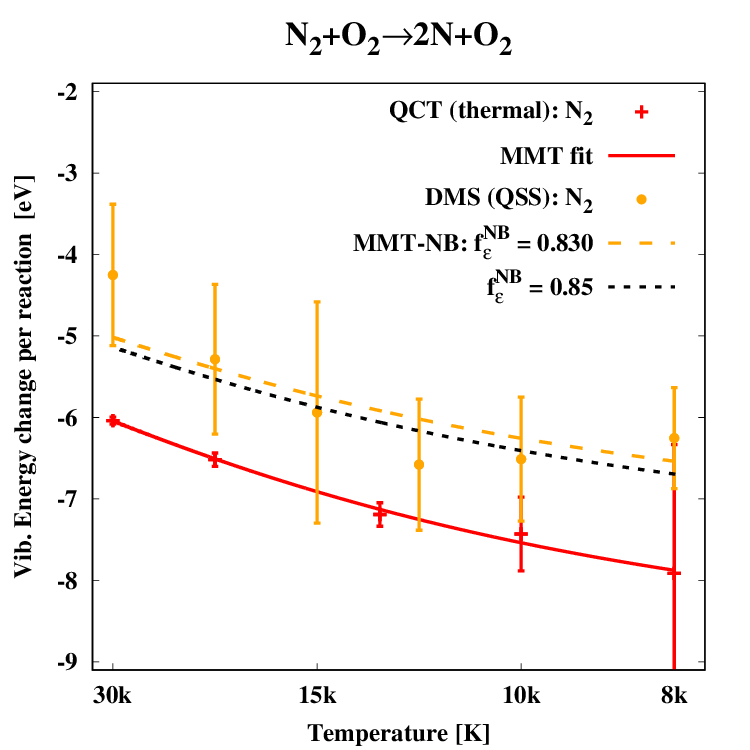}}
 
 \caption{$\boldsymbol{\mathrm{N_2 + O_2 \rightleftharpoons 2N + O_2}}$}
 \label{fig:N2O2_N2_diss}
\end{figure}

\begin{figure}
 \centering

 \subfloat[Rate coefficient $\mathrm{\, [cm^3 \cdot mol^{-1} \cdot s^{-1}]}$]{\label{fig:N2-N_diss_arrhenius_fits2}
 \includegraphics[width=0.49\textwidth]{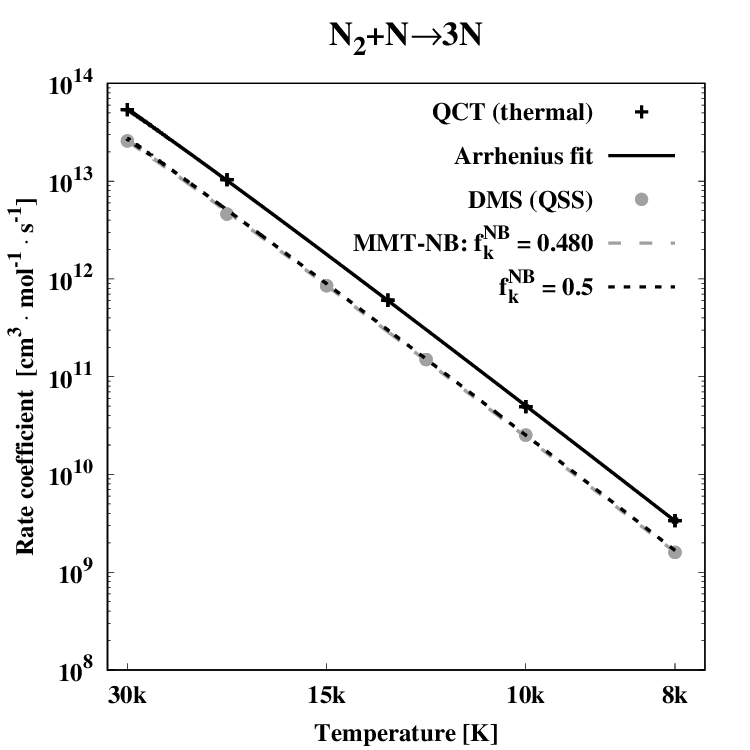}}~
 \subfloat[Vibrational energy change per reaction $\mathrm{\, [eV]}$]{\label{fig:mmt_evib_fit_N2-N_diss2}
 \includegraphics[width=0.49\textwidth]{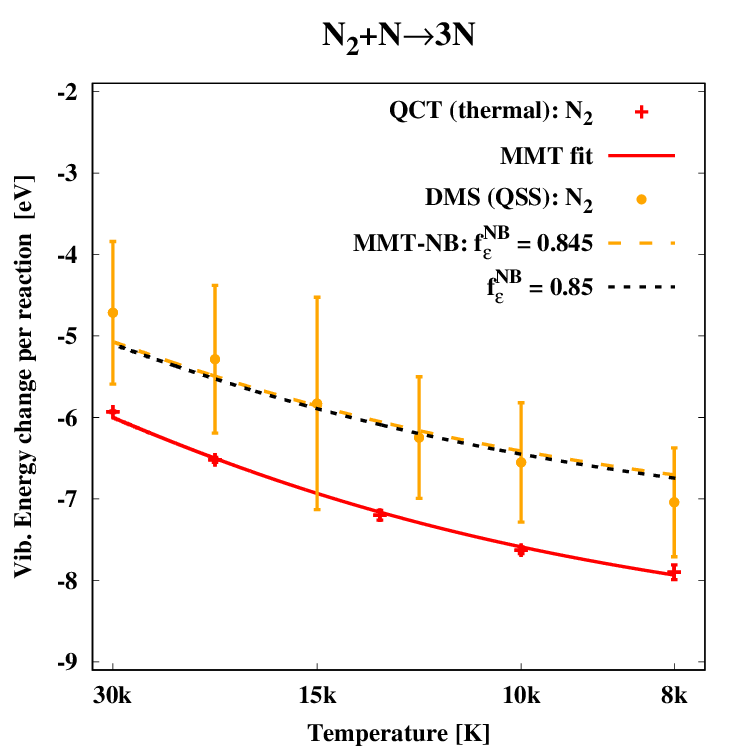}}
 
 \caption{$\boldsymbol{\mathrm{N_2 + N \rightleftharpoons 3N}}$}
 \label{fig:N2-N_diss}
\end{figure}

\begin{figure}
 \centering

 \subfloat[Rate coefficient $\mathrm{\, [cm^3 \cdot mol^{-1} \cdot s^{-1}]}$]{\label{fig:N2-O_diss_arrhenius_fits}
 \includegraphics[width=0.49\textwidth]{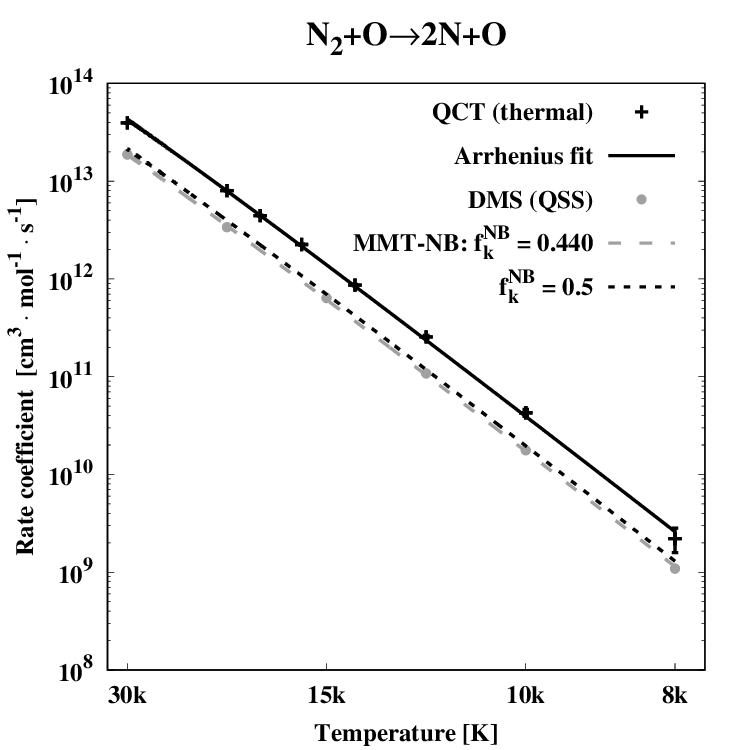}}~
 \subfloat[Vibrational energy change per reaction $\mathrm{\, [eV]}$]{\label{fig:mmt_evib_fit_N2-O_diss}
 \includegraphics[width=0.49\textwidth]{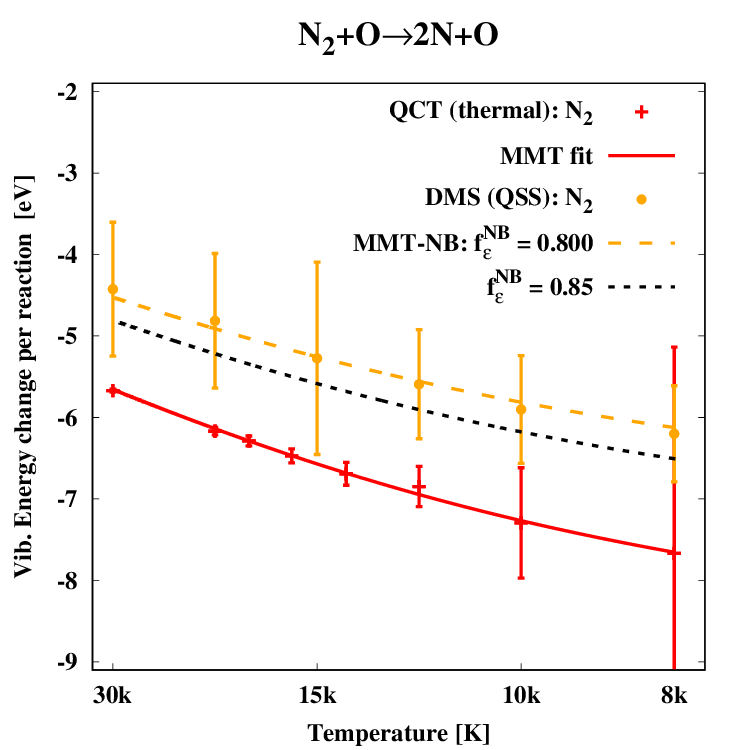}}
 
 \caption{$\boldsymbol{\mathrm{N_2 + O \rightleftharpoons 2N + O}}$}
 \label{fig:N2_O_diss}
\end{figure}

\begin{figure}
 \centering
 
 \subfloat[Rate coefficient $\mathrm{\, [cm^3 \cdot mol^{-1} \cdot s^{-1}]}$]{\label{fig:N2O2_O2_Diss_arrhenius_fits}
 \includegraphics[width=0.49\textwidth]{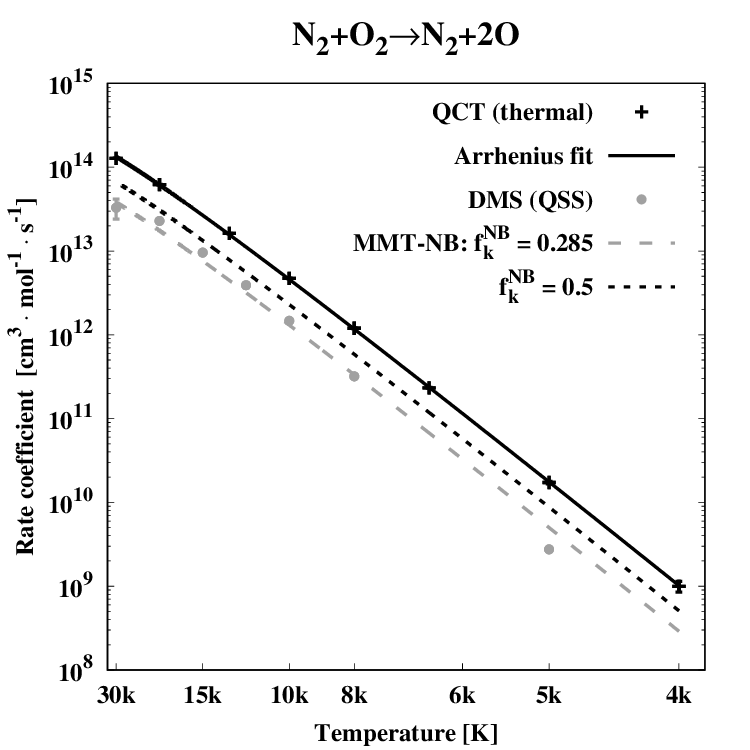}}~
 \subfloat[Vibrational energy change per reaction $\mathrm{\, [eV]}$]{\label{fig:mmt_evib_fit_N2O2_O2_diss}
 \includegraphics[width=0.49\textwidth]{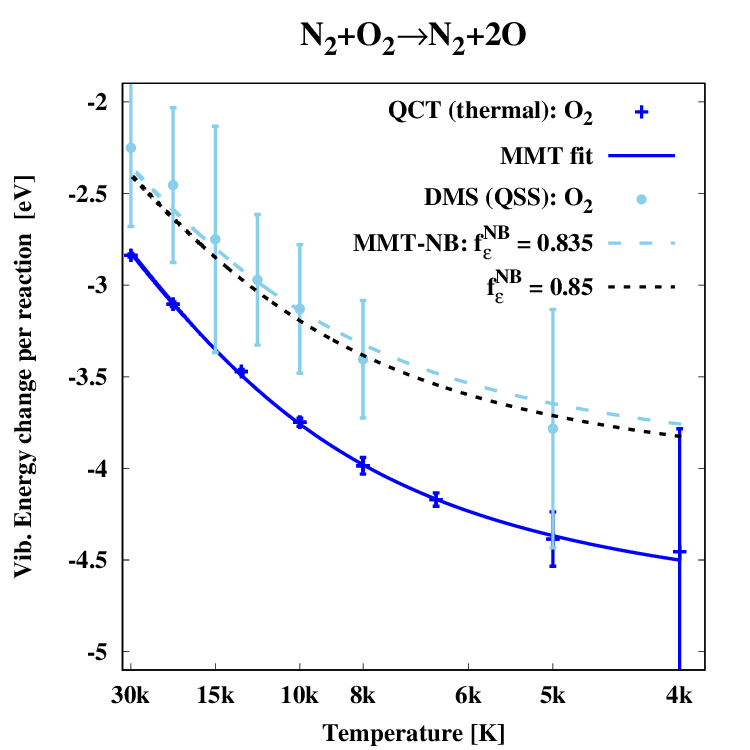}}
 
 \caption{$\boldsymbol{\mathrm{N_2 + O_2 \rightleftharpoons N_2 + 2O}}$}
 \label{fig:N2O2_O2_Diss}
\end{figure}

\begin{figure}
 \centering
 
 \subfloat[Rate coefficient $\mathrm{\, [cm^3 \cdot mol^{-1} \cdot s^{-1}]}$]{\label{fig:O2-O2_diss_arrhenius_fits2}
 \includegraphics[width=0.49\textwidth]{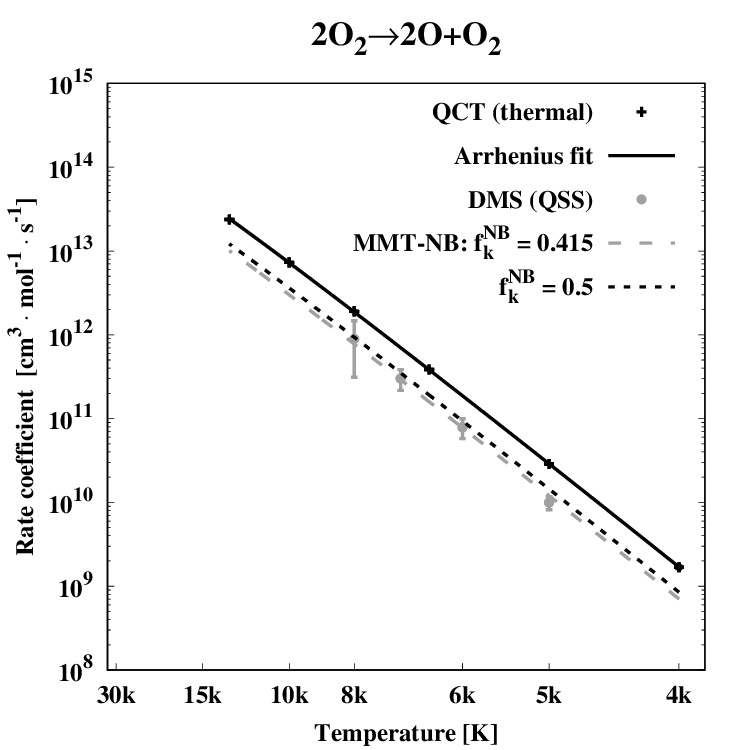}}~
 \subfloat[Vibrational energy change per reaction $\mathrm{\, [eV]}$]{\label{fig:mmt_evib_fit_O2-O2_diss2}
 \includegraphics[width=0.49\textwidth]{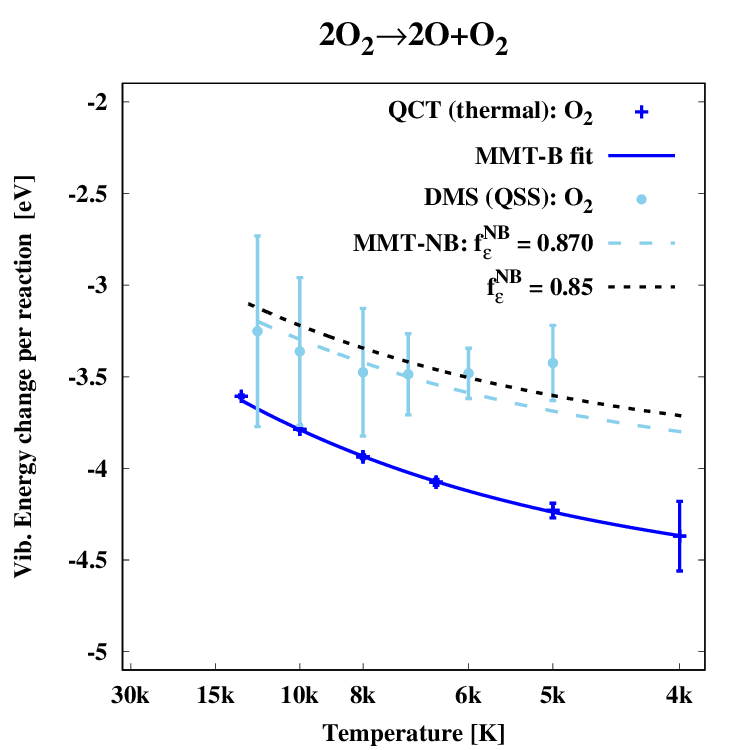}}
 
 \caption{$\boldsymbol{\mathrm{2O_2 \rightleftharpoons 2O + O_2}}$}
 \label{fig:O2-O2_diss}
\end{figure}

\begin{figure}
 \centering

 \subfloat[Rate coefficient $\mathrm{\, [cm^3 \cdot mol^{-1} \cdot s^{-1}]}$]{\label{fig:O2-N_diss_arrhenius_fits}
 \includegraphics[width=0.49\textwidth]{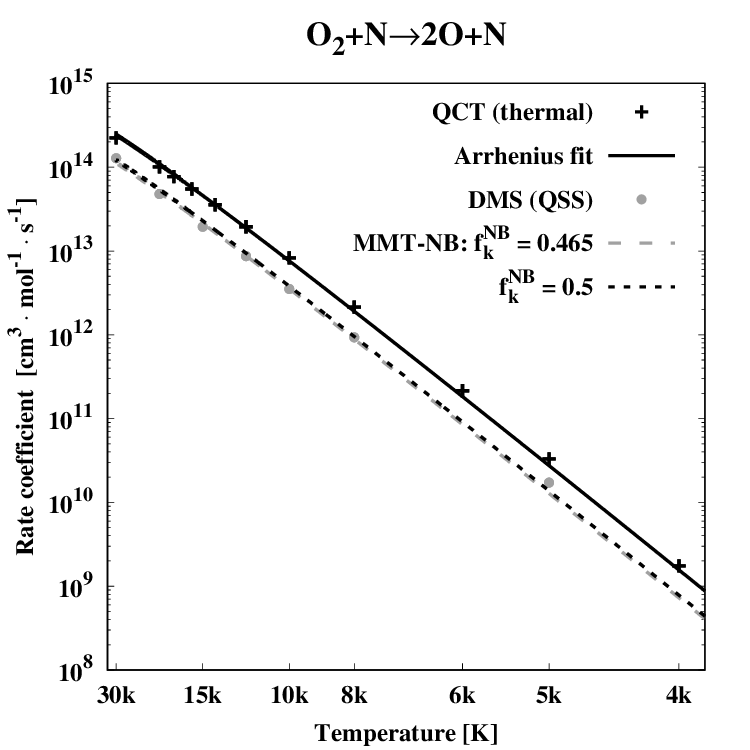}}~
 \subfloat[Vibrational energy change per reaction $\mathrm{\, [eV]}$]{\label{fig:mmt_evib_fit_O2-N_diss}
 \includegraphics[width=0.49\textwidth]{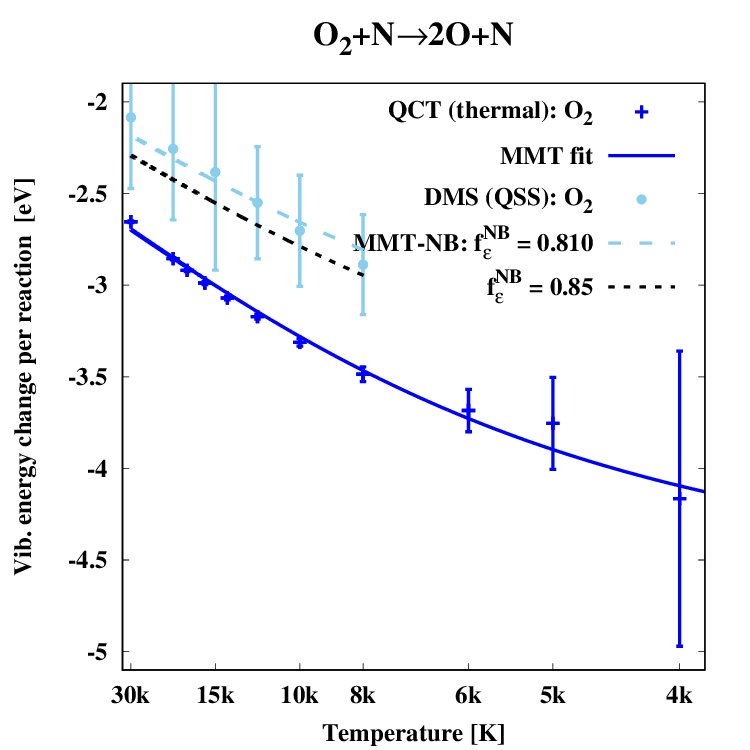}}
 
 \caption{$\boldsymbol{\mathrm{O_2 + N \rightleftharpoons 2O + N}}$}
 \label{fig:O2_N_diss}
\end{figure}

\begin{figure}
 \centering

 \subfloat[Rate coefficient $\mathrm{\, [cm^3 \cdot mol^{-1} \cdot s^{-1}]}$]{\label{fig:O2-O_diss_arrhenius_fits2}
 \includegraphics[width=0.49\textwidth]{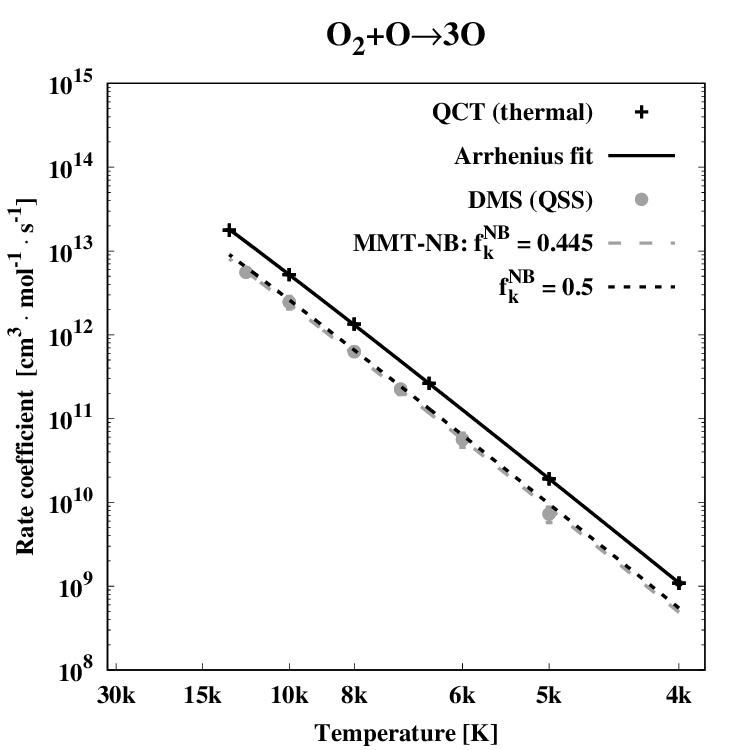}}~
 \subfloat[Vibrational energy change per reaction $\mathrm{\, [eV]}$]{\label{fig:mmt_evib_fit_O2-O_diss2}
 \includegraphics[width=0.49\textwidth]{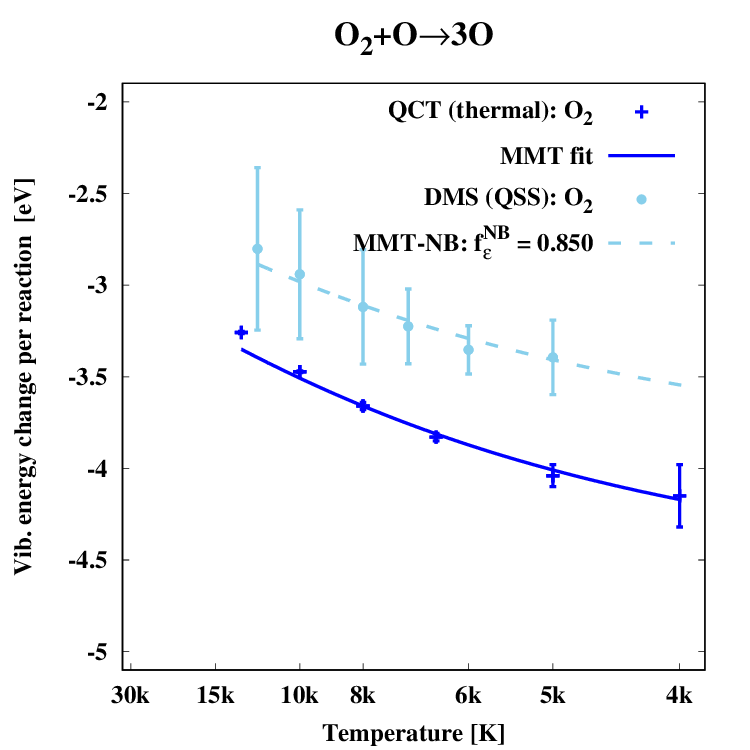}}
  
 \caption{$\boldsymbol{\mathrm{O_2 + O \rightleftharpoons 3O}}$}
 \label{fig:O2_O_diss}
\end{figure}

\begin{figure}
 \centering
 
 \subfloat[Rate coefficient $\mathrm{\, [cm^3 \cdot mol^{-1} \cdot s^{-1}]}$]{\label{fig:NONO_Diss_combined_arrhenius_fits}
 \includegraphics[width=0.49\textwidth]{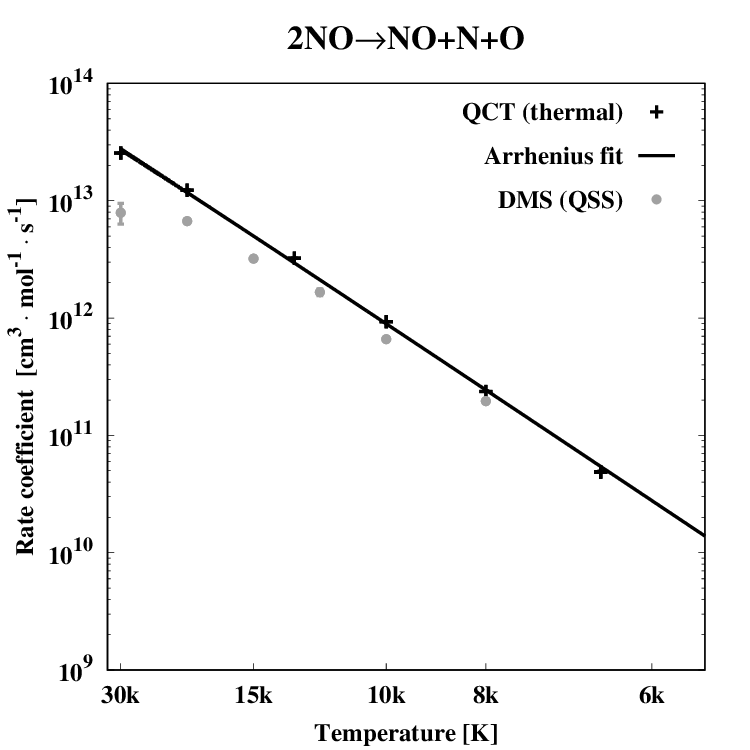}}~
 \subfloat[Vibrational energy change per reaction $\mathrm{\, [eV]}$]{\label{fig:mmt_evib_fit_NONO_Diss_combined}
 \includegraphics[width=0.49\textwidth]{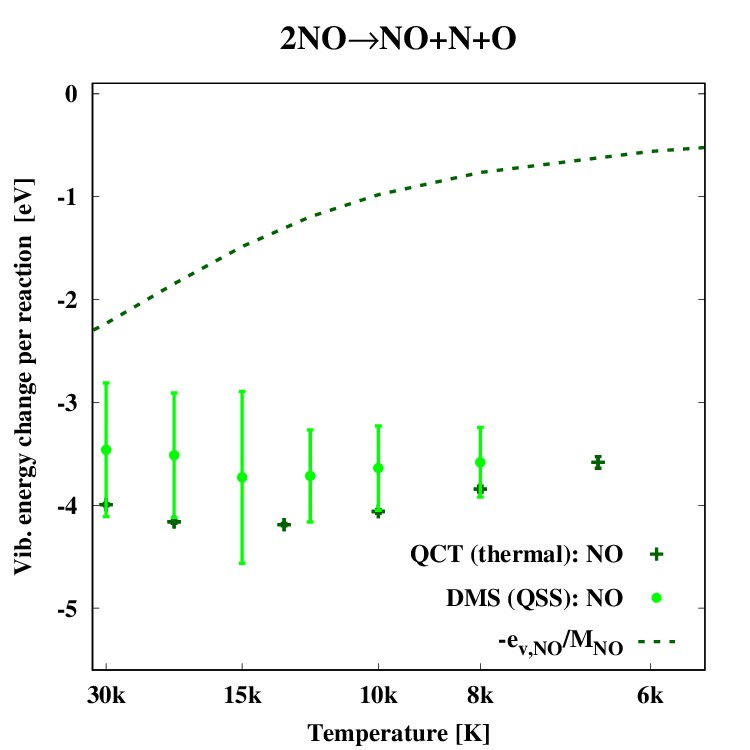}}
 
 \caption{$\boldsymbol{\mathrm{2NO \rightleftharpoons NO + N + O}}$}
 \label{fig:NONO_Diss}
\end{figure}

\begin{figure}
 \centering
 
 \subfloat[Rate coefficient $\mathrm{\, [cm^3 \cdot mol^{-1} \cdot s^{-1}]}$]{\label{fig:NO-N_diss_arrhenius_fits}
 \includegraphics[width=0.49\textwidth]{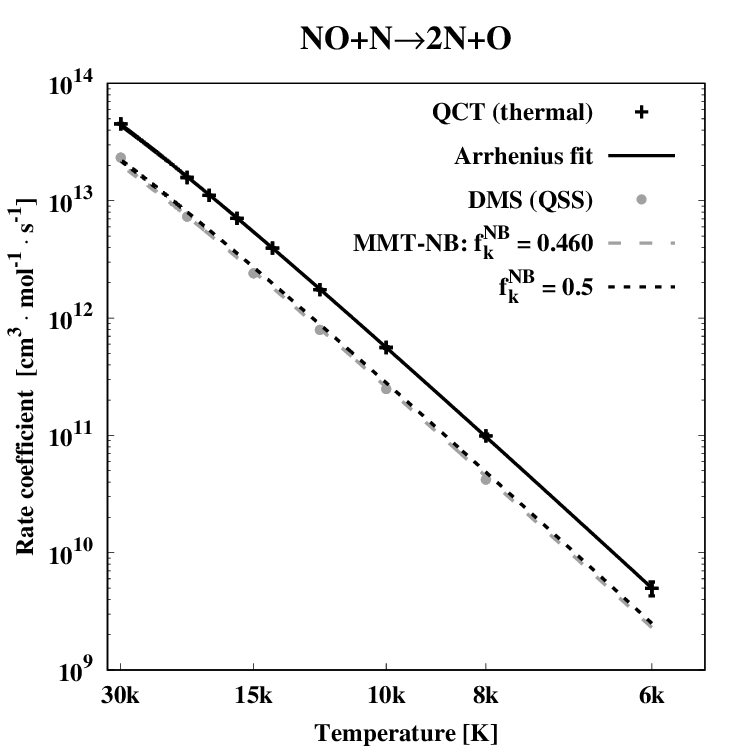}}~
 \subfloat[Vibrational energy change per reaction $\mathrm{\, [eV]}$]{\label{fig:mmt_evib_fit_NO-N_diss}
 \includegraphics[width=0.49\textwidth]{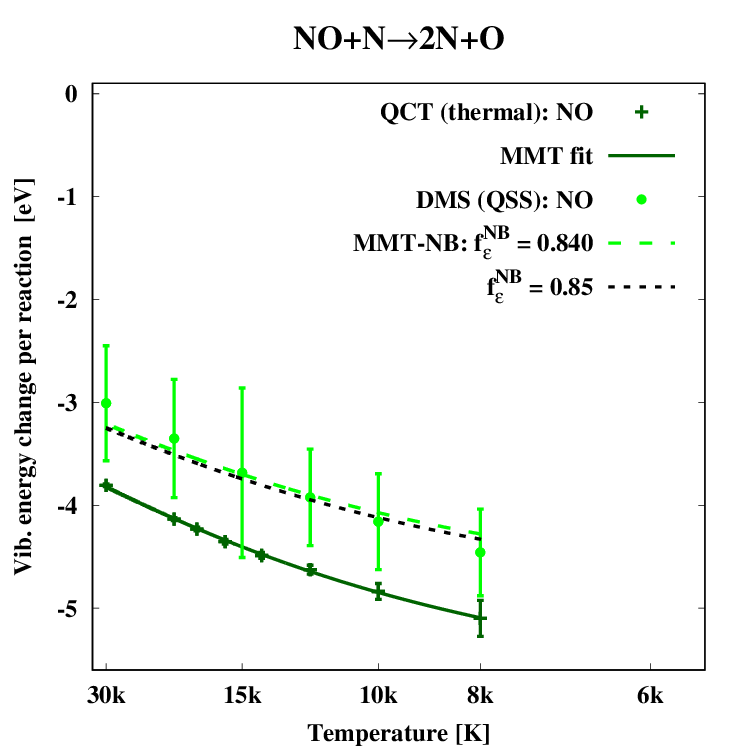}}
 
 \caption{$\boldsymbol{\mathrm{NO + N \rightleftharpoons 2N + O}}$}
 \label{fig:NO-N_diss}
\end{figure}

\begin{figure}
 \centering
 
 \subfloat[Rate coefficient $\mathrm{\, [cm^3 \cdot mol^{-1} \cdot s^{-1}]}$]{\label{fig:NO-O_diss_arrhenius_fits}
 \includegraphics[width=0.49\textwidth]{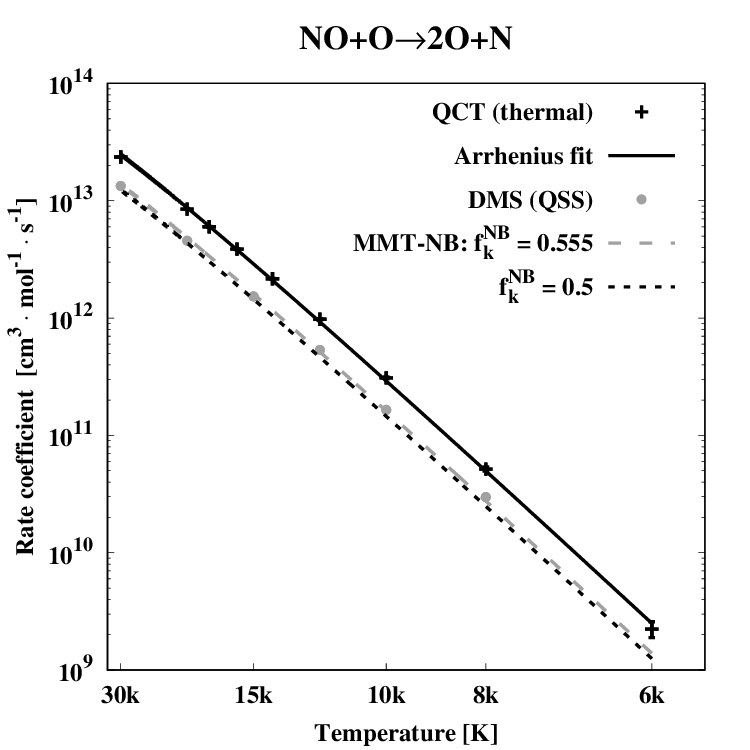}}~
 \subfloat[Vibrational energy change per reaction $\mathrm{\, [eV]}$]{\label{fig:mmt_evib_fit_NO-O_diss}
 \includegraphics[width=0.49\textwidth]{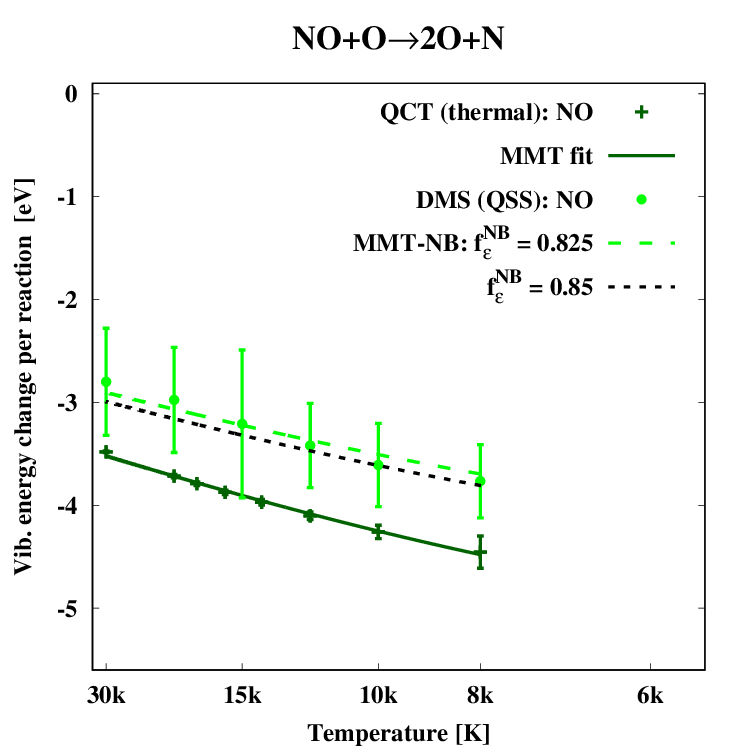}}
 
 \caption{$\boldsymbol{\mathrm{NO + O \rightleftharpoons 2O + N}}$}
 \label{fig:NO-O_diss}
\end{figure}

\begin{figure}
 \centering
 
 \subfloat[Rate coefficient $\mathrm{\, [cm^3 \cdot mol^{-1} \cdot s^{-1}]}$]{\label{fig:N2-O_exch_arrhenius_fits}
 \includegraphics[width=0.49\textwidth]{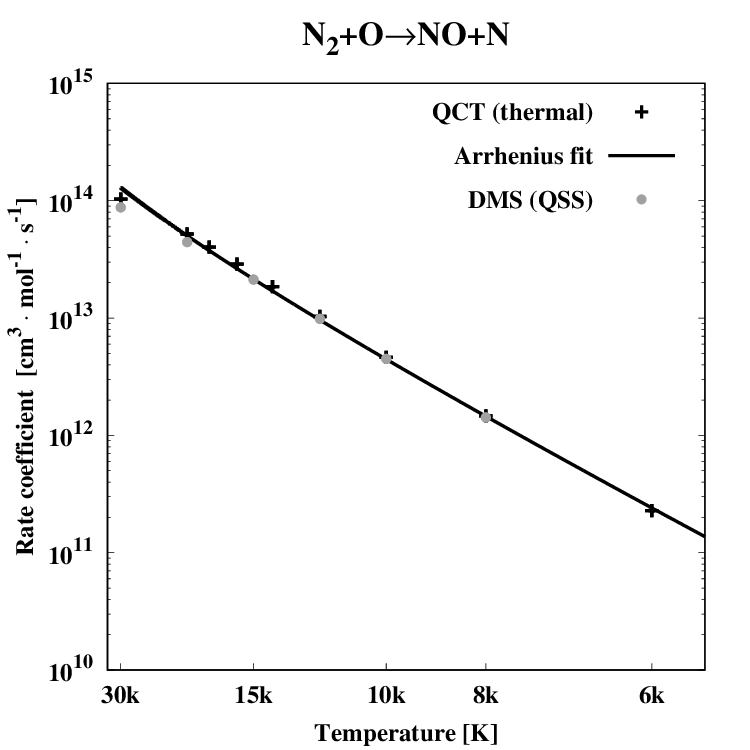}}~
 \subfloat[Vibrational energy change per reaction $\mathrm{\, [eV]}$]{\label{fig:mmt_evib_fit_N2-O_exch}
 \includegraphics[width=0.49\textwidth]{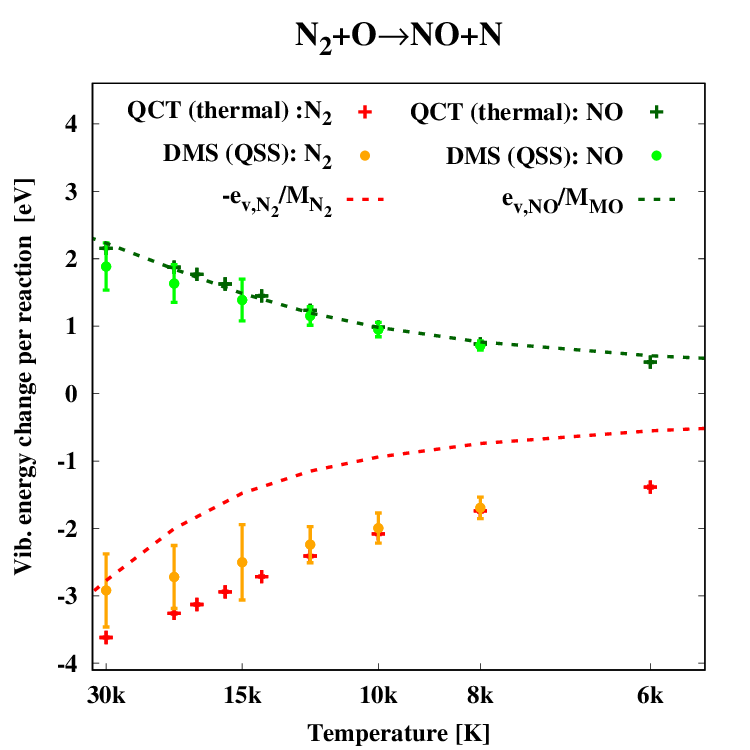}}
 
 \caption{$\boldsymbol{\mathrm{N_2 + O \rightleftharpoons NO + N}}$}
 \label{fig:N2-O_exch}
\end{figure}

\begin{figure}
 \centering
 
 \subfloat[Rate coefficient $\mathrm{\, [cm^3 \cdot mol^{-1} \cdot s^{-1}]}$]{\label{fig:NO-O_exch_arrhenius_fits}
 \includegraphics[width=0.49\textwidth]{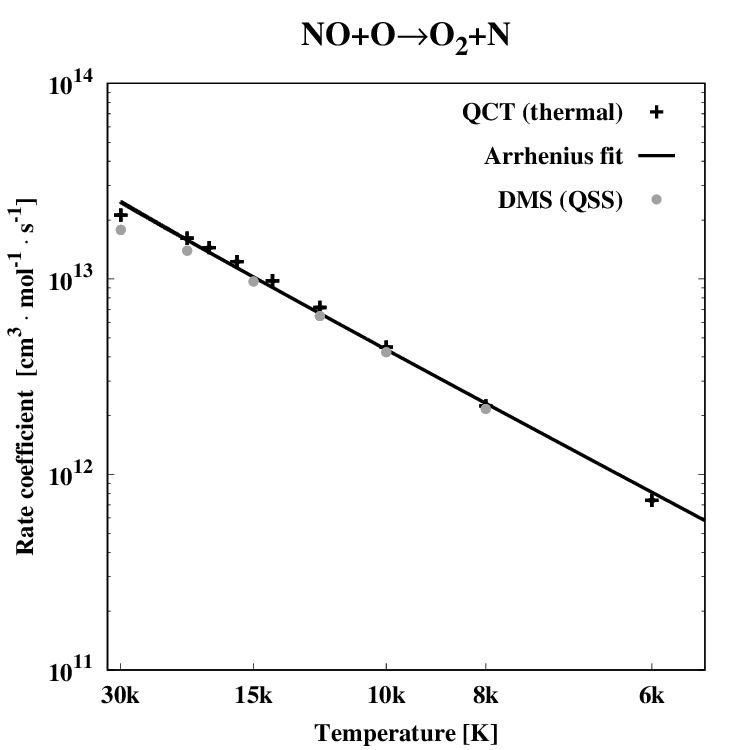}}~
 \subfloat[Vibrational energy change per reaction $\mathrm{\, [eV]}$]{\label{fig:mmt_evib_fit_NO-O_exch}
 \includegraphics[width=0.49\textwidth]{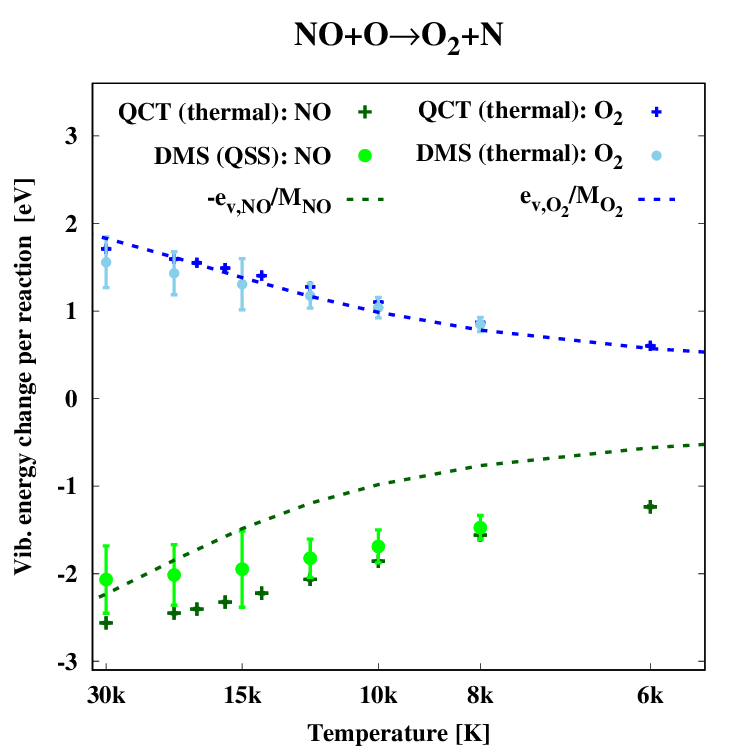}}
 
 \caption{$\boldsymbol{\mathrm{NO + O \rightleftharpoons O_2 + N}}$}
 \label{fig:NO-O_exch}
\end{figure}

\begin{figure}
 \centering
 
 \subfloat[Rate coefficient $\mathrm{\, [cm^3 \cdot mol^{-1} \cdot s^{-1}]}$]{\label{fig:N2O2_Double_NO_Form_arrhenius_fits}
 \includegraphics[width=0.49\textwidth]{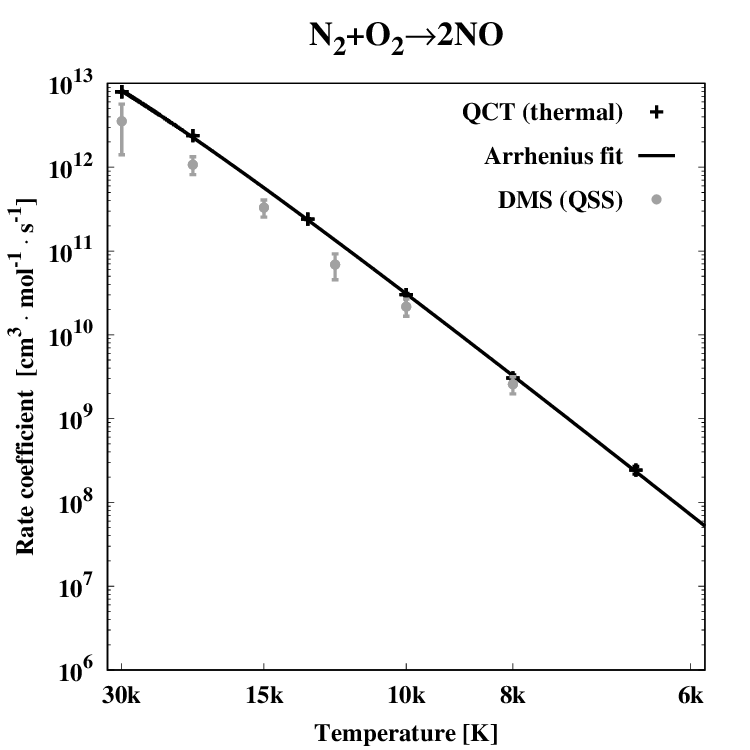}}~
 \subfloat[Vibrational energy change per reaction $\mathrm{\, [eV]}$]{\label{fig:mmt_evib_fit_N2O2_Double_NO_Form}
 \includegraphics[width=0.49\textwidth]{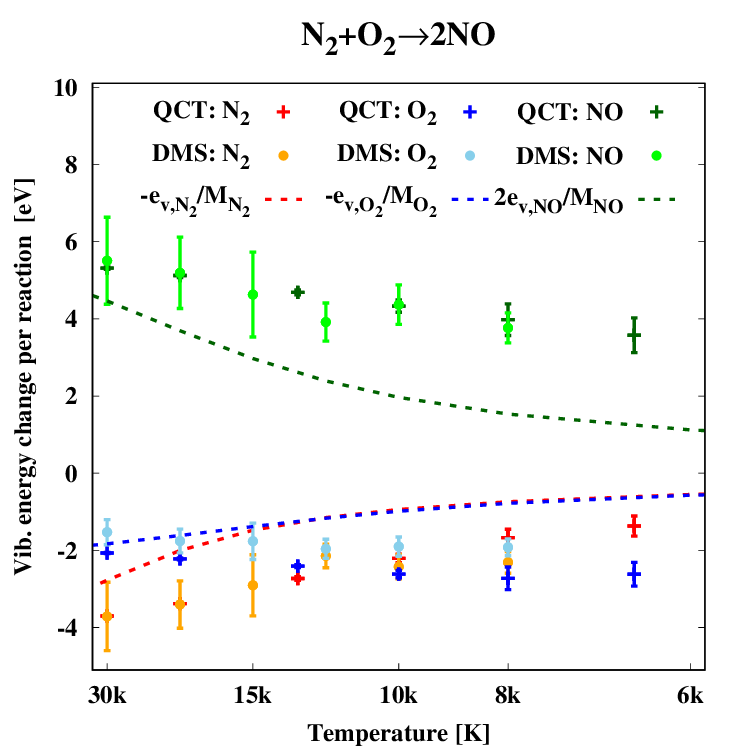}}
 
 \caption{$\boldsymbol{\mathrm{N_2 + O_2 \rightleftharpoons 2NO}}$}
 \label{fig:N2O2_Double_NO_Form}
\end{figure}

\begin{figure}
 \centering

 \subfloat[Rate coefficient $\mathrm{\, [cm^3 \cdot mol^{-1} \cdot s^{-1}]}$]{\label{fig:NONO_Diss_Exch_N2_Product_arrhenius_fits}
 \includegraphics[width=0.49\textwidth]{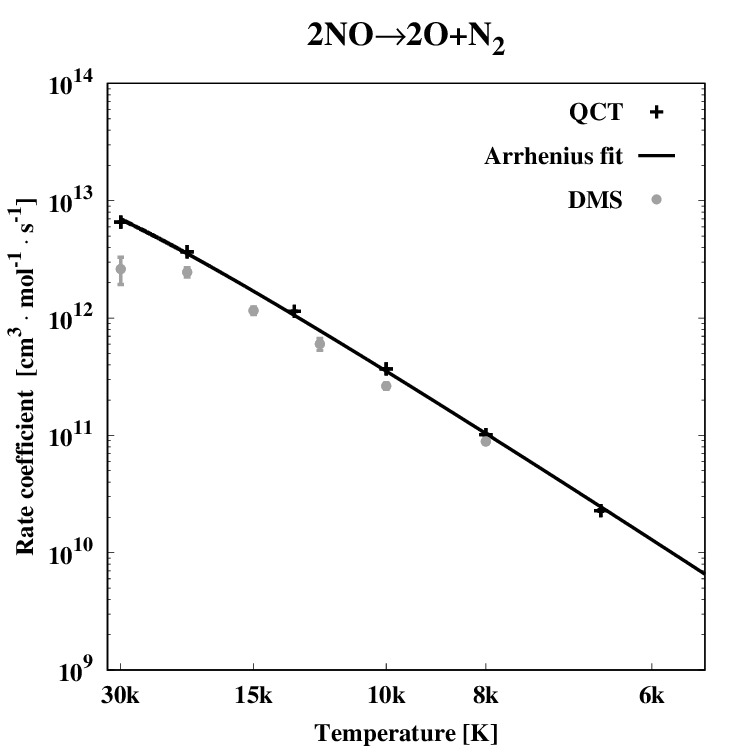}}~
 \subfloat[Vibrational energy change per reaction $\mathrm{\, [eV]}$]{\label{fig:mmt_evib_fit_NONO_Diss_Exch_N2_Product}
 \includegraphics[width=0.49\textwidth]{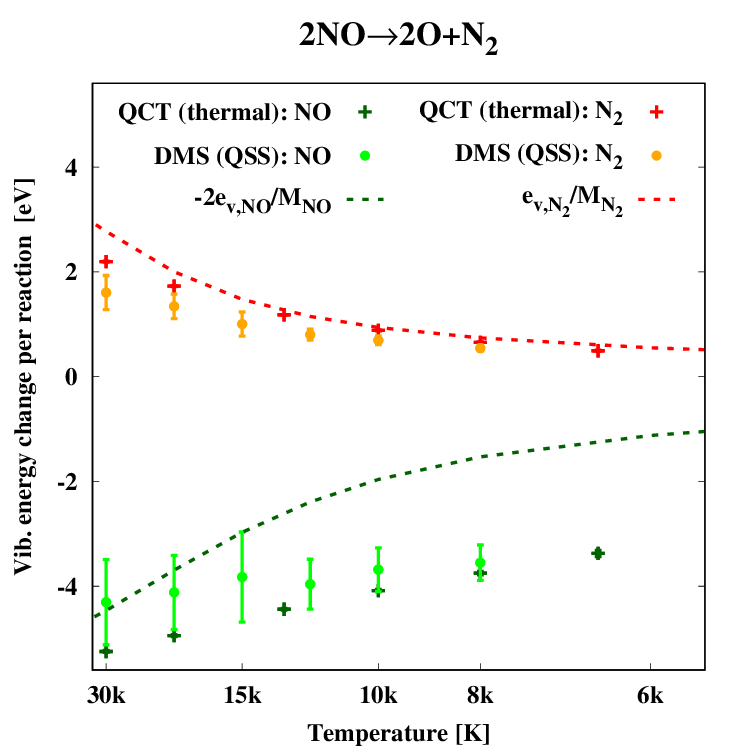}}
 
 \caption{$\boldsymbol{\mathrm{2NO \rightleftharpoons 2O + N_2}}$}
 \label{fig:NONO_Diss_Exch_N2_Product}
\end{figure}

\begin{figure}
 \centering

 \subfloat[Rate coefficient $\mathrm{\, [cm^3 \cdot mol^{-1} \cdot s^{-1}]}$]{\label{fig:NONO_Diss_Exch_O2_Product_arrhenius_fits}
 \includegraphics[width=0.49\textwidth]{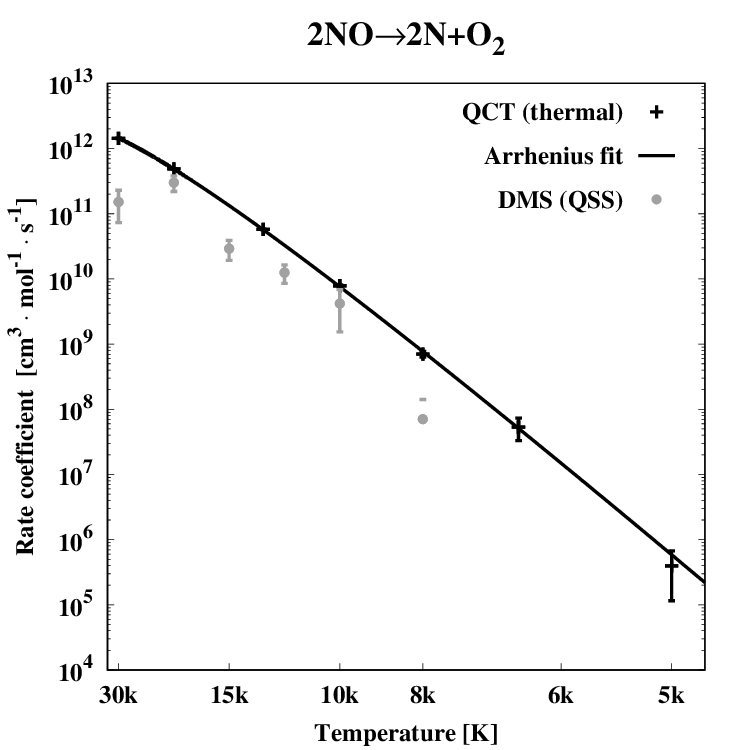}}~
 \subfloat[Vibrational energy change per reaction $\mathrm{\, [eV]}$]{\label{fig:mmt_evib_fit_NONO_Diss_Exch_O2_Product}
 \includegraphics[width=0.49\textwidth]{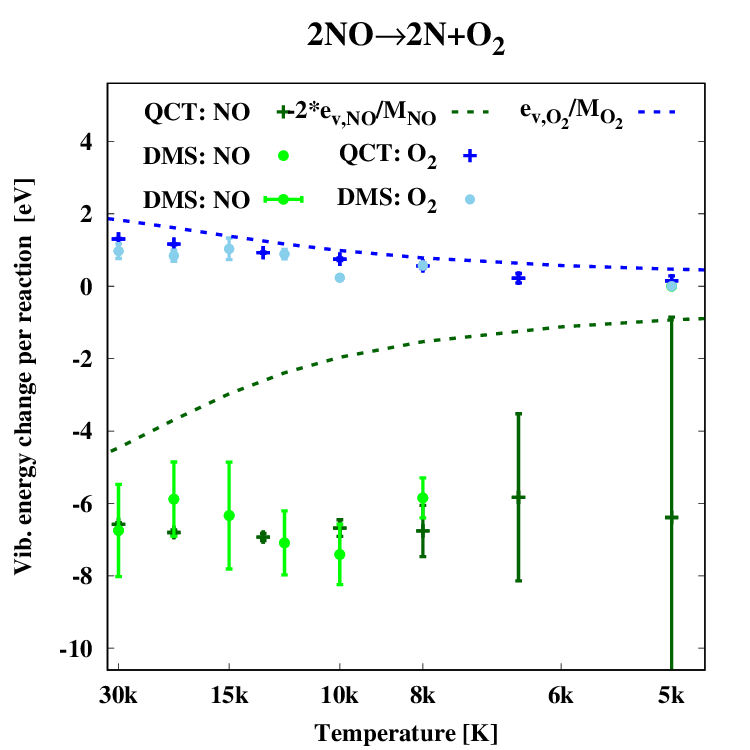}}

 \caption{$\boldsymbol{\mathrm{2NO \rightleftharpoons 2N + O_2}}$}
 \label{fig:NONO_Diss_Exch_O2_Product}
\end{figure}

\begin{figure}
 \centering

 \subfloat[Rate coefficient $\mathrm{\, [cm^3 \cdot mol^{-1} \cdot s^{-1}]}$]{\label{fig:N2O2_NO_Form_arrhenius_fits}
 \includegraphics[width=0.49\textwidth]{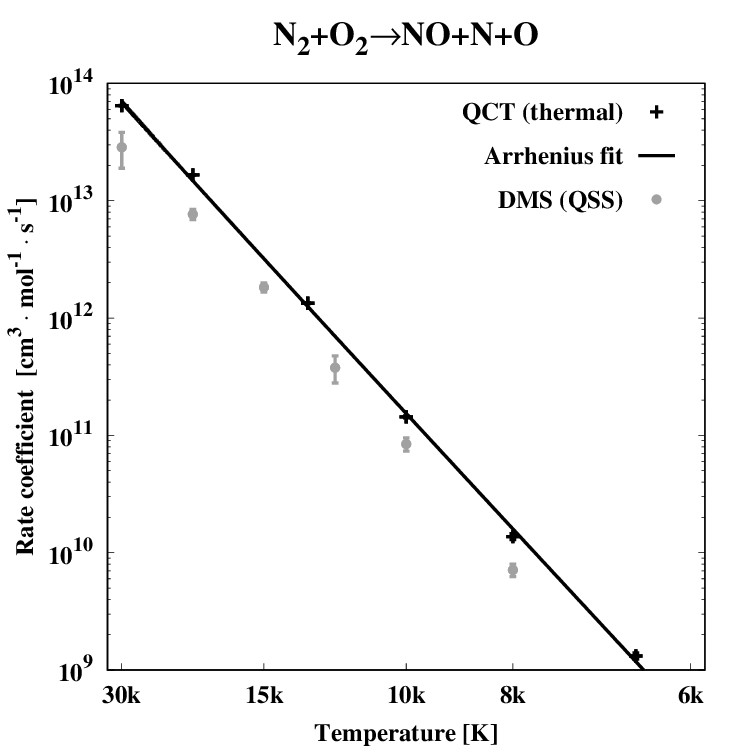}}~
 \subfloat[Vibrational energy change per reaction $\mathrm{\, [eV]}$]{\label{fig:mmt_evib_fit_N2O2_NO_Form}
 \includegraphics[width=0.49\textwidth]{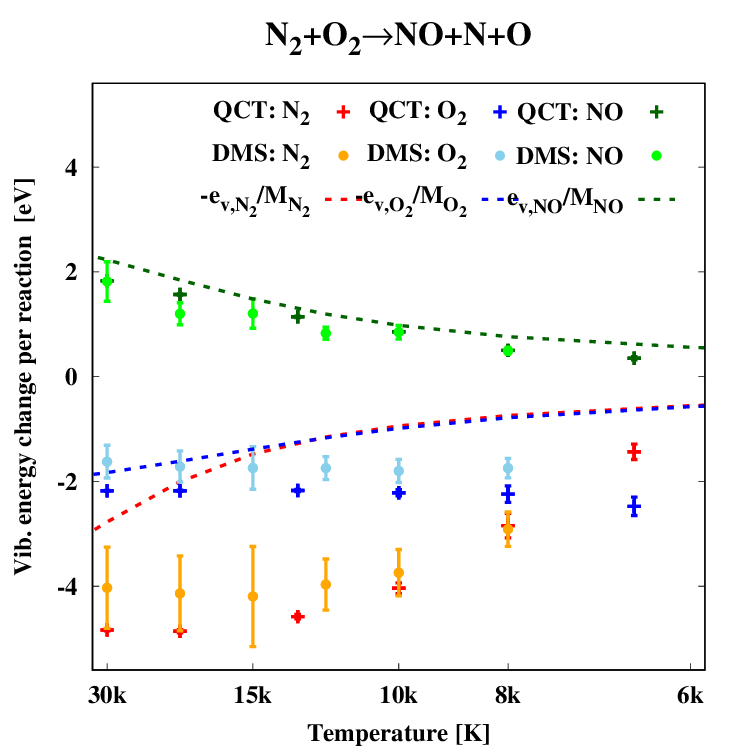}}

 \caption{$\boldsymbol{\mathrm{N_2 + O_2 \rightleftharpoons NO + N + O}}$}
 \label{fig:N2O2_NO_Form}
\end{figure}

This section contains a pair of plots for every reaction for which ab initio QCT and DMS data are available, see Tables~1-3 of the main article. For each reaction, sub-figure (a) shows an Arrhenius plot with the rate coefficient and sub-figure (b) the corresponding vibrational energy \emph{change} per reaction, i.e $\langle \Delta \varepsilon_{\mathrm{v},s} \rangle_r = (\nu_{s,r}^b - \nu_{s,r}^f) \, \langle \varepsilon_{\mathrm{v},s} \rangle_r$. These figures are listed in the same order as in the tables, with plots for the four available nitrogen dissociation reactions appearing first (reactions 1,2,4 and 5) in Figs.~\ref{fig:N2-N2_diss}-\ref{fig:N2_O_diss}, followed by those for the four available oxygen dissociation reactions (reactions 6,7,9 and 10) in Figs.~\ref{fig:N2O2_O2_Diss}-\ref{fig:O2_O_diss} and those for the three available nitric oxide dissociation reactions (reactions 13,14 and 15) shown in Figs.~\ref{fig:NONO_Diss}-\ref{fig:NO-O_diss}. For the three exchange-type reactions (reactions 16-18) the plots are shown in Figs.~\ref{fig:N2-O_exch}-\ref{fig:N2O2_Double_NO_Form} and plots for the remaining three, ``mixed dissociation-exchange'' type reactions (reactions 19-21) are shown in Figs.~\ref{fig:NONO_Diss_Exch_N2_Product}-\ref{fig:N2O2_NO_Form}. 

The QCT-derived thermal-equilibrium rate coefficients in sub-figures (a) are plotted as black crosses with their respective error bars representing one standard deviation. A continuous black line represents the modified Arrhenius fit, evaluated at temperature $T$ according to Eq.~(12) of the main article and the parameters in Table~2 of the main article. Gray circles represent the DMS-derived rate coefficients extracted during the QSS dissociation regime. For most of the dissociation-type reactions these values lie noticeably and consistently below their thermal-equilibrium QCT counterparts. The error bars correspond to the standard error, calculated as described in Sec.~IV of Ref.~\cite{torres24a}. For those dissociation reactions handled by the MMT functional form (again, see Tables~1-3 of the main article), a gray dashed line marks the MMT rate coefficient evaluated according to Eq.~(11) of the main article when $T_\mathrm{v}\!=\!T$ and the reaction-specific non-Boltzmann factor $f_k^\mathrm{NB}$ from Table~3 of the main article is used. A corresponding dotted black line represents the same rate coefficient, but evaluated for $f_k^\mathrm{NB} = 0.5$. These last two lines are absent in Arrhenius plots for reactions not handled by the MMT functional form.

In sub-figures (b), the QCT-derived vibrational energy change per reaction event at thermal-equilibrium temperature $T$ is plotted using red, dark blue and dark green crosses for $\mathrm{N_2}$, $\mathrm{O_2}$ and $\mathrm{NO}$ respectively. This quantity will be negative if the diatomic species in question is being destroyed, or positive if it is being produced as a result of the given reaction. Depending on the reaction in question, the vibrational energy change for multiple species may appear in these plots. Corresponding DMS-derived values for vibrational energy change during the QSS dissociation regime are plotted using orange, light blue and light green circles for $\mathrm{N_2}$, $\mathrm{O_2}$ and $\mathrm{NO}$ respectively. For those reactions whose vibrational energy change is evaluated by the MMT functional form, continuous lines labeled ``MMT-fit'' show the result of evaluating $\langle \varepsilon_{\mathrm{v},s} \rangle_r$ according to Eq.~(19) in the main article at $T_\mathrm{v}\!=\!T$, together with the parameters from Table~3 of the main article, but with the non-Boltzmann factor $f_\varepsilon^\mathrm{NB}$ set to unity. For these reactions two additional lines are shown. The dashed orange/light blue/light green lines represent evaluating $\langle \varepsilon_{\mathrm{v},s} \rangle_r$ using the same MMT functional form and parameters just mentioned, except that now the reaction-specific non-Boltzmann factor from Table~3 of the main article is used. This yields the closest fit to the available DMS data. The third line, dotted red/dark blue or dark green respectively represents the same MMT curve, but using the common non-Boltzmann factor of 0.85. The latter is the one actually employed in all the MMT calculations presented in the main article. For all other reactions, i.e. where $\langle \varepsilon_{\mathrm{v},s} \rangle_r$ was not fit to the MMT functional form, the three lines just mentioned are absent. Instead, the dotted red/dark blue/dark green line(s) appearing in those plots represent the vibrational energy change estimate for non-preferential coupling, i.e. $\langle \Delta \varepsilon_{\mathrm{v},s} \rangle_r^\mathrm{non-pref.} \approx (\nu_{s,r}^b - \nu_{s,r}^f) \, e_{\mathrm{v},s} / M_s$. Non-preferential coupling is assumed for all reactions that did not exhibit the characteristic vibrational bias of the main dissociation reactions.

\clearpage

 \bibliography{rgd}
 
\end{document}